\documentclass[11pt,preprintnumbers,aps,amssymb,nofootinbib,amsmath,superscriptaddress]
{revtex4}
\usepackage{epsfig,epsf}
\usepackage{bm} 
\newcommand{\beq}{\begin{equation}}
\newcommand{\beql}[1]{\begin{equation}\label{#1}}
\newcommand{\eeq}{\end{equation}}
\newcommand{\eq}[1]{(\ref{#1})}
\newcommand{\fig}[1]{Fig.~\ref{#1}}
\renewcommand{\sec}[1]{Sec.~\ref{#1}}
\newcounter{topiccounter}
\renewcommand{\b}[1]{{\bm #1}} 
\newcommand{\unit}[1]{\hat {{\bm #1}}} 

\newcommand{\as}{\alpha_s}

\newcommand{\e}{\varepsilon}
\newcommand{\aver}[1]{\left\langle #1 \right\rangle}
\newcommand{\jpsi}{J\mskip -2mu/\mskip -0.5mu\psi}

\begin{document}

\title{Particle production in strong electromagnetic fields in relativistic heavy-ion collisions}

\author{Kirill Tuchin}

\affiliation{
Department of Physics and Astronomy, Iowa State University, Ames, IA 50011}

\date{\today}


\begin{abstract}
I review the origin and properties of electromagnetic fields produced in heavy ion collisions. The field strength immediately after a collision  is proportional to the collision energy and reaches $\sim m_\pi^2$ at RHIC and $\sim 10 m_\pi^2$ at LHC. I demonstrate by explicit analytical calculation that after dropping by about one--two orders of magnitude during the first fm/c of plasma expansion, it freezes out and 
lasts for as long as quark-gluon plasma lives as a consequence of finite electrical conductivity of the plasma. 
Magnetic field breaks spherical symmetry in the direction perpendicular to the reaction  plane and therefore all kinetic coefficients are anisotropic. I examine viscosity of QGP  and show that magnetic field induces azimuthal anisotropy on plasma flow even in spherically symmetric geometry. 
Very strong electromagnetic field has an important impact  on particle production. I discuss the problem of energy loss and polarization of fast fermions due to synchrotron radiation, consider  photon decay induced by magnetic field,  elucidate $\jpsi$ dissociation via Lorentz ionization mechanism and examine electromagnetic radiation by plasma. I conclude that \emph{all} processes in QGP are affected by strong electromagnetic field and call for experimental investigation.

\end{abstract}

\maketitle

\newpage

\tableofcontents

\setcounter{equation}{0}
\newpage
\section{Origin and properties of electromagnetic field}\label{sec:a}

\subsection{Origin of magnetic field}\label{sec:aa}

We can understand the origin of magnetic field in heavy-ion collisions by considering collision of two ions of radius $R$ with electric charge $Ze$ ($e$ is the magnitude of electron charge) at impact parameter $\b b$. 
According to the Biot and Savart law they create magnetic field that in the center-of-mass frame has magnitude 
\beql{aa0}
B\sim \gamma \, Ze\, \frac{b}{R^3}\,
\eeq
and points in the direction perpendicular to the reaction plane (span by the momenta of ions). Here $\gamma=\sqrt{s_{NN}}/2m_N$ is the Lorentz factor. At RHIC heavy-ions are collided at 200~GeV per nucleon, hence $\gamma=100$. Using $Z=79$ for Gold and $b\sim R_A\approx 7$~fm we estimate $eB\approx m_\pi^2\sim 10^{18}$~G. To appreciate how strong is this field, compare it with the following numbers: the strongest magnetic field created on Earth in a form of electromagnetic shock wave is $\sim 10^7$~G \cite{StrongestMagField}, magnetic field of a neutron star is estimated to be $10^{10}-10^{13}$~G, that of a magnetar up to $10^{15}$~G \cite{Kouveliotou:2003tb}. It is perhaps the strongest magnetic field that have ever existed in nature.

It has been known for a long time that classical electrodynamics breaks down at the critical (Schwinger) field strength $F= m_e^2/e$ \cite{Schwinger:1951nm}. In cgs units the corresponding magnetic field is $10^{13}$~G. Because $m_\pi/m_e= 280$, electromagnetic fields created at RHIC and LHC are well above the critical value. This offers a unique opportunity to study the super-strong electromagnetic fields in laboratory. The main challenge is to identify experimental observables that are sensitive to such fields. The problem is that nearly all observables studied in heavy-ion collisions are strongly affected  both by the strong color forces acting in quark-gluon plasma (QGP) \emph{and} by electromagnetic fields often producing qualitatively similar effects. An outstanding experimental problem thus is to separate the two effects. In sections \ref{sec:visc}--\ref{sec:y} I examine several processes strongly affected by intense magnetic fields and discuss their phenomenological significance. But first, in this section,  let me 
derive a quantitative estimate of electromagnetic field. 

\begin{figure}[ht]
      \includegraphics[height=4.5cm]{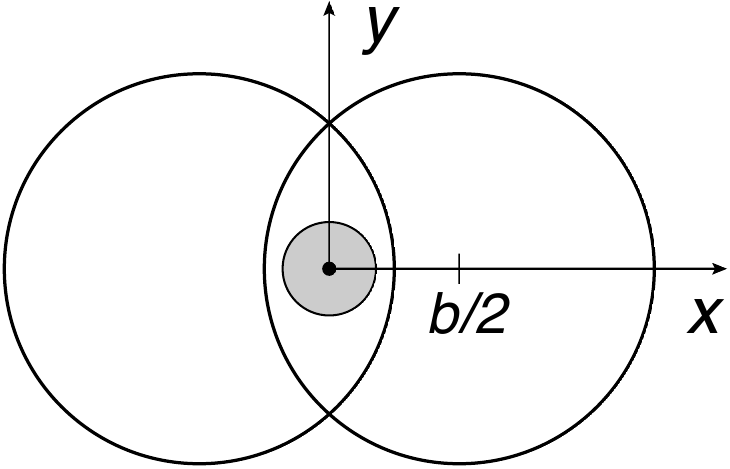} 
  \caption{ Heavy-ion collision geometry as seen along the collision axis $z$. Adapted from \cite{Bzdak:2011yy}.}
\label{fig:aa1}
\end{figure}

Throughout this article, the heavy-ion collision axis is denoted by  $z$. Average magnetic field then points in the $y$-direction, see \fig{fig:aa1} and \fig{geom}. Plane $xz$ is the reaction plane and $b$ is the impact parameter. 

\subsection{Magnetic field in vacuum}\label{ac}

\addtocontents{toc}{\protect\setcounter{tocdepth}{2}}

\subsubsection{Time-dependence}

To obtain a quantitative estimate of magnetic field we need to take into account a realistic distribution of protons in a nucleus. This has been first done in \cite{Kharzeev:2007jp}\footnote{In the case of high-energy $pp$ collisions, magnetic field was first estimated in \cite{Ambjorn:1990jg} who also pointed out a possibility of formation of $W$-condensate \cite{Ambjorn:1990jg,Olesen:2012zb}.}.
  Magnetic field at point $\b r$ created by two heavy ions moving in the positive or negative $z$-direction can be calculated using the Li\'enard-Wiechert potentials as follows
\begin{align}
e\b {E}(t,\b r) &=\alpha _{\mathrm{em}}\sum_{a}\frac{
(1-v_{a}^{2})\b R_a}{ R_a^{3}\left[ 1-(\b R_a\times \b{v}_{a})^{2}/R_a^{2}
\right] ^{3/2}}\,,\label{aa1} \\
e\b {B}(t,\b r) &=\alpha _{\mathrm{em}}\sum_{a}\frac{
(1-v_{a}^{2})(\b{v}_{a}\times \b R_a)}{ R_a^{3}\left[ 1-(\b R_a\times \b{v}_{a})^{2}/ R_a^{2}
\right]^{3/2}}\,,  \label{aa2}
\end{align}
with $\b R_a=\b r-\b r_{a}(t)$, where sums run over all $Z$ protons in each nucleus, their positions and velocities  being $\b r_{a}$ and $\b{v}_{a}$. The magnitude of velocity $v_a$ is determined by the
collision energy $\sqrt{s_{NN}}$ and the proton mass $m_{p}$, $v_{a}^{2}=1-\left(
2m_{p}/\sqrt{s_{NN}}\right) ^{2}$. These formulas are derived in the eikonal approximation, assuming that protons travel on straight lines before and after the scattering. This is a good approximation  since baryon stopping is  a small effect at high energies. 
Positions of protons in heavy-ions can be determined by one of the standard models of the nuclear charge density $\rho(\b r_a)$.  
Ref.\cite{Kharzeev:2007jp} employed the ``hard sphere" model, while 
\cite{Bzdak:2011yy} used a bit more realistic Wood-Saxon distribution.

\begin{figure}[ht]
\begin{tabular}{cc}
      \includegraphics[height=6.5cm]{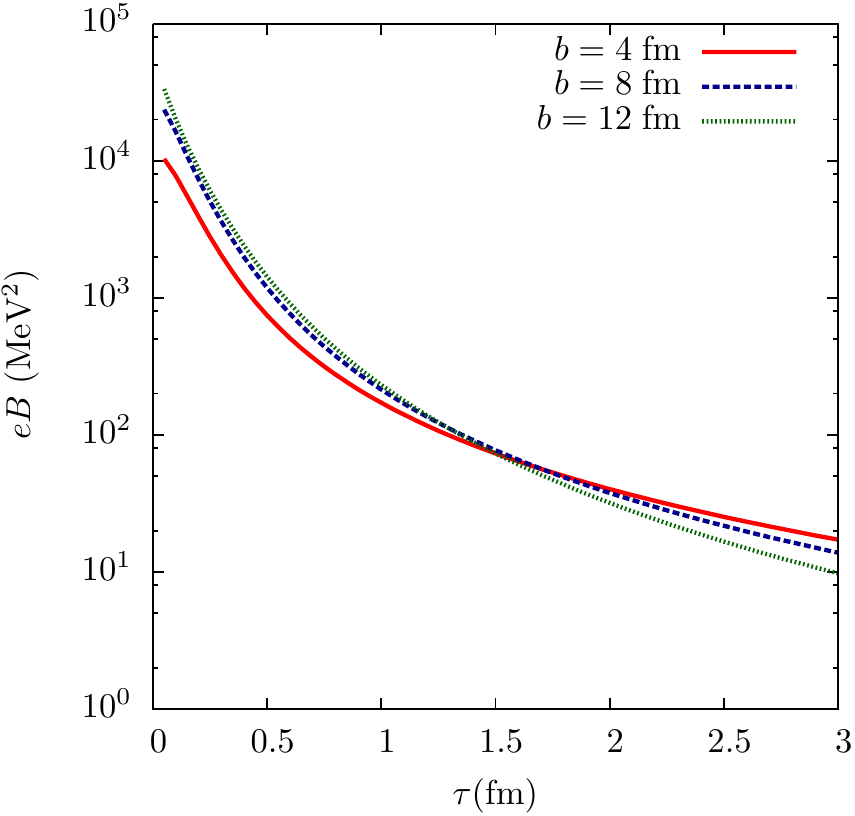} &
      \includegraphics[height=6.5cm]{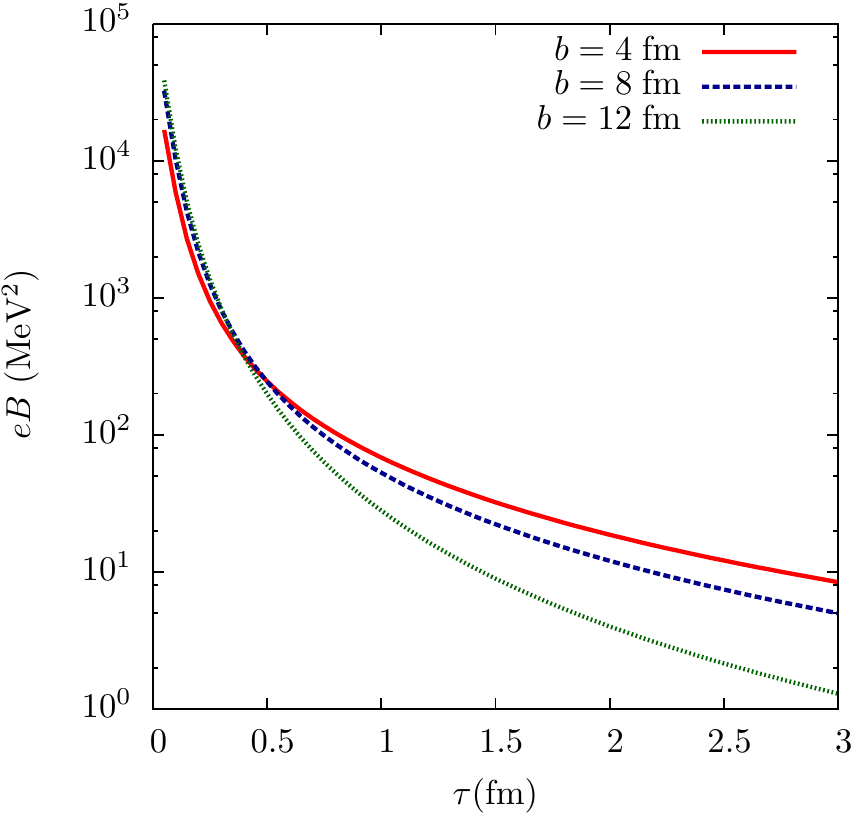}\\
      $(a)$ & $(b)$ 
      \end{tabular}
  \caption{Magnetic field $\b B= B\unit y$ (multiplied by $e$) at the origin  $ \b r = 0$ produced in collision of two gold ions  at beam energies (a) $\sqrt{s_{NN}}=62$~GeV and (b) $\sqrt{s_{NN}}=200$~GeV. Adapted from \cite{Kharzeev:2007jp}. Note, that $eB$  is the same in Gauss and Lorentz-Heaviside units in contrast to $B$.}
\label{fig:aa3}
\end{figure}

 Numerical integration in \eq{aa2} including  small contribution from baryon stopping 
yields for magnetic field the result shown in \fig{fig:aa3} as a function of the proper time $\tau = (t^2 - z^2)^{1/2}$. Evidently, magnetic field rapidly decreases as a power of  time, so that after first 3 fm it drops by more than three orders of magnitude.  

\subsubsection{Event-by-event fluctuations in proton positions}

\begin{figure}[ht]
\begin{tabular}{cc}
      \includegraphics[height=5.5cm]{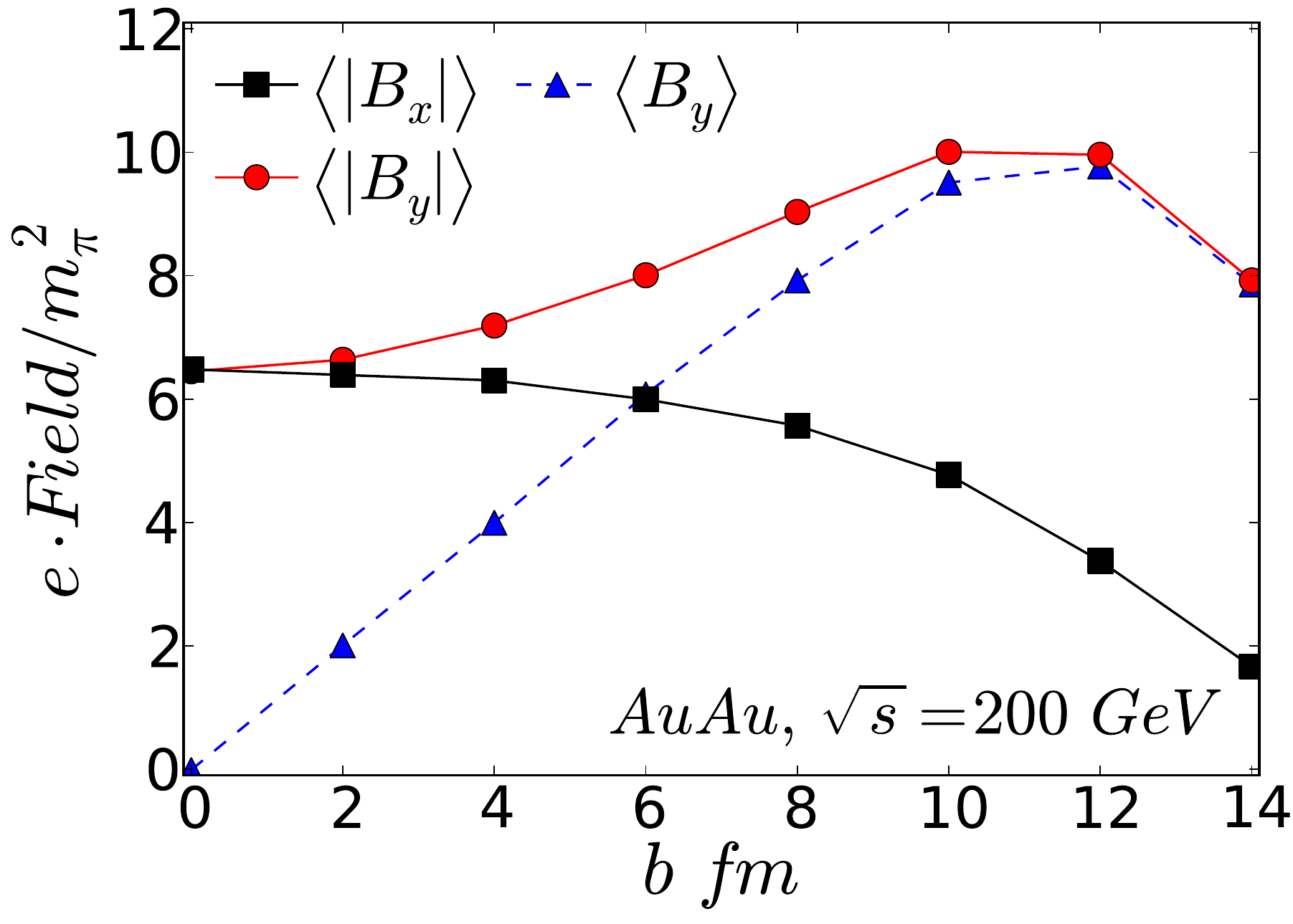} &
      \includegraphics[height=5.5cm]{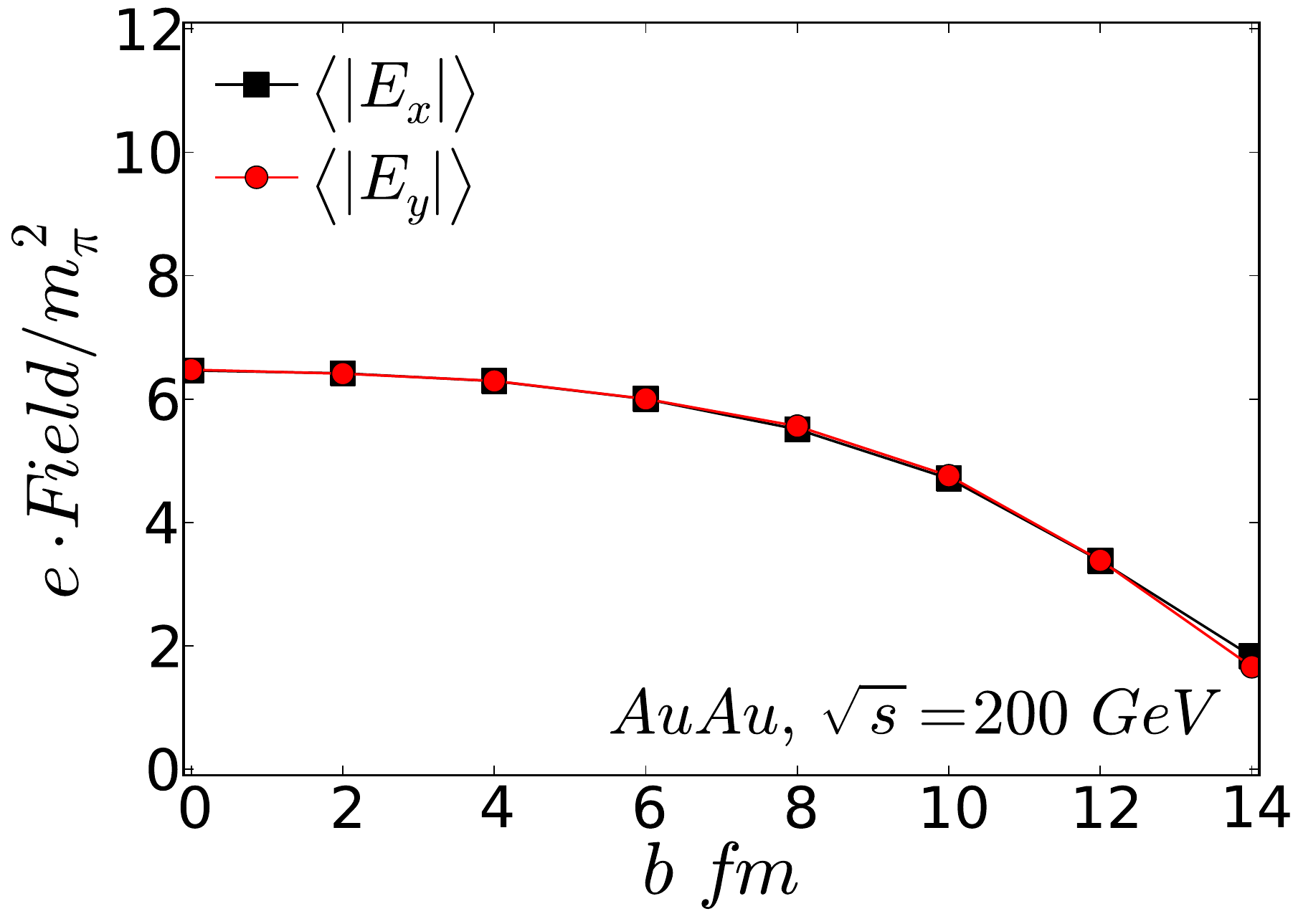}\\
      $(a)$ & $(b)$ 
      \end{tabular}
\caption{The mean absolute value of (a) magnetic field and (b) electric field at $t=0$ and $\b r=0$ as a function of impact parameter $b$ for $AuAu$ collision at $\sqrt{s_{NN}}=200$ GeV. }
\label{fig:aa5}
\end{figure}

Nuclear charge density $\rho$ provides only event-averaged distribution of protons. The actual distribution in a given event is different form $\rho$ implying that in a single event there is not only magnetic field along the $y$-direction, but also other components of electric and magnetic fields. This leads to event-by-event fluctuations of electromagnetic field  \cite{Bzdak:2011yy}. Shown in \fig{fig:aa5} are electric and magnetic field components at $t=0$ at the origin (denoted by a
black dot in \fig{fig:aa1}) in $AuAu$ collisions at $\sqrt{s_{NN}}=200$ GeV. 

\fig{fig:aa5} clearly shows that although on average the only non-vanish component of the field is $B_y$, which is also clear from the symmetry considerations, other components are finite in each event and are of the same order of magnitude 
\beql{aa8}
\left\langle |B_{x}|\right\rangle \approx \left\langle |E_{x}|\right\rangle
\approx \left\langle |E_{y}|\right\rangle \,.
\eeq
To appreciate the magnitude of electric field produced in heavy-ion collisions note that $E\sim m_\pi^2= 10^{21}$~V/cm. The corresponding intensity is $10^{39}$~W/cm$^2$ which is instructive to compare with the power generated by the most powerful state-of-the-art lasers: $10^{23}$~W/cm$^2$.

Electromagnetic fields created in heavy-ion collisions were also examined in more elaborated approaches  in \cite{Skokov:2009qp,Voronyuk:2011jd,Deng:2012pc}. They  yielded qualitatively similar results on electromagnetic field strength and its relaxation time.

\subsection{Magnetic field in quark-gluon plasma}\label{sec:z}

\subsubsection{Li\'enard-Wiechert  potentials in static medium}\label{sec:za}

In the previous section I discussed  electromagnetic field in vacuum. A more realistic estimate must include medium effects. Indeed, state-of-the art phenomenology of quark-gluon plasma (QGP) indicates that strongly interacting medium is formed at as early as  0.5~fm/c. Even before this time, strongly interacting medium exist in a form of \emph{Glasma} \cite{Lappi:2006fp,Blaizot:2012qd}. Therefore, a calculation of magnetic field must involve response of medium determined by its electrical conductivity.  
It has been found in the lattice calculations that the gluon contribution to electrical conductivity of static quark-gluon plasma is \cite{Ding:2010ga}
\beql{za0}
\sigma= (5.8\pm 2.9)\,\frac{T}{T_c} \,\text{MeV}\,,
\eeq
where $T$ is plasma temperature and $T_c$ its critical temperature. This agrees with \cite{Aarts:2007wj}, but is at odds with an earlier calculation \cite{Gupta:2003zh}. It is not clear whether \eq{za0} adequately describes the electromagnetic response of realistic quark-gluon plasma because it neglects quark contribution and assumes that medium is static. Theoretical calculations are of little help at  the temperatures of interest since the perturbation theory is not applicable. In absence of a sensible alternative I will use \eq{za0}  as a best estimate of  electrical conductivity. If medium is static then $T$ is constant as a function of time $t$. The static case is considered in this section, while in the next section I consider expanding medium.

In medium, magnetic field created by a charge $e$ moving in $z$-direction with velocity $v$ is a solution of the following equations
\begin{align}
&\b\nabla \cdot \b B=0\,, &\b \nabla\times \b E= -\frac{\partial\b E}{\partial t}\,,\label{za-13}\\
& \b \nabla \cdot \b E= e\delta(z-vt)\delta(\b b)\,, & \b\nabla \times\b B = \frac{\partial \b E}{\partial t}+ \sigma \b E + ev\unit z \delta(z-vt)\delta(\b b)\,,\label{za-12}
\end{align}
where we used the Ohm's law $\b j= \sigma \b E$ to describe currents induced in the medium. Position of the observation point is specified by the longitudinal and transverse coordinates $z$ and $\b b$, $\b r= z\unit z+\b b$. Taking curl of the second equation in \eq{za-12} and substituting eqs.~\eq{za-13} we get
\beql{za-10}
\nabla^2\b B= \frac{\partial^2 \b B}{\partial t^2}+\sigma \frac{\partial \b B}{\partial t}-ev\b\nabla\times [\unit z\delta(z-vt)\delta(\b b)]\,.
\eeq
The particular solution reads
\beql{za-8}
\b B(z,\b b, t)= \int_0^t dt' \int_{-\infty}^\infty  dz\int d^2\b b\, G(z-z',\b b-\b b', t-t')\, ev\b\nabla'\times [\unit z\delta(z'-vt')\delta(\b b')]\,,
\eeq
where the Green's function $G(z-z',\b b-\b b', t-t')$ satisfies the following equation
\beql{za-6}
\nabla^2G- \frac{\partial^2 G}{\partial t^2}-\sigma \frac{\partial G}{\partial t}= -\delta(z-z')\delta(\b b-\b b')\delta (t-t')\,,
\eeq
which is solved by
\begin{align}
\label{za-4}
&G(z-z',\b b-\b b', t-t')\nonumber\\
&=\int \frac{d^2k_\bot}{(2\pi)^2}e^{i(\b b-\b b')\cdot \b k_\bot}\int_{-\infty}^\infty \frac{dk_z}{2\pi}e^{ik_z (z-z')}\int_{-\infty}^\infty \frac{d\omega}{2\pi}e^{-i\omega (t-t')}\, 
\frac{1}{k_z^2+ k_\bot^2-\omega^2-i\omega \sigma}\,,
\end{align}
where $\b k= k_z\unit z+\b k_\bot$.
Plugging this into \eq{za-8} and substituting for the expression in the square brackets in \eq{za-8} its Fourier image we obtain
\begin{align}
\label{za-3}
\b B(z,\b b, t)&=2\pi ev\int \frac{d^2k_\bot}{(2\pi)^2}e^{i\b b\cdot \b k_\bot}\int_{-\infty}^\infty \frac{dk_z}{2\pi}e^{ik_z z}\int_{-\infty}^\infty \frac{d\omega}{2\pi}e^{-i\omega t}\,
\frac{i\b k\times \unit z}{k_z^2+ k_\bot^2-\omega^2-i\omega \sigma}\delta(\omega- k_zv)\,,
\\
\label{za-2}
&= e\int \frac{d^2k_\bot}{(2\pi)^2}e^{i\b b\cdot \b k_\bot}\int_{-\infty}^\infty \frac{d\omega}{2\pi}e^{-i\omega t}\,e^{i\omega z/v}
\frac{i\b k_\bot \times \unit z}{\omega^2/v^2+ k_\bot^2-\omega^2-i\omega \sigma}\,.
\end{align}
We are interested in the $y$-component of the field. Noting that $(\b k_\bot \times \unit z)\cdot \unit y= -k_\bot \cos\phi$, where $\phi$ is the azimuthal angle in the transverse plane, and integrating over $d^2k_\bot$ we derive
\beql{za1}
e B_y = \frac{\alpha}{\pi}\,\int_{-\infty}^\infty s(\omega)\, K_1(s(\omega)b)\,e^{i\omega (z/v-t)}\, d\omega\,,
\eeq
where we introduced notation 
\beql{za3}
s(\omega) = \omega\sqrt{\frac{1}{v^2}-\epsilon(\omega)}\,.
\eeq
$\epsilon(\omega)$ is the dielectric constant of the plasma with the following frequency dependence
\beql{za5}
\epsilon(\omega) = 1+\frac{i\sigma}{\omega}\,.
\eeq 

Eq.~\eq{za1} is actually valid for any functional form of $\epsilon(\omega)$ \cite{Jackson's_text}, which can be easily verified by using electric displacement $\b D$ instead of $\b E$ in eqs.~\eq{za-12}. In this case \eq{za5} can be viewed as a low frequency expansion of $\epsilon(\omega)$. Magnetic field in this approximation is quasi-static. Therefore, we could have neglected the second time derivative in \eq{za-10}  and then keeping only the leading powers of $\omega$ we would have derived \eq{za1} with $s^2= i\omega \sigma$. After integration over $\omega$ this gives \eq{za15}. Let us take notice of the fact that neglecting the second time derivative in \eq{za-10} yields  \emph{diffusion equation} for magnetic field in plasma. 

It is instructive to compare  time-dependence of magnetic field created by moving charges in vacuum and in plasma. In vacuum, setting $\sigma =0$ in \eq{za-2} and integrating first over  $\omega$ and then over $\b k_\bot$ gives 
\beql{za7}
e\b B = \unit y\, \alpha_\text{em}\frac{b\gamma }{(b^2+\gamma^2 (t-z)^2)^{3/2}}\,,
\eeq
where we used $v\approx 1$. This coincides with \eq{aa2} for a single proton when we take $\b R_a = \b b +(z-vt)\unit z$. Consider field strength \eq{za7} at the origin $z=0$. At times $t<b/\gamma$ the field is constant, while at $t\gg b/\gamma$  it decreases as $B_\infty \propto 1/t^3$. At the time $t\approx b$ the ratio between these two  is 
\beql{za9}
\frac{B_0}{B_\infty}=\frac{1}{\gamma^3}\ll 1\,,
\eeq
which is a very small number ($\sim 10^{-6}$ at RHIC).

\begin{figure}[ht]
      \includegraphics[height=5.5cm]{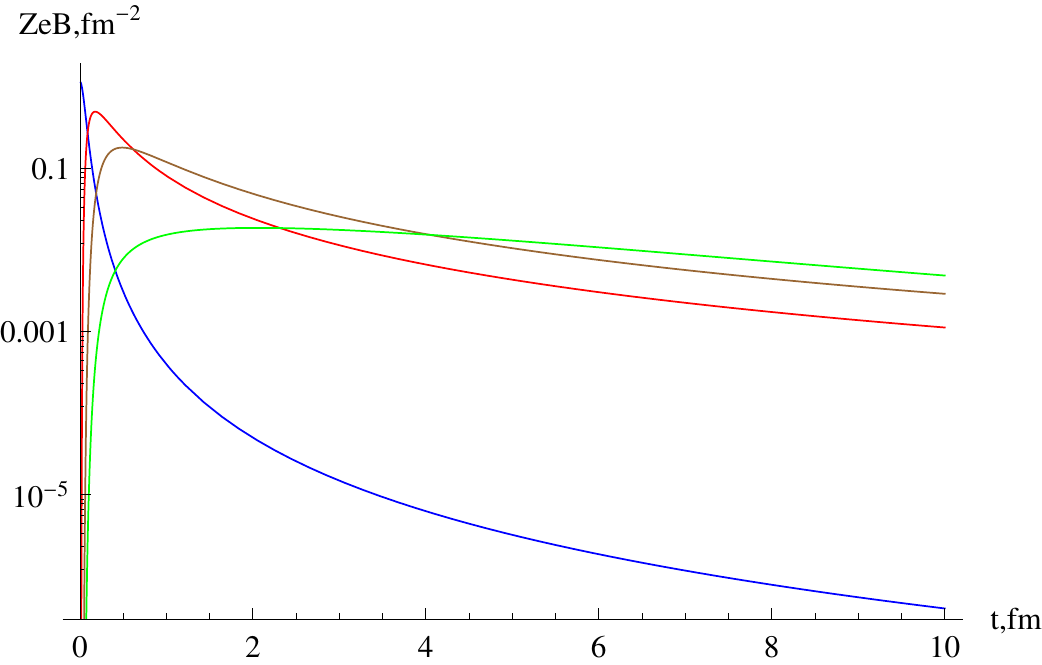} 
\caption{ Relaxation of magnetic field at $z=0$ in vacuum (blue), in static conducting medium at $\sigma=5.8$~MeV (red) and  at $\sigma=16$~MeV (brown) and in the expanding medium (green). Units of $B$ is $\text{fm}^{-2}\approx 2m_\pi^2$. $b=7$~fm, $Z=79$ (Gold nucleus), $\gamma=100$ (RHIC).}
\label{fig:za1}
\end{figure}

In matter $\sigma>0$. Let me write the modified Bessel function appearing in \eq{za1} as follows
\beql{za11}
s K_1(sb) = \int_0^\infty \frac{J_1(x b)x^2dx}{x^2+s^2}\,.
\eeq
Substituting \eq{za11} into \eq{za1} and using \eq{za5} we have ($v=1$)
\begin{align}\label{za13}
e\b B=\frac{\alpha_\text{em}}{\pi}\unit y\int_0^\infty dx  \int_{-\infty}^\infty d\omega\, \frac{J_1(xb)x^2}{x^2-i\omega \sigma}\,e^{i\omega (z-t)} \,
\end{align}
Closing the contour in the lower half-plane  of complex $\omega$ picks a pole at $\omega = -ix^2/\sigma$. We have
\beql{za15}
 e\b B= \frac{2\alpha_\text{em}}{\sigma}\unit y\,\int_0^\infty dx x^2 J_1(xb) e^{-\frac{x^2}{\sigma}(t-z)} =  \unit y\,\frac{\alpha_\text{em}b\,\sigma}{2(t-z)^2}e^{-\frac{b^2\sigma}{4(t-z)}}\,.
 \eeq
At $z=0$ this function vanishes at $t=0$ and $t\to \infty$ and has maximum at the time instant $t= b^2\sigma/8$ which is much larger than $b/\gamma$. The value of the magnetic field at this time is
\beql{za17}
eB_\text{max} = \frac{32e^{-2}\alpha_\text{em}}{b^3\sigma}\,.
\eeq
(Here $e$ is the base of natural logarithm). This is smaller that the maximum field in vacuum
\beql{za19}
\frac{B_\text{max}}{B_0}= \frac{32e^{-2}}{\sigma b\gamma}\,
\eeq
but is still a huge field.  We compare the two solutions \eq{za7} and \eq{za15}  in \fig{fig:za1}. We see that in a conducting medium magnetic field stays for a long time. 

One essential component is still missing in our arguments -- time-dependence of plasma properties due to its expansion. Let us now turn to this problem. 

\subsubsection{Magnetic field in expanding medium}\label{sec:zb}

So far I treated quark-gluon plasma as a static medium. In expanding medium temperature and hence conductivity are functions of time. In Bjorken scenario \cite{Bjorken:1982qr}, expansion is isentropic, i.e.\ $nV= \text{const}$, where $n$ is the particle number density and $V$ is plasma volume. Since $n\sim T^3$ and at early times expansion is one-dimensional $V\sim t$ it follows that    $T\propto t^{-1/3}$. (Eventually, we will consider the midrapidity region $z=0$, therefore distinction between the proper time and $t$ is not essential). Eq.~\eq{za0} implies that $\sigma\sim t^{-1/3}$. I will parameterize conductivity as follows
\beql{zb1}
\sigma(t) =\sigma_0 \left(\frac{ t_0}{t_0+t}\right)^{1/3}\,,
\eeq
where I took $t_0\approx 0.5$~fm to be  the initial time (or longitudinal size) of plasma evolution. Suppose that plasma lives for 10 fm/c and then undergoes phase transition to hadronic gas at $T_c$. Then employing \eq{za0} we estimate $\sigma_0\approx 16$~MeV. Let me  define another parameter  that I will need in the forthcoming calculation:
\beql{zb3}
\beta = \frac{4\sigma_0}{3t_0}\approx 43 \, \frac{\text{MeV}}{\text{fm}}\,.
\eeq

Magnetic field in expanding medium is still governed by \eq{za-10}. As was explained in the preceding subsection, time-evolution of magnetic field is quasi-static, which allows me to neglect the second time derivative. Let me introduce a new ``time" variable $\rho$ as follows
\beql{zb5}
\rho = (1+t/t_0)^{4/3}-1\,.
\eeq
Field $\b B(z,\b b, \rho)$ satisfies equation
\beql{zb7}
\nabla^2\b B= \beta\frac{\partial \b B}{\partial \rho}-ev\b\nabla\times \big\{\unit z\delta[z-vt(\rho)]\delta(\b b)\big\}\,,
\eeq
where 
\beql{zb8}
t(\rho)= t_0[(\rho+1)^{3/4}-1]\,.
\eeq
Its solution can be written as
\beql{zb9}
\b B(z,\b b, \rho)= \int_0^\rho d\rho' \int_{-\infty}^\infty  dz\int d^2\b b\, \mathfrak{G}(z-z',\b b-\b b', \rho-\rho')\, ev\b\nabla\times \big\{\unit z\delta[z-vt(\rho)]\delta(\b b)\big\}\,,
\eeq
in terms of the Green's function $\mathfrak{G}(z-z',\b b-\b b', \rho-\rho')$ satisfying 
\beql{zb12}
\nabla^2 \mathfrak{G}-\beta\frac{\partial \mathfrak{G}}{\partial \rho}=-\delta(z-z')\delta(\b b-\b b')\delta(\rho-\rho')\,,
\eeq
To solve this equation we represent $\mathfrak{G}$ as three-dimensional Fourier integral with respect to the space coordinates and Laplace transform with respect to the ``time" coordinate:
\beql{zb15}
\mathfrak{G}(z,\b b, \rho)= \int \frac{d^3k}{(2\pi)^3}e^{i(\b k_\bot\cdot \b b+k_z z)}\int _\text{C}\frac{d\lambda}{2\pi i}e^{\lambda\rho}\frac{1}{k_\bot^2+k_z^2+\beta\lambda}\,,
\eeq
with the contour C running  parallel to the imaginary axis to the right of all integrand singularities. 
Now I would like to write the expression in the curly brackets in \eq{zb9} also as Fourier-Laplace expansion. To this end we calculate
\begin{align}\label{zb17}
f_{\b k, \lambda}&= \int d^2\b b\int_{-\infty}^\infty dz \int_0^\infty d\rho\, e^{-i(\b k_\bot\cdot \b b+k_z z)}e^{-\lambda\rho}\, \delta(z-vt(\rho))\,\delta(\b b)\\
\label{zb18} &=\int_0^\infty d\rho \, e^{-ik_z v t_0[ (\rho+1)^{3/4}-1]}\,e^{-\lambda\rho}\,.
\end{align}
 Therefore,
\beql{zb21}
ev\b\nabla\times \big\{\unit z\delta[z-vt(\rho)]\delta(\b b)\big\}=
ev\int \frac{d^3k}{(2\pi)^3}e^{i(\b k_\bot\cdot \b b+k_z z)}\int _\text{C}\frac{d\lambda}{2\pi i}e^{\lambda\rho}\, i\b k_\bot \times \unit z\, f_{\b k,\lambda}\,.
\eeq 
 Substituting \eq{zb15} and \eq{zb21} into \eq{zb9} we obtain upon integration over the volume and time
\begin{align}\label{zb23}
 \b B(z,\b b, \rho)&= \int \frac{d^3k}{(2\pi)^3}e^{i(\b k_\bot\cdot \b b+k_z z)}\int _\text{C}\frac{d\lambda}{2\pi i}e^{\lambda\rho}\frac{ev\, i\b k_\bot\times \unit z}{k_\bot^2+k_z^2+\beta\lambda}\, f_{\b k,\lambda}\, \theta(\rho)\,,
\end{align} 
 where $\theta$ is the step-function. Taking consequent integrals over $\lambda$ and $k_z$   gives
 \begin{align}
 \label{zb25}
 \b B(0,\b b, \rho)&=\frac{ev}{\beta}\int \frac{d^2k_\bot}{(2\pi)^2}e^{i\b k_\bot\cdot \b b}
 \int_{-\infty}^\infty \frac{dk_z}{2\pi} i\b k_\bot\times \unit z\int_0^\rho d\rho' 
 \, e^{-ik_z v t_0[ (\rho'+1)^{3/4}-1]}\, e^{-\frac{k_\bot^2+k_z^2}{\beta}(\rho-\rho')}\\
 \label{zb26}
 &=\frac{ev}{\beta}\int \frac{d^2k_\bot}{(2\pi)^2}e^{i\b k_\bot\cdot \b b}
 \frac{1}{2\pi} i\b k_\bot\times \unit z\int_0^\rho d\rho' 
 \, e^{-\frac{k_\bot^2}{\beta}(\rho-\rho')}\frac{\sqrt{\pi\beta}}{\sqrt{\rho-\rho'}}
 e^{-\frac{v^2t_0^2\beta [ (\rho'+1)^{3/4}-1]^2}{4(\rho-\rho')}}\,.
\end{align}  
Consider now $B_y$. Integrating  over azimuthal angle $\phi$  and then over 
$k_\bot$ as in \eq{za-2},\eq{za1} yields
\beql{zb30}
eB_y(0,\b b, \rho)=\frac{\alpha_\text{em}vb\beta^{3/2}}{2\sqrt{\pi}}\int_0^\rho d\zeta \,\zeta^{-5/2}\, e^{-\frac{\b b^2\beta}{4\zeta}}\, 
 e^{-\frac{v^2t_0^2\beta [ (\rho-\zeta+1)^{3/4}-1]^2}{4\zeta}}\,,
\eeq
where $\zeta = \rho-\rho'$. 

The results of a numerical calculation of \eq{zb30} are shown in \fig{fig:za1}. We see that expansion of plasma tends to increase the relaxation time, although this effect is rather modest. We conclude that due to finite electrical conductivity of QGP, magnetic field essentially freezes in the plasma for as long as plasma exists. Similar phenomenon, known as skin--effect,  exists in good conductors placed in time-varying magnetic field: conductors expel time dependent magnetic fields form conductor volume confining them into a thin layer of width $\delta\sim \omega^{-1/2}$ on the surface.

\subsubsection{Diffusion of magnetic field in QGP}

The dynamics of  magnetic field relaxation in conducting plasma plasma can be understood in a simple model \cite{Tuchin:2010vs}. Suppose at some initial time $t=0$ magnetic field $\b B(0,\b r)$ permits the plasma. The problem is to find the time-dependence of the field at $t>0$. 
In this model, the field sources turn off at $t=0$ and do not at all contribute to the field at $t>0$. 
Electromagnetic field is governed by the following equations
\begin{align}
\b\nabla\times \b B&= \b j\,, &\b j = \sigma \b E\label{aa20}\\  
\b\nabla \times \b E&= -\frac{\partial \b B}{\partial t }\,, &\b\nabla\cdot \b B=0\,,\label{aa21}
\end{align}
that lead to the diffusion equation for $\b B$, after we neglect the second time derivative as discussed before
\begin{align}\label{aa23}
\nabla^2 \b B= \sigma\frac{\partial \b B}{\partial t }.
\end{align}
For simplicity we treat electrical   conductivity $\sigma$ as constant. Initial condition at $t=0$ reads 
\beql{aa25}
\b B(0,\b r)=  \b B_0\, e^{-\frac{\b b^2}{R^2}}\,,
\eeq
where the Gaussian profile is chosen for illustration purposes and $R$ is the nuclear radius. Solution to the problem \eq{aa23},\eq{aa25} is
\beql{aa27}
\b B(t,\b r) =\int dV' \b B(0,\b r')\,G(t,\b r-\b r')\,.
\eeq
where the Green's function is
\beql{aa29}
G(t,\b r)= \frac{1}{(4\pi t/\sigma)^{3/2}}\exp\left[ -\frac{\b r^2}{4t/\sigma}\right]\,.
\eeq
Integrating over the entire volume   we derive
\beql{aa31}
\b B(t,\b r) = \b B_0\,\frac{R^2}{R^2+4t/\sigma}\,\exp\left[ -\frac{\b b^2}{R^2+4t/\sigma}\right]\,.
\eeq
It follows from \eq{aa31} that as long as $t\ll t_\text{relax}$, where $t_\text{relax}$ is a characteristic time $t_\text{relax} = R^2\sigma/4$
magnetic field $\b B$ is approximately time-independent. This estimate is the same as the one we arrived at after \eq{za15}.

\bigskip

{\bf In summary},  magnetic field in quark-gluon plasma appears to be extremely strong and slowly varying function of time for most of the plasma life-time. At RHIC it decreases from $eB\approx (2.5 m_\pi)^2$ right after  the collision to $eB\approx (m_\pi/4)^2$ at $t\approx 5$~fm, see \fig{fig:za1}. This has a profound impact on all the processes occurring in QGP.

\subsubsection{Schwinger mechansim}\label{sec:ab}

Schwinger mechanism of pair production \cite{Schwinger:1951nm} is operative if electric field exceeds the critical value of $m^2/e$, where $m$ is mass of lightest electrically charged particle. Indeed,  in order to excite a fermion out of the Dirac sea, electric force $e\b E$ must do work along the path $d\b \ell$ satisfying 
\beql{ab5}
\int_0^\ell e\b E\cdot d\b \ell' >2m\,.
\eeq
If $\b E= \text{const}$, then $E\gtrsim m/\ell e$. The maximal value of $\ell$ is the fermion  Compton's  wavelength $\ell \sim \lambdabar=1/m$ implying that the minimum (or critical) value of electric field is 
\beql{ab6}
E_c= \frac{m^2}{e}\,.
\eeq
Notice that in stronger fields  $\ell\sim m/eE<\lambdabar$.  
\fig{fig:aa5} indicates that electron-positron pairs are certainly produced at RHIC. An important question then is the role of these pairs in the electromagnetic field relaxation in plasma. 
There are two associated effects: (i) before $e^-e^+$ pairs  thermalize, they contribute to the Foucault currents,  (ii) after they thermalize, their density contributes to the polarization of plasma in electric field, and hence to its conductivity. 

Since  space dimensions of  QGP are much less than $\lambdabar_e=380$~fm, it may seem inevitable that space-dependence of electric field (in addition to its time-dependence)  has a significant impact on the Schwinger process in heavy-ion collisions. However, his conclusion  is premature. Indeed, suppose that electric field is a slow function of  coordinates. Then $\b E(\b r)\approx \b E(0)+\b r\cdot \b \nabla \b E(0)$. Work done by electric field 
is 
\beql{ab7} 
\int_0^\ell e\b E(0)\cdot d\b\ell'+ \int_0^\ell (\b r\cdot \b \nabla) e\b E(0)\cdot d\b\ell'\sim eE\ell + \frac{e\ell^2 E(0)}{\lambda}\,,
\eeq
where $\lambda$ is length scale describing space variation of electric field. In order that contribution of space variation to work be negligible, the second term in the r.h.s.\ of \eq{ab7} must satisfy $e\ell^2 E(0)/\lambda\ll m$. Employing the estimate $\ell \sim m/eE(0)$ that we obtained after \eq{ab6} implies $m/eE(0)\lambda\ll 1$.  Following \cite{Dunne:2005sx} I define  the  inhomogeneity parameter 
\beql{ab11}
\tilde \gamma = \frac{m}{\lambda eE}\,
\eeq
that describes the effect of spatial variation of electric field on the pair production rate. For electrons $m=0.5$~MeV in QGP $\lambda\sim 0.5$~fm at $eE\sim m_\pi^2$ we have $\tilde \gamma \sim 0.01$. Therefore, somewhat counter-intuitively , electric field can be considered as spatially homogeneous. The same conclusion can be derived from results of \cite{Wang:1988ct}. Schwinger mechanism in spatially-dependent electric fields was also discussed in \cite{Martin:1988gr,Kim:2007pm}.

\begin{figure}[ht]
      \includegraphics[width=15cm]{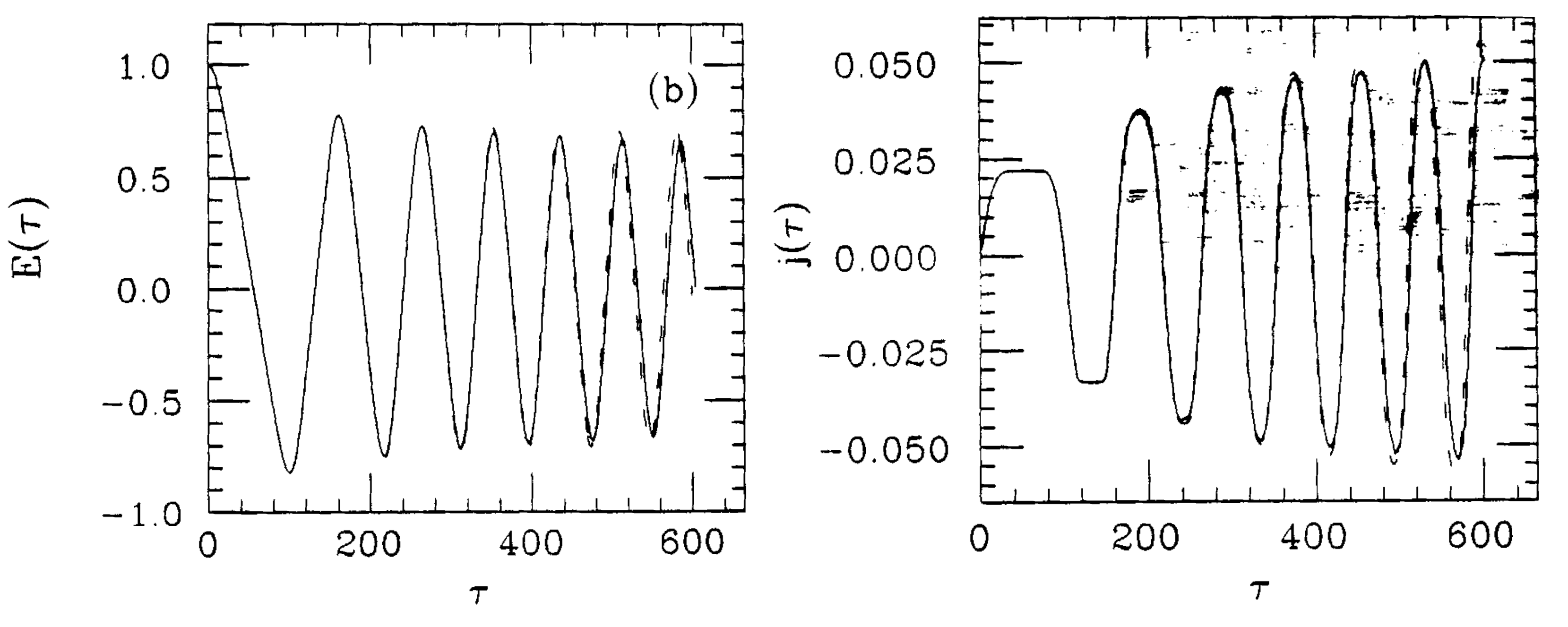}  
\caption{Time-dependence of electric field due to the Schwinger mechanism back-reaction and the corresponding electric current density of Schwinger pairs.  Dimensionless time variable is defined as $\tau = t/\lambdabar$. For electrons $\lambdabar = 380$~fm. Plasma  undergoes phase transition at about $\tau=1/38$. Adapted from  \cite{Kluger:1992gb}.}
\label{fig:ab1}
\end{figure}

In view of smallness of $\tilde\gamma$ one can employ the extensive literature on Schwinger effect in time-dependent spatially-homogeneous electric fields. As far as heavy-ion physics is concerned, the most comprehensive study has been done in \cite{Kluger:1992gb,Cooper:1992hw,Kluger:1991ib} who developed an approach to include  the effect of back reaction. They argued that time-evolution of electric field can be studied in  adiabatic approximation and used the kinetic approach to study the time-evolution. Their results are exhibited in \fig{fig:ab1}. Similar results were obtained in \cite{Tanji:2008ku}. We observe that response time of the current density of Schwinger pairs $\sim 10^4$~fm/c is much larger than the plasma lifetime $\sim 10$~fm/c and therefore no sizable electric current is generated.

\bigskip

{\bf In summary}, strong electric field is generated in heavy ion collisions in every event, but averages to zero in a large event ensemble.  This field exceeds the critical value for electrons and light  $u,d$ quarks. However, during the plasma lifetime no significant current of Schwinger pairs is generated.

\setcounter{equation}{0}
\newpage
\section{Flow of quark-gluon plasma in strong magnetic field}\label{sec:visc}

\subsection{Azimuthal asymmetry}\label{sec:ca}

Magnetic field is known to have a profound influence on  kinetic properties of plasmas. 
Once the spherical symmetry is broken, distribution of particles in plasma is only axially symmetric with respect to the magnetic field direction. This symmetry however is not manifest in the plane span by magnetic field and the impact parameter vectors, viz.\ $xy$-plane in \fig{fig:aa1}.  Charged particles moving along the magnetic field direction $y$ are not influenced by the magnetic Lorentz force while those moving the $xz$-plane (i.e.\ the reaction plane) are affected the most. The result is azimuthally anisotropic flow of expanding plasma in the $xy$-plane even when initial plasma geometry is completely spherically symmetric. The effect of weak magnetic field on quark-gluon plasma flow was first considered in 
 \cite{Mohapatra:2011ku} who argued that magnetic field is able to  enhance 
 the azimuthal anisotropy  of produced particles up to $30\%$. This conclusion was reached by utilizing a solution of the magneto-hydrodynamic equations in weak magnetic field. 
 
 A characteristic feature of the viscous pressure tensor in magnetic field is its azimuthal anisotropy. This anisotropy is the result of  suppression of the momentum transfer in QGP in the direction perpendicular to the magnetic field. Its macroscopic manifestation is decrease of the viscous pressure tensor components in the plane perpendicular to the magnetic field, which coincides with the reaction plane in the heavy-ion phenomenology. Since Lorentz force vanishes in the direction parallel to the field,  viscosity along that direction is not affected at all. In fact, the viscous pressure tensor component in the reaction plane is twice as small as the one in the field direction. As the result,  transverse flow of QGP develops azimuthal anisotropy in presence of the magnetic field. Clearly, this anisotropy is completely different from the one generated by the anisotropic pressure gradients and exists even if the later are absent. In fact, 
 because spherical symmetry in magnetic field is broken, viscous effects in plasma cannot 
 be described by only two parameters: shear $\eta$ and bulk viscosity $\zeta$. Rather
 the viscous pressure tensor of  magnetoactive plasma is characterized by seven viscosity coefficients,  among which five are shear viscosities and two are bulk ones.

\subsection{Viscous pressure in strong magnetic field}\label{sec:cb}
\subsubsection{Viscosities from kinetic equation}

Generally, calculation of the viscosities requires knowledge of the strong interaction dynamics of the QGP components. However, in strong magnetic field  these interactions can be considered as a perturbation and viscosities can be analytically calculated using the kinetic equation \cite{LLX,Erkelens:1977}.  To apply this approach to QGP  in strong magnetic field we start with  kinetic equation for the distribution function $f$ of a quark flavor of charge $ze$ is 
\beq\label{kinetic0}
p^\mu \partial_\mu f= zeB^{\mu\nu}\frac{\partial f}{\partial u^\mu}u_\nu +\mathcal{C}[ f,\dots]
\eeq
where $\mathcal{C}$ is the collision integral and $B^{\mu\nu}$ is the electro-magnetic tensor, which contains only magnetic field components in the laboratory frame. Ellipsis in the argument of $\mathcal{C}$ indicates  the distribution  functions of other quark flavors and gluons (I will omit them below). 
The equilibrium  distribution reads
\beq\label{f0}
f_0= \frac{\rho}{4\pi m^3 T K_2(\beta m)}e^{ - \beta\, p\cdot U(x)}
\eeq
where $U(x)$ is the macroscopic velocity of the fluid, $p^\mu= m u^\mu$ is particle momentum, $\beta = 1/T$ and $\rho$ is the mass density. 
Since $\frac{\partial f_0}{\partial u^\mu}\propto u_\mu$, the first term on the r.h.s.\ of \eq{kinetic0} as well as the collision integral vanishe in equilibrium. Therefore, we can write the kinetic equation as an equation for $\delta\! f$
\beq\label{kinetic}
p^\mu \partial_\mu f_0= zeB^{\mu\nu}\frac{\partial (\delta\! f)}{\partial u^\mu}u_\nu +\mathcal{C}[\delta\! f]
\eeq
where $\delta\! f$ is a deviation from equilibrium. Differentiating  \eq{f0} we find 
\beq\label{derf0}
\partial_\mu f_0= -f_0\frac{1}{T}\,p^\lambda\partial_\mu U_\lambda(x)
\eeq
Since $U^\lambda= (\gamma_V, \gamma_V \b V)$ and $p^\lambda= (\e, \b p)=(\gamma_v m,\gamma_v m\b v)$ it follows that
\beq
p\cdot U= \frac{m}{\sqrt{1-v^2}\sqrt{1-V^2}}(1-\b v\cdot \b V)
\eeq
Thus, in the comoving frame
\beql{z1}
\partial_\mu f_0|_{\b V=0}= f_0 \,\frac{1}{T}\, p_\nu \partial_\mu V^\nu
\eeq
Substituting \eq{z1} in \eq{kinetic} yields
\beq\label{kinetic2}
-\frac{f_0}{T}p^\mu p^\nu V_{\mu\nu}= zeB^{\mu\nu}\frac{\partial (\delta\! f)}{\partial u^\mu}u_\nu +\mathcal{C}[\delta\! f]
\eeq
where I defined
\beq \label{Vdef}
V_{\mu\nu}= \frac{1}{2}(\partial_\mu V_\nu+\partial_\nu V_\mu)
\eeq
and used $u^\mu u^\nu \partial_{\mu}V_\nu = u^\mu u^\nu V_{\mu\nu}$.

 Since the time-derivative of $f_0$ is irrelevant for the calculation of the viscosity I will drop it from the kinetic equation. All indices thus become the usual three-vector ones. To avoid confusion we will label them by the Greek letters from the beginning of the alphabet.  Introducing $b_{\alpha\beta}= B^{-1}\varepsilon_{\alpha\beta\gamma}B_\gamma$ we cast \eq{kinetic2} in the form
\beq\label{ke}
\frac{1}{T}p^\alpha u^\beta V_{\alpha\beta}f_0=  - zeB b_{\alpha\beta} v^\beta \frac{\partial (\delta\! f)}{\partial v_\alpha} \frac{1}{\e}-\mathcal{C}[\delta\! f] \,.
\eeq

The viscous pressure  generated by a deviation from equilibrium is given by the tensor
\beq\label{stress-def}
-\Pi_{\alpha\beta}= \int p_\alpha p_\beta\, \delta\! f \,\frac{d^3p}{\e}
\eeq
Effectively it can be parameterized in terms of the viscosity coefficients as follows (we neglect the bulk viscosities)
\beq\label{visc.tens1}
\Pi_{\alpha\beta} = \sum_{n=0}^4 \eta_n\, V^{(n)}_{\alpha\beta}
\eeq
where the linearly independent tensors $V^{(n)}_{\alpha\beta}$ are given by
\begin{subequations}\label{vns}
\begin{eqnarray}\label{visc.tens}
V^{(0)}_{\alpha\beta} &=& \left(3b_\alpha b_\beta-\delta_{\alpha\beta}\right) \left(b_\gamma b_\delta V_{\gamma\delta}-\frac{1}{3}\nabla\cdot \b V\right) \\
V^{(1)}_{\alpha\beta}&=& 2V_{\alpha\beta}+
\delta_{\alpha\beta}V_{\gamma\delta}b_\gamma b_\delta  - 2V_{\alpha\gamma}b_\gamma b_\beta- 2V_{\beta\gamma}b_\gamma b_\alpha + (b_\alpha b_\beta-\delta_{\alpha\beta}) \nabla\cdot \b V +b_\alpha b_\beta V_{\gamma\delta} b_\gamma b_\delta\\
V^{(2)}_{\alpha\beta}&=& 2(V_{\alpha\gamma} b_{\beta\gamma} +V_{\beta\gamma} b_{\alpha\gamma}- V_{\gamma\delta} b_{\alpha\gamma}b_\beta b_\delta )\\
V^{(3)}_{\alpha\beta}&=& V_{\alpha\gamma}b_{\beta \gamma} + V_{\beta\gamma} b_{\alpha\gamma} - V_{\gamma\delta} b_{\alpha\delta} b_{\alpha\gamma} b_\beta b_\delta - V_{\gamma\delta} b_{\beta \gamma} b_\alpha b_\delta\\
V^{(4)}_{\alpha\beta}&=& 2( V_{\gamma\delta} b_{\alpha\delta} b_{\alpha\gamma} b_\beta b_\delta+
V_{\gamma\delta} b_{\beta \gamma} b_\alpha b_\delta)\,.
\end{eqnarray}
\end{subequations}
For the calculation of the shear viscosities $\eta_n$, $n=1,\ldots,4$ we can set $\nabla\cdot \b V=0$ and $V_{\alpha\beta}b_\alpha b_\beta=0$.

Let us expand $\delta\! f$  to the second order in velocities in terms of the tensors $V_{\alpha\beta}^{(n)}$ as follows
\beq\label{delf}
\delta\! f = \sum_{n=0}^4 g_n V^{(n)}_{\alpha\beta} v^\alpha v^\beta
\eeq
Then, substituting \eq{delf} into \eq{visc.tens1} and requiring consistency of \eq{stress-def} and \eq{visc.tens1} yields 
\beq\label{visc}
\eta_n = -\frac{2}{15}\int \e v^4 g_n d^3 p
\eeq 
This gives the viscosities in the magnetic field in terms of the deviation of the distribution function from equilibrium. Transition to the  non-relativistic limit in \eq{visc} is achieved by the replacement $\e\to m$ \cite{LLX}.


\subsubsection{Collisionless plasma}\label{sec:B}

In strong magnetic field we can determine $g_n$ by the method of consecutive approximations.
Writing $\delta\! f = \delta\! f^{(1)}+\delta\! f^{(2)}$ and substituting into \eq{ke} we find
\beq\label{ke-delta}
\frac{1}{T}p^\alpha v^\beta V_{\alpha\beta}f_0=  - zeB b_{\alpha\beta} v^\beta \frac{\partial (\delta\! f^{(1)}+\delta\! f^{(2)})}{\partial v_\alpha} \frac{1}{\e}+\mathcal{C}[\delta\! f^{(1)}] \,.
\eeq
Here I assumed that  the deviation from equilibrium due to the strong magnetic field is much larger than due to the particle  collisions. The explicit form of $\mathcal{C}$ is determined by the strong interaction dynamics, but drops off the equation in the leading oder. 
The first correction to the equilibrium distribution obeys the equation
\beq\label{deltaf1}
\frac{1}{T}p_\alpha v_\beta V_{\alpha\beta}f_0=  - zeB b_{\alpha\beta} v_\beta \frac{\partial \delta\! f^{(1)}}{\partial v_\alpha} \frac{1}{\e} \,.
\eeq
Using \eq{delf} we get
\beql{w1}
b_{\alpha\beta} v_\beta\frac{\partial \delta\! f^{(1)}}{\partial v_\alpha} = 2b_{\alpha\beta} v_\beta\sum_{n=0}^4 g_n\, V^{(n)}_{\alpha\gamma}v_\gamma\,.
\eeq
Substituting \eq{w1} into \eq{deltaf1} and using \eq{vns} yields:
\begin{eqnarray}\label{cond1}
\frac{\e}{T zeB}\,p_\alpha v_\beta V_{\alpha\beta}f_0&=& - 2b_{\beta\nu} v_\alpha v_\nu [ g_1(2V_{\alpha\beta}-2 V_{\beta\gamma}b_\gamma b_\alpha)+2g_2 V_{\beta\gamma} b_\gamma b_\alpha\nonumber\\
&& +g_3 (V_{\alpha\gamma} b_{\beta\gamma}+V_{\beta\gamma}b_{\alpha\gamma}-V_{\gamma\delta}b_\alpha b_\delta)+2g_4 V_{\gamma\delta} b_{\beta\gamma} b_\alpha b_\delta)]\,,
\end{eqnarray}
where I used the following identities $b_{\alpha\beta}b_\alpha =b_{\alpha\beta}b_\beta=b_{\alpha\beta}v_{\alpha}v_\beta=0$. Clearly, \eq{cond1} is satisfied only if $g_1=g_2=0$. Concerning the other two coefficients, we use the identities
\begin{subequations}
\begin{eqnarray}
b_{\alpha\beta}b_{\beta\gamma}&=& b_\gamma b_\alpha - \delta_{\alpha\gamma}b^2\,,\\ 
  \varepsilon_{\alpha\beta\gamma}\varepsilon_{\delta\epsilon\zeta}    &  =& \delta_{\alpha\delta}\left( \delta_{\beta\epsilon}\delta_{\gamma\zeta} - \delta_{\beta\zeta}\delta_{\gamma\epsilon}\right) - \delta_{\alpha\epsilon}\left( \delta_{\beta \delta}\delta_{\gamma\zeta} - \delta_{\beta\zeta}\delta_{\gamma\delta} \right) + \delta_{\alpha\zeta} \left( \delta_{\beta\delta}\delta_{\gamma\epsilon} - \delta_{\beta\zeta}\delta_{\gamma\delta} \right) \,
\end{eqnarray}
\end{subequations}
that we substitute into \eq{cond1} to derive
\beql{prom1}
-\frac{\e}{2T zeB}\,p^\alpha v^\beta V_{\alpha\beta}f_0=  g_3 [2V_{\alpha\beta} b_\alpha b_{\beta}-4V_{\alpha\beta}v_\alpha b_\beta (\b b\cdot\b  v)]+2g_4 V_{\alpha\beta}v_\alpha b_\beta (\b b\cdot\b  v)\,.
\eeq
Since $p_\alpha = \e v_\alpha$ we obtain
\beql{g3g4}
g_3 = \frac{g_4}{2}= -\frac{\e^2 f_0}{4T zeB}\,.
\eeq

Using \eq{f0}, \eq{g3g4} in \eq{visc} in the comoving frame (of course $\eta_n$'s do not depend on the frame choice) and integrating using 3.547.9 of \cite{GR} we derive \cite{Tuchin:2011jw}
\beql{visc3}
\eta_3= \frac{K_3(\beta m)}{K_2(\beta m)}\frac{\rho T}{2  zeB}\,.
\eeq
The non-relativistic limit corresponds to $m\gg T$ in which case we get
\beql{eta3NR}
\eta_3^\mathrm{NR}=\frac{\rho T}{2zeB}\,.
\eeq
In the opposite ultra-relativistic case $m\ll T$ (high-temperature plasma)
\beql{eta3UR}
\eta_3^\mathrm{UR}=\frac{2 n T^2   }{zeB}\,,
\eeq
where $ n= \rho/m$ is  the number density.

\subsubsection{Contribution of collisions}

In the relaxation-time approximation we can write the collision integral as 
\beql{rel.time}
\mathcal{C}[\delta\! f]= -\nu\, \delta\! f\,,
\eeq
where $\nu$ is an effective collision rate. Strong field limit means that
\beql{con}
\omega_B\gg \nu\,,
\eeq
where $\omega_B = zeB/\e$ is the synchrotron frequency. Whether $\nu$ itself is function of the field depends on the relation between the Larmor radius $r_B=v_T/\omega_B$, where $v_T$ is the particle velocity in the plane orthogonal to $\b B$ and the  Debye radius $r_D$. If 
\beql{w}
r_B\gg r_D\,,
\eeq
then the effect of the field on the collision rate  $\nu$ can be neglected  \cite{LLX}. Assuming that \eq{w} is satisfied the collision rate reads
\beql{nu}
\nu = nv\sigma_t\,,
\eeq
where $\sigma_t$ is the transport cross section,  which is a function of the saturation momentum $Q_s$ \cite{Gribov:1984tu,Blaizot:1987nc}. We estimate $\sigma_t\sim \as^2 /Q_s^2$, with $Q_s\sim 1$~GeV and $n= P/T$ with pressure $\as^2P\sim 1$~GeV/fm$^3$ we get $\nu\sim 40$~MeV. Inequality \eq{con} is well satisfied since $eB\simeq m_\pi^2$  \cite{Kharzeev:2007jp,Skokov:2009qp} and $m$ is  in the range between the current and the constituent quark masses. On the other hand, applicability of the condition \eq{w} is marginal and is very sensitive to the interaction details. In this section we assume that \eq{w} holds in order to obtain the analytic solution. Additionally, the general condition for the applicability of the hydrodynamic approach $\ell = 1/\nu \ll L$, where $\ell$ is the mean free path and $L$ is the plasma size is assumed to hold. Altogether we have $r_D\ll r_B \ll \ell \ll L$.
 
Equation for the second correction to the equilibrium distribution $\delta\! f^{(2)}$ follows from \eq{ke-delta} after substitution \eq{rel.time}
\beql{second-cor}
\frac{zeB}{\e} b_{\alpha\beta} v_\beta \frac{\partial \delta\! f^{(2)}}{\partial v_\alpha} = - \nu \delta\! f^{(1)}\,.
\eeq
Now, plugging 
\begin{subequations}\label{df1}
\begin{eqnarray}
\delta\! f^{(1)}&=& [g_3 V_{\alpha\beta}^{(3)}+g_4 V_{\alpha\beta}^{(4)}] v_\alpha v_\beta\,,\\
\delta\! f^{(2)}&=& [g_1 V_{\alpha\beta}^{(1)}+g_2 V_{\alpha\beta}^{(2)}] v_\alpha v_\beta\,,
\end{eqnarray}
\end{subequations}
into \eq{second-cor} yields
\begin{eqnarray}
\frac{2zeB}{\e}\left\{ g_1[ 2V_{\beta\alpha}b_{\alpha\gamma}v_\beta v_\gamma - 
2V_{\beta\alpha}b_{\alpha\gamma}v_\beta v_\gamma (\b v\cdot \b b)]+2g_2 V_{\beta\alpha}b_{\alpha\gamma}v_\beta v_\gamma (\b v\cdot \b b)\right\}&&\nonumber\\
=-\nu g_3\left \{ -2 V_{\beta\alpha}b_{\alpha\gamma}v_\beta v_\gamma - 6 V_{\beta\alpha}b_{\alpha\gamma}v_\beta v_\gamma (\b v\cdot \b b)\right\}\,,
\end{eqnarray}
where I used $g_4=2g_3$. It follows that 
\beql{g1g2}
g_1= \frac{g_2}{4}= \frac{\nu\gamma_v g_3}{2\omega_B} \,.
\eeq
With the help of \eq{nu},\eq{f0},\eq{visc} we obtain \cite{Tuchin:2011jw}
\beql{visc1}
\eta_1=\frac{\eta_2}{4}= \frac{8}{5\sqrt{2\pi}}\frac{\rho^2\sigma_t \,T^{3/2}}{(zeB)^2 m^{1/2}}\frac{K_{7/2}(\beta m)}{ K_2(\beta m)}\,.
\eeq

\subsection{Azimuthal asymmetry of transverse flow: a simple model}\label{sec:flow}

To illustrate the effect of the magnetic field on the viscous flow of the electrically charged component of the quark-gluon plasma I will assume that the flow is non-relativistic and use the Navie-Stokes equations that read
\beql{n-s}
\rho \left( \frac{\partial V_\alpha}{\partial t}+V_\beta\frac{\partial V_\alpha}{\partial x_\beta}\right)= -\frac{\partial P}{\partial x_\alpha}+\frac{\partial \Pi_{\alpha\beta}}{\partial x_\beta}\,,
\eeq
where $\Pi_{\alpha\beta}$ is the viscous pressure tensor, $\rho = mn$ is mass-density and $P$ is pressure. I will additionally 
assume that the flow is non-turbulent and that the plasma is non-compressible. The former assumption amounts to dropping the terms non-linear in velocity, while the later implies vanishing divergence of velocity
\beql{non-comp}
\b\nabla\cdot \b V=0\,,
\eeq 
Because of the approximate boost invariance of the heavy-ion collisions, we can restrict our attention to the two dimensional flow in the $xy$ plane corresponding to the central rapidity region. 

The viscous pressure tensor in vanishing magnetic field is isotropic in the $xy$-plane and is given by 
\beql{P0}    
\Pi_{\alpha\beta}^0= \eta \left( \frac{\partial V_\alpha}{\partial x_\beta}+\frac{\partial V_\beta}{\partial x_\alpha}\right) = 2\eta \left(
\begin{array}{cc}
V_{xx} & V_{xy}\\
V_{yx} & V_{yy}
\end{array}
\right)\,,
\eeq
where the superscript $0$ indicates absence of the magnetic field. In the opposite case of very strong magnetic field the viscous pressure tensor has a different form \eq{visc.tens1}. Neglecting all $\eta_n$ with $n\ge 1$ we can write
\beql{PB}
\Pi_{\alpha\beta}^\infty = \eta_0\left(
\begin{array}{cc}
-V_{yy} & 0\\
0 & 2V_{yy}
\end{array}
\right) = 
2\eta_0\left(
\begin{array}{cc}
\frac{1}{2}V_{xx} & 0\\
0 & V_{yy}
\end{array}
\right)\,,
\eeq
where we also used \eq{non-comp}. 
Notice that $\Pi_{xx}^\infty=\frac{1}{2}\Pi_{yy}^\infty=\frac{1}{2}\Pi_{xx}^0$ indicating that the plasma flows in the direction perpendicular to the magnetic field with twice as small viscosity as in the direction of the field. The later is not affected by the field at all, because the Lorentz force vanishes in the field direction.
Substituting \eq{PB} into \eq{n-s} we derive the following two equations characterizing the plasma velocity in the strong magnetic field  \cite{Tuchin:2011jw}
\begin{align}\label{eq1}
\rho \frac{\partial V_x}{\partial t}= -\frac{\partial P}{\partial x}+\eta_0\frac{\partial ^2 V_x}{\partial x^2}\,,\qquad 
\rho \frac{\partial V_y}{\partial t}= -\frac{\partial P}{\partial y}+2\eta_0\frac{\partial ^2 V_y}{\partial y^2}\,.
\end{align}
Additionally, we need to set the initial conditions 
\begin{align}\label{ic1}
V_x\big|_{t=0}= \varphi_1(x,y)\,, \qquad V_y\big|_{t=0}= \varphi_2(x,y)\,.
\end{align}
The solution to the the problem \eq{eq1},\eq{ic1} is
\begin{subequations}
\begin{align}\label{sol1}
&V_x(x,y,t)= \int_{-\infty}^\infty dx' \varphi_1(x',y)G_{\frac{1}{2}}(x-x',t)-\frac{1}{\rho}\int_0^tdt'\int _{-\infty}^\infty dx' G_{\frac{1}{2}}(x-x',t-t')\frac{\partial P(x',y,t') }{\partial x'}\,,\\
&V_y(x,y,t)= \int_{-\infty}^\infty dy' \varphi_2(x,y')G_1(y-y',t)-\frac{1}{\rho}\int_0^tdt'\int _{-\infty}^\infty dy' G_1(y-y',t-t')\frac{\partial P(x,y',t') }{\partial y'}\,.\label{sol2}
\end{align}
\end{subequations}
Here the Green's function is given by 
\beql{Green}
G_k(y,t)= \frac{1}{\sqrt{4\pi a^2k t}}e^{-\frac{y^2}{4a^2kt}}
\eeq
and the diffusion coefficient by
\beql{dif}
a^2= \frac{2\eta_0}{\rho}\,.
\eeq

Suppose that the pressure is isotropic, i.e.\ it depends on the coordinates $x$,$y$ only via the radial coordinate $r= \sqrt{x^2+y^2}$; accordingly we pass from the integration variables $x'$  and $y'$ to $r'$  in \eq{sol1} and \eq{sol2} correspondingly.  At later times we can expand the Green's function \eq{Green} in inverse powers of $t$. The first terms in the r.h.s.\ of \eq{sol1} and \eq{sol2} are subleading  and we obtain
\begin{subequations}
\begin{align}\label{q1}
V_x(x,y,t)&\approx  -\frac{1}{\rho}\int_0^tds \int _{-\infty}^\infty dr\, \frac{1}{\sqrt{2\pi a^2 s}}\frac{\partial P(r,t-s) }{\partial r}\nonumber\\
&=-\frac{1}{\rho}\int_0^tds  \frac{1}{\sqrt{2\pi a^2 s}}\left[ P(R(s),t-s)-P(0,t-s)\right] \,,
\end{align}
and by the same token 
\begin{align}\label{q2}
&V_y(x,y,t)\approx -\frac{1}{\rho}\int_0^tds \, \frac{1}{\sqrt{4\pi a^2 s}}\left[ P(R(s),t-s)-P(0,t-s)\right]\,,
\end{align}
\end{subequations}
where $R(t)$ denotes the boundary beyond which the density of the plasma is below the critical value. 
We observe that $V_x/V_y= \sqrt{2}$.
Consequently, the azimuthal anisotropy of the hydrodynamic flow is \cite{Tuchin:2011jw}
\beql{anisotropy}
\frac{V_x^2-V_y^2}{V_x^2+V_y^2}= \frac{1-\frac{1}{2}}{1+\frac{1}{2}}=\frac{1}{3}\,.
\eeq
Since I assumed that the  initial conditions and  the pressure are isotropic, the azimuthal asymmetry \eq{anisotropy} is generated exclusively by the magnetic field. 

We see that at later times after the heavy-ion collision, flow velocity is proportional to  $\eta_0^{-1/2}$, where $\eta_0$ is the finite shear viscosity coefficient, see  \eq{sol1} and \eq{sol2}. If the system is such that in absence of the magnetic field it were azimuthally symmetric, then the magnetic field induces azimuthal asymmetry of 1/3, see \eq{anisotropy}.  The effect of the magnetic field on flow is strong and must be taken into account in phenomenological applications. 
Neglect of the contribution by the magnetic field  leads to underestimation of the phenomenological value of viscosity extracted from the data \cite{Song:2007fn,Romatschke:2007mq,Dusling:2007gi}.  In other words, the more viscous QGP in magnetic field produces the same azimuthal anisotropy as a less viscous QGP in vacuum. 


A model that I considered in this section to illustrate the effect of the magnetic field on the azimuthal anisotropy of a viscous fluid flow does not take into account many important features of a realistic heavy-ion collision. 
To be sure, a comprehensive approach must involve numerical solution of the relativistic magnetohydrodynamic equations with a realistic geometry. 
A potentially important effect that I have not considered here, is plasma instabilities \cite{Weibel,Mrowczynski:1994xv},  which warrant further investigation. 

The structure of the viscous stress tensor in  very strong magnetic field \eq{PB} is general, model independent.  However, as explained, the precise amount of the azimuthal anisotropy that it generates cannot be determined without taking into account many important effects.  Even so, I  draw the reader's attention to the fact that analysis of \cite{Mohapatra:2011ku} using quite different arguments arrived at  similar conclusion. Although a more quantitive numerical calculation is certainly required before a final conclusion can be made, it looks very plausible that the QGP viscosity is significantly higher than the presently accepted value extracted without taking into account the magnetic field effect \cite{Song:2007fn,Romatschke:2007mq,Dusling:2007gi} and is perhaps closer to the value calculated using the  perturbative theory \cite{Arnold:2003zc,Baym:1990uj}.

\setcounter{equation}{0}
\newpage
\section{Energy loss and polarization due to synchrotron radiation}\label{sec:d}

\subsection{Radiation of fast quark in magnetic field}\label{sec:da}

General problem of charged fermion radiation  in   external magnetic field was solved in \cite{Nikishov:1964zza,Sokolov:1963zn,Ritus-dissertation}.  It has important applications in collider physics, see e.g.\ \cite{Berestetsky:1982aq,Jackson:1975qi}. In heavy-ion phenomenology, synchrotron radiation provides one of the mechanisms of  energy loss in quark-gluon plasma, which  is an important probe of QGP \cite{Gyulassy:1993hr,Baier:1994bd}.\footnote{Synchrotron radiation in chromo-magnetic fields was discussed in \cite{Shuryak:2002ai,Kharzeev:2008qr,Zakharov:2008uk}.}

\begin{figure}[ht]
      \includegraphics[height=4cm]{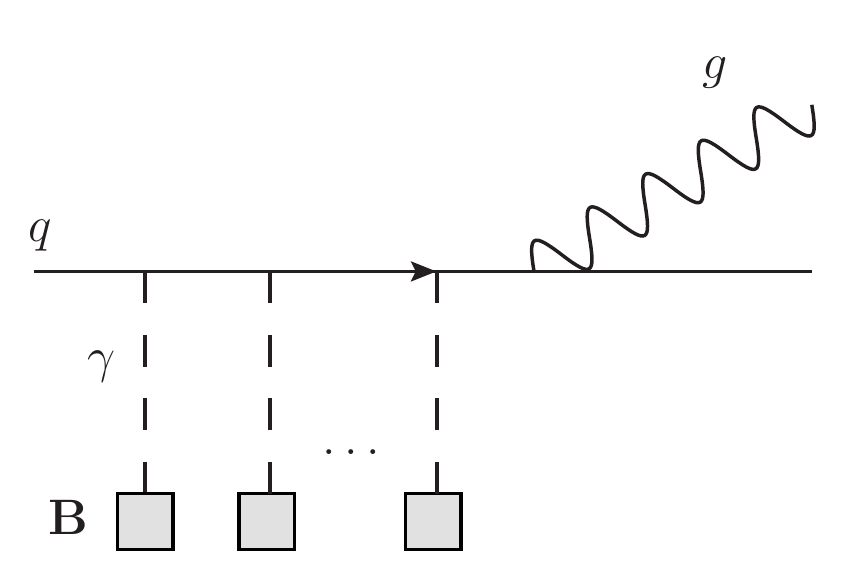} 
  \caption{A typical diagram contributing to the synchrotron radiation by a quark. }
\label{bremss}
\end{figure}

A typical diagram contributing to the synchrotron radiation, i.e.\ radiation in external magnetic field, by a quark is shown in \fig{bremss} \cite{Tuchin:2010vs}. This diagram is proportional to $(eB)^n$, where $n$ is the number of external field lines. In strong field,  powers of $eB$ must be summed up,  which may be accomplished by exactly solving the Dirac equation for  the relativistic fermion and then calculating the matrix element for the transition $q\to q+g$. 
Such calculation has been done in QED for some special cases including the  homogeneous constant field and can be readily generalized for gluon radiation. 
Intensity of the radiation can be expressed via the invariant parameter $\chi$ defined as 
\beq\label{chi}
\chi^2 =-\frac{\alpha_\mathrm{em}Z_q^2\hbar^3}{m^6}\,(F_{\mu\nu}p^\nu)^2 = \frac{\alpha_\mathrm{em}Z_q^2\hbar^3}{m^6}(\b p\times \b B)^2
\eeq
where the initial quark 4-momentum is $p^\mu=(\e,\b p)$, $Z_q$ is the quark charge in units of the absolute value of the electron charge $e$. At high energies,
\beql{chi-appr}
\chi \approx \frac{Z_qB\e}{B_cm}\,.
\eeq
The regime of weak fields corresponds to $\chi\ll 1$, while in strong fields $\chi \gg 1$. In our case, $eB/eB_c \approx (m_\pi/m_u)^2\gg 1$ (at RHIC) and therefore $\chi\gg 1$.  
In terms of $\chi$, spectrum of radiated gluons of frequency $\omega$ can be written as \cite{Nikishov:1964zza}
\beq\label{glue-spec}
\frac{dI}{d\omega} = -\as C_F \,\frac{m^2 \,\omega}{\e^2}\left\{ 
\int_x^\infty \mathrm{Ai}(\xi)\,d\xi +
\left( \frac{2}{x}+\frac{\omega}{\e}\,\chi \, x^{1/2}\right)\mathrm{Ai}'(x)
\right\}\,,
\eeq
where $I$ is the intensity,
$$
x= \left( \frac{\hbar\omega}{\e'\chi}\right)^{2/3}\,,
$$
and $\e'$ is the quark's energy in the final state.    $\mathrm{Ai}$ is the Ayri function. Eq.~\eq{glue-spec} is valid under the assumption that the initial quark remains ultra-relativistic, which implies that the energy loss due to the synchrotron radiation $\Delta\e$ should be small compared to the quark energy itself $\Delta \e\ll \e$.

Energy loss by a relativistic quark per unit length  is given by \cite{Berestetsky:1982aq}
\beql{enegyloss}
\frac{d\e}{dl}=- \int_0^\infty d\omega \frac{dI}{d\omega} = \alpha_s C_F\,\frac{m^2\,\chi^2}{2}\,\int_0^\infty\frac{4+5\chi\, x^{3/2}+4\chi^2\,x^3}{(1+\chi\, x^{3/2})^4}\,\mathrm{Ai}'(x)\,x\,dx\,.
\eeq
In two interesting limits, energy loss behaves quite differently. At $\eta=\varphi=0$ we have \cite{Berestetsky:1982aq}
\begin{subequations}
 \begin{eqnarray}\label{est}
\frac{d\e}{dl}&=&-\frac{2\,\alpha_s\,\hbar\, C_F\,(Z_qeB)^2\e^2}{3m^4}\,,\quad \chi\ll 1\,,\\
\frac{d\e}{dl}&=&-0.37\,\as \,\hbar^{-1/3}\,C_F\,\left( Z_qeB\, \e \right)^{2/3}\,,\quad \chi\gg 1\,.
\end{eqnarray}
\end{subequations}
In the strong field limit energy loss is independent of the quark mass, whereas in the weak field case it decreases as $m^{-4}$. 
Since $\chi\propto \hbar$, limit of $\chi\ll 1$ corresponds to the classical energy loss. 

\begin{figure}[ht]
      \includegraphics[height=9cm]{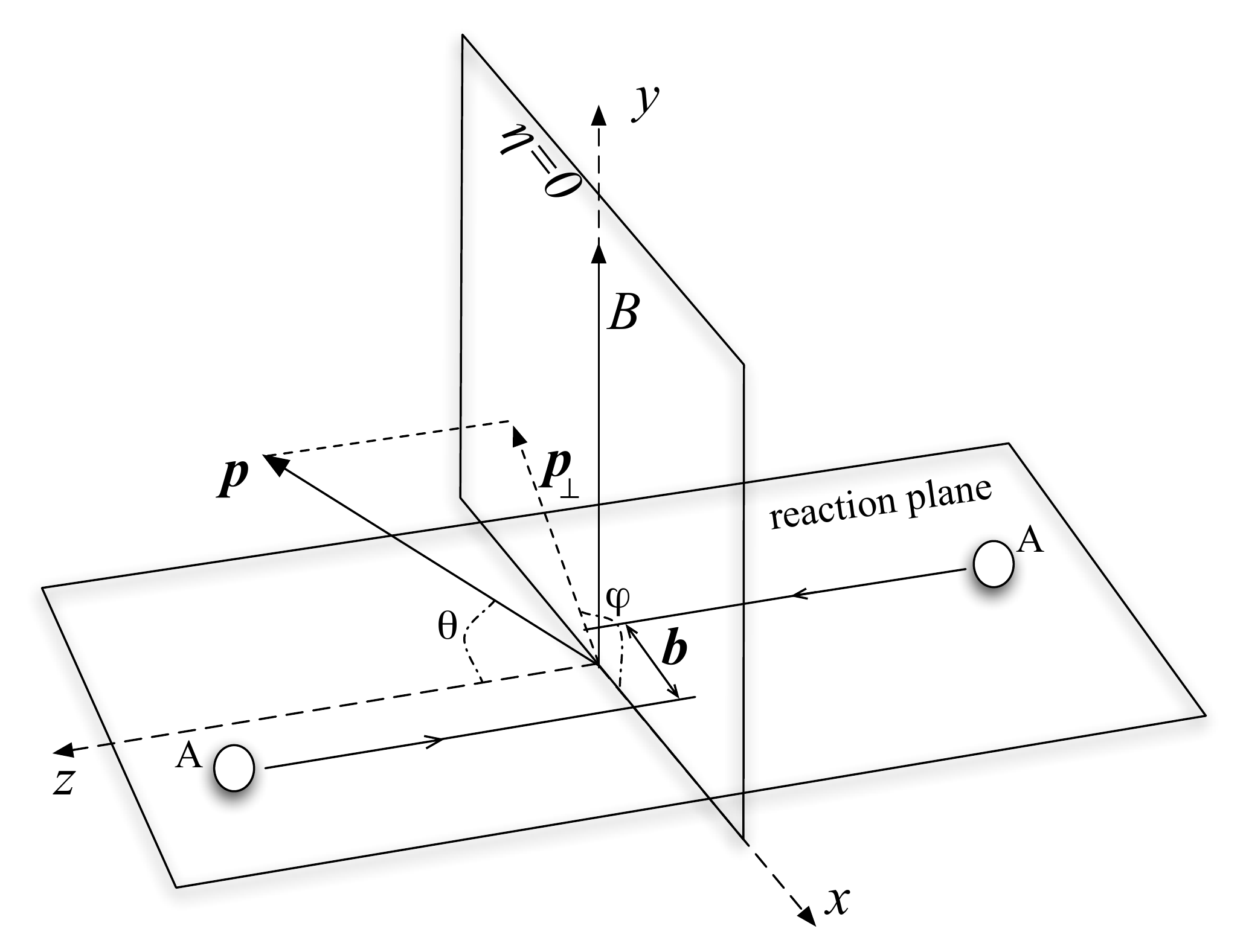} 
  \caption{Geometry of a heavy-ion collision. $\b p$ denotes momentum of a  fast quark.  Two orthogonal planes are the reaction plane  span by the initial heavy-ion momenta, and the mid-rapidity plane $\theta=\pi/2$, which is labeled as $\eta=0$. }
\label{geom}
\end{figure}

To apply this result to heavy-ion collisions we need to write down the invariant $\chi$ in a suitable kinematic variables. The geometry of a heavy-ion collision is depicted in \fig{geom}. Magnetic field $\b B$ is orthogonal to the reaction plane span by the impact parameter vector $\b b$ and  the collision axis ($z$-axis). For a quark of momentum $\b p$ we define the polar angle $\theta$ with respect to the $z$-axis and azimuthal angle $\varphi$ with respect to the reaction plane. 
In this notation, $\b B= B\,\unit y$ and $\b p = p_z\unit z+p_\bot (\unit x\, \cos\varphi+\unit y\sin\varphi)$, where $ p_\bot = |\b p|\sin\theta\approx \e \sin\theta$. Thus, $(\b B\times\b p)^2= B^2(p_z^2+p_\bot^2\cos^2\varphi)$. Conventionally, one expresses the longitudinal momentum and energy using the rapidity $\eta$ as $\e = m_\bot \cosh\eta$ and $p_z= m_\bot \sinh\eta$, where $m_\bot^2= m^2+p_\bot^2$.
We have
\beql{chi-hi}
\chi^2= \frac{\hbar^2(e B)^2}{m^6}\,p_\bot^2 (\sinh^2\eta+\cos^2\varphi)\,.
\eeq

\begin{figure}[ht]
\begin{tabular}{cc}
      \includegraphics[height=1.8in]{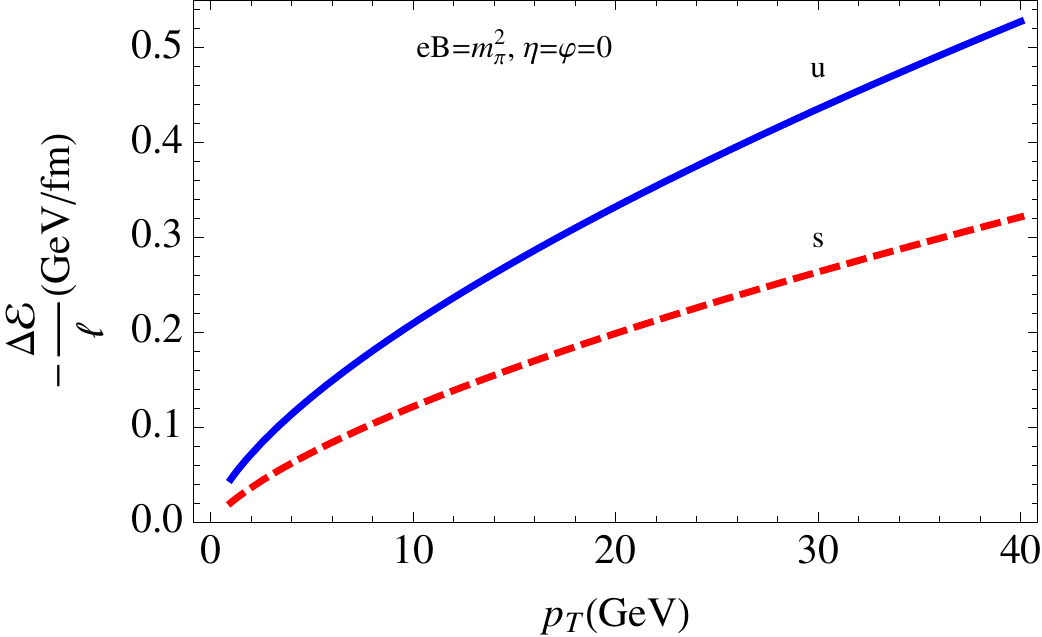} &
      \includegraphics[height=1.8in]{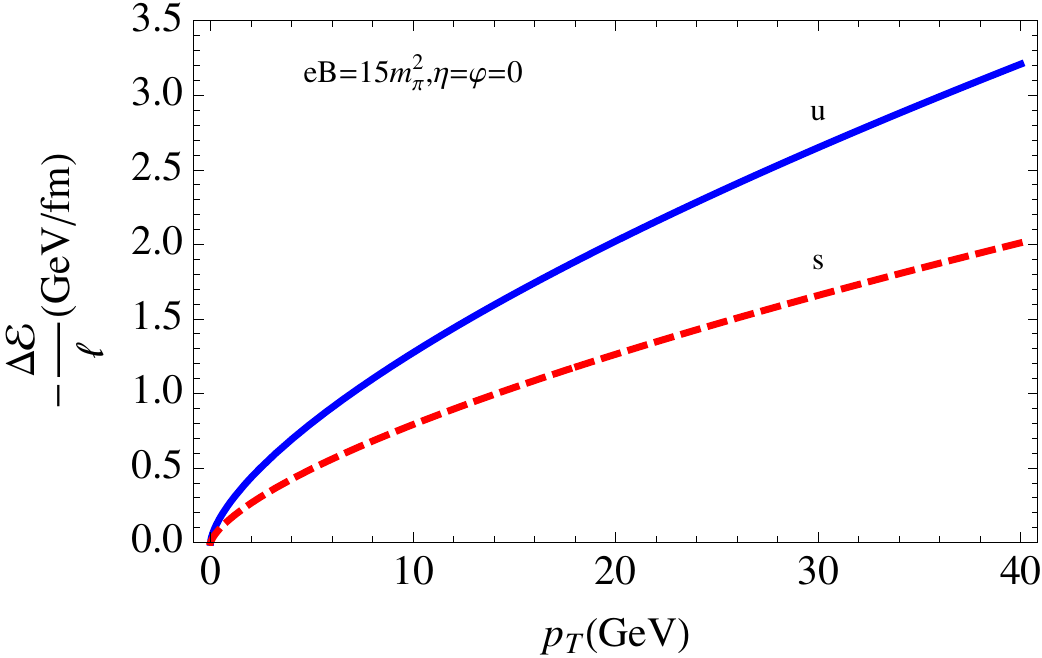}\\
      $(a)$ & $(b)$  \\[0.2in]
        \includegraphics[height=1.8in]{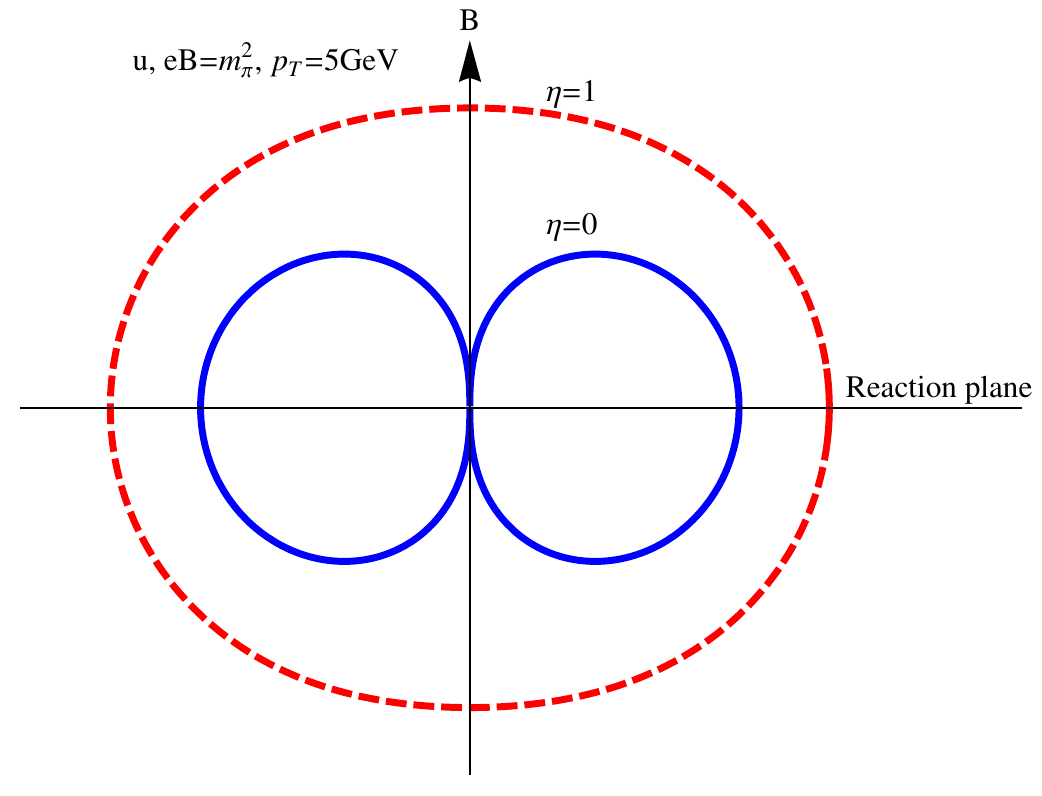} &
      \includegraphics[height=1.8in]{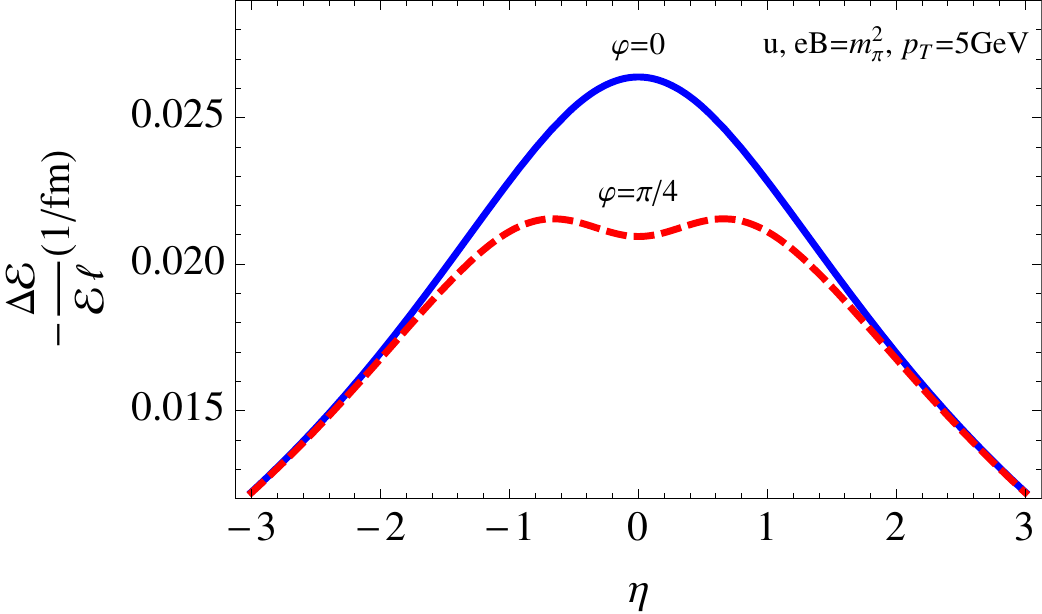}\\
        $(c)$ & $(d)$
      \end{tabular}
  \caption{Energy loss \emph{per unit length} by  quarks moving in constant external magnetic field  as a function of (a)  $p_T$ at RHIC at $\eta=\varphi=0$, (b)  $p_T$ at LHC at $\eta=\varphi=0$, (c)  azimuthal  angle $\varphi$ with respect to the reaction plane for $p_T=5$~GeV and $\eta=0,1$. (d)  Fractional energy loss vs rapidity $\eta$ at $p_T=5$~GeV and $\varphi= 0,\pi/4$.  }
\label{fig:loss-const}
\end{figure}

In \fig{fig:loss-const} a numerical calculation of the energy loss \emph{per unit length} in a constant magnetic field using \eq{enegyloss} and \eq{chi-hi} is shown \cite{Tuchin:2010vs}. We see that energy loss  of a $u$ quark  with $p_\bot = 10$~GeV is about 0.2~GeV/fm at RHIC and 1.3~GeV/fm at LHC. This corresponds  to the loss of 10\% and 65\% of its initial energy  after traveling 5~fm at RHIC and LHC respectfully. Therefore, energy loss due to the synchrotron radiation at LHC gives a phenomenologically important contribution to the total energy loss. 

Energy loss due to the synchrotron radiation has a very non-trivial azimuthal angle and rapidity dependence that comes from the corresponding dependence of the $\chi$-parameter \eq{chi-hi}. As can be seen in  \fig{fig:loss-const}(c), energy loss has a minimum at 
$\varphi=\pi/2$ that corresponds to quark's transverse momentum $\b p_\bot$ being parallel (or anti-parallel) to the field direction. It has a maximum  at $\varphi= 0,\pi$ when $\b p_\bot$ is perpendicular to the field direction and thus lying in the reaction plane. It is obvious from \eq{chi-hi} that at mid-rapidity $\eta=0$ the azimuthal angle dependence is much stronger pronounced than in the forward/backward direction. 
Let me emphasize, that  the energy loss \eq{enegyloss} divided by $m^2$, i.e.\ $d\e/(dl\, m^2)$ scales with $\chi$. In turn, $\chi$ is a function of magnetic field, quark mass, rapidity and azimuthal angle. Therefore,  all the features seen in \fig{fig:loss-const} follow from this scaling behavior. 

\subsection{Azimuthal asymmetry of  gluon spectrum}\label{sec:db}

In magnetic field  gluon spectrum  is azimuthally asymmetric. It is customary to describe this asymmetry by Fourier  coefficients of intensity defined as 
\beql{db1}
I(\varphi)= \bar I\left(1+2\sum_{n=1}^\infty v_n\,\cos(n\varphi)\right)\,,\quad v_n= \frac{1}{2\pi\bar I}\int_{-\pi}^{\pi}I(\varphi)\,\cos(n\varphi)\,d\varphi\,.
\eeq
Azimuthal averaged intensity is $\bar I = \int_{-\pi}^\pi I(\varphi)d\varphi/2\pi$. In strong fields $\chi\gg 1$ and we can write
\beql{db3}
I(\varphi) = 0.37\,\as C_F m^2 \chi^{2/3}= 0.37\, \as C_F (eBp_\bot)^{2/3}\, (\sinh^2\eta+\cos^2\varphi)^{1/3}\,.
\eeq
We have
\beql{db5}
v_n = \frac{\int_{-\pi}^\pi I(\varphi) \cos(n\varphi)d\varphi}{\int_{-\pi}^\pi I(\varphi) d\varphi}=
\frac{\int_{-\pi}^\pi (\sinh^2\eta+\cos^2\varphi)^{1/3} \cos(n\varphi)d\varphi}{\int_{-\pi}^\pi (\sinh^2\eta+\cos^2\varphi)^{1/3} d\varphi}\,.
\eeq
At $\eta =0$ the Fourier coefficients $w_n$ can be calculated analytically using formula 3.631.9 of \cite{GR}
\begin{align}\label{fcoeff}
v_{n}=\frac{B\left(\frac{4}{3},\frac{4}{3} \right)}{B\left(\frac{4}{3}+\frac{n}{2},\frac{4}{3}-\frac{n}{2} \right)}\,,\quad \text{if}\quad n\,\in\, \text{even}\,;\qquad 
v_n=0\,, \quad \text{if}\quad n\,\in\, \text{odd}\,,
\end{align}
where $B$ is the Euler's Beta-function. The corresponding numerically values of the lowest harmonics are $v_2=0.25$, $v_4= -0.071$, $v_6=0.036$, $v_8=-0.022$, $v_{10}=0.015$.

\subsection{Polarization of light quarks }\label{sec:dc}

 Synchrotron radiation leads to polarization of   electrically charged fermions; this is  known as the Sokolov-Ternov effect \cite{Sokolov:1963zn}.  It was applied to heavy-ion collisions in \cite{Tuchin:2010vs}.
Unlike energy loss that I discussed so far, this is a purely quantum effect. 
It arises because the probability of the spin-flip transition depends on the orientation of the quark spin with respect to the direction of the magnetic field and on the sign of fermion's electric charge. The spin-flip probability per unit time reads  
\cite{Sokolov:1963zn}
\beql{wst}
w=\frac{5\sqrt{3}\as C_F}{16}\frac{\hbar^2}{m^2}\left( \frac{\e}{m}\right)^5\,\left( \frac{Z_qe\,|\b v\times \b B|}{\e}\right)^3\,\left( 1-\frac{2}{9}\,(\b\zeta\cdot \b v)^2-\frac{8\sqrt{3}}{15}\mathrm{sign}\,(e_q)\,(\b \zeta\cdot \unit z)\right)\,,
\eeq
where 
$\b\zeta$ is a unit  axial vector that coincides with the direction of  quark spin in its rest frame, $\b v= \b p/\e$ is the initial fermion velocity. 

The nature of this spin-flip transition is transparent in the non-relativistic case, where it  is induced by the interaction Hamiltonian \cite{Jackson:1975qi}
\beql{nr}
\mathcal{H}=-\b \mu\cdot \b B= -\left( \frac{geZ_q\hbar}{2m}\right)\,\b s\cdot \b B\,,
\eeq
It is seen, that negatively charged quarks and anti-quarks  (e.g.\ $\bar u$ and $ d$) tend to align against the field, while the positively charged ones (e.g.\ $ u$ and $\bar d$) align parallel to the field. 

Let $n_{\uparrow(\downarrow)}$ be the number of fermions or anti-fermions with given momentum and spin direction parallel (anti-parallel) to the field produced in a given event. At initial time $t=t_0$ the spin-asymmetry defined as 
\beql{as1}
 A = \frac{n_\uparrow-n_\downarrow}{n_\uparrow+n_\downarrow}\,
\eeq
vanishes. Eq.~\eq{wst} implies that at later times, a beam of  non-polarized  fermions develops a finite asymmetry   
given by \cite{Sokolov:1963zn}
\beql{as2}
A= \frac{8\sqrt{3}}{15}\left(1-e^{-\frac{t-t_0}{\tau}}\right)\,,
\eeq
 where
 \beql{tauP}
 \tau= \frac{8}{5\sqrt{3}\, m\, \alpha_s C_F}\,\left(\frac{m}{\e}\right)^2\,\left( \frac{m^2}{Z_qe|\b v\times \b B|}\right)^3\,
\eeq
is the characteristic time  over which the maximal possible asymmetry is achieved. 
This time is extremely small on the scale of $t_0$. For example, it takes only $\tau=6\cdot 10^{-8}$~fm for a $d$ quark of momentum $p_\bot =1$~GeV at $\eta=\varphi=0$ at RHIC to achieve the maximal asymmetry of $A_m= \frac{8}{5\sqrt{3}}=92$\%. Therefore, quarks and anti-quarks are polarized almost instantaneously after being released from their wave functions. However, subsequent interaction with QGP and fragmentation wash out the polarization of quarks. 

A more sensitive probe are leptons  weakly interacting with QGP and not  undergoing a fragmentation process. Thus, their polarization can present a direct experimental evidence for the existence and strength of magnetic field. In case of muons we can estimate $\tau$ by replacing $\as C_F\to \alpha_\mathrm{em}$. For muons we get $\tau =0.3 $~fm/c, which is still much smaller than magnetic field life-time. Observation of such a lepton polarization asymmetry is perhaps the most definitive proof of existence of the strong magnetic field at  early times after a heavy-ion collision  regardless of its later time-dependence.

\bigskip

{\bf In summary}, energy loss per unit length for a light quark with $p_T=15$~GeV is about 0.27~GeV/fm at RHIC and 1.7~GeV/fm at LHC, which is comparable to the losses due to interaction with the plasma. Thus, the synchrotron radiation alone is able to account for quenching of jets at LHC with $p_\bot$ as large as 20~GeV. Synchrotron radiation is definetely one of missing pieces in the puzzle of the jet energy loss in heavy ion collisions. Quarks and leptons are expected to be strongly polarized in plasma in the direction parallel or anti-parallel to the magnetic field depending on the sign of their electric charge.

\setcounter{equation}{0}
\newpage
\section{Photon decay}\label{sec:f}

In this section I consider pair-production by photon in external magnetic field \cite{Tuchin:2010gx}, which is  a cross-channel of synchrotron radiation discussed in the previous section. 
Specifically, we are interested to determine photon decay rate $w$ in the process $\gamma B\to f\bar f B$, where  $f$ stands for a charged fermion, as a function of photon's transverse momentum $k_T$, rapidity $\eta$ and azimuthal angle $\varphi$. Origin of these photons in heavy-ion collisions will not be of interest to us in this section.

Characteristic frequency of a fermion  of species $a$ of mass  $m_a$ and charge $z_ae$ ($e$ is the absolute value of electron charge) moving in external magnetic field $B$ (in a plane perpendicular to the field direction) is 
\beql{char}
\hbar\omega_B= \frac{z_aeB}{\e}\,,
\eeq
where $\e$ is the fermion energy. Here -- in the spirit of the adiabatic approximation -- $B$ is a slow function of time. Calculation of the photon decay probability significantly simplifies if motion of electron is quasi-classical, i.e.\ quantization of fermion motion in the magnetic field can be neglected. This condition is fulfilled if $\hbar \omega_B\ll \e$. This implies that
\beql{est1}
\e\gg  \sqrt{zeB}\,.
\eeq
For RHIC it is equivalent to $\e\gg m_\pi$, for LHC $\e\gg 4m_\pi$.

Photon decay rate was calculated in \cite{Nikishov:1964zza} and, using the quasi-classical method, in \cite{Baier:1964}. It reads
\beql{main}
w=-\sum_a\frac{\alpha_\mathrm{em}\,z_a^3 \,e B }{m_a \varkappa_a}\int^\infty_{(4/ \varkappa_a)^{2/3}}\frac{2(x^{3/2}+1/ \varkappa_a)\,\text{Ai}'(x)}{x^{11/4}(x^{3/2}-4/ \varkappa_a)^{3/2}}\,,
\eeq
where summation is over fermion species and the invariant parameter $\varkappa$ is defined as 
\beq\label{kappa}
 \varkappa_a^2 =-\frac{\alpha_\mathrm{em}z_a^2\hbar^3}{m_a^6}\,(F_{\mu\nu}k^\nu)^2 = \frac{\alpha_\mathrm{em}z_a^2\hbar^3}{m_a^6}(\b k\times \b B)^2\,,
\eeq
with the initial photon 4-momentum  $k^\mu=(\hbar \omega,\b k)$.  With notations of \fig{geom}, $\b k = k_z\unit z+k_\bot (\unit x\, \cos\varphi+\unit y\sin\varphi)$, where $ k_\bot = |\b k|\sin\theta=\hbar \omega  \sin\theta$. Thus, $(\b B\times\b k)^2= B^2(k_z^2+k_\bot^2\cos^2\varphi)$. Employing $\hbar \omega = k_\bot \cosh\eta$ and $k_z= k_\bot \sinh\eta$ we write
\beql{kappa-hi}
 \varkappa_a= \frac{\hbar(z_a e B)}{m_a^3}\,k_\bot \sqrt{\sinh^2\eta+\cos^2\varphi}\,.
\eeq

\begin{figure}[ht]
\begin{tabular}{cc}
      \includegraphics[height=5cm]{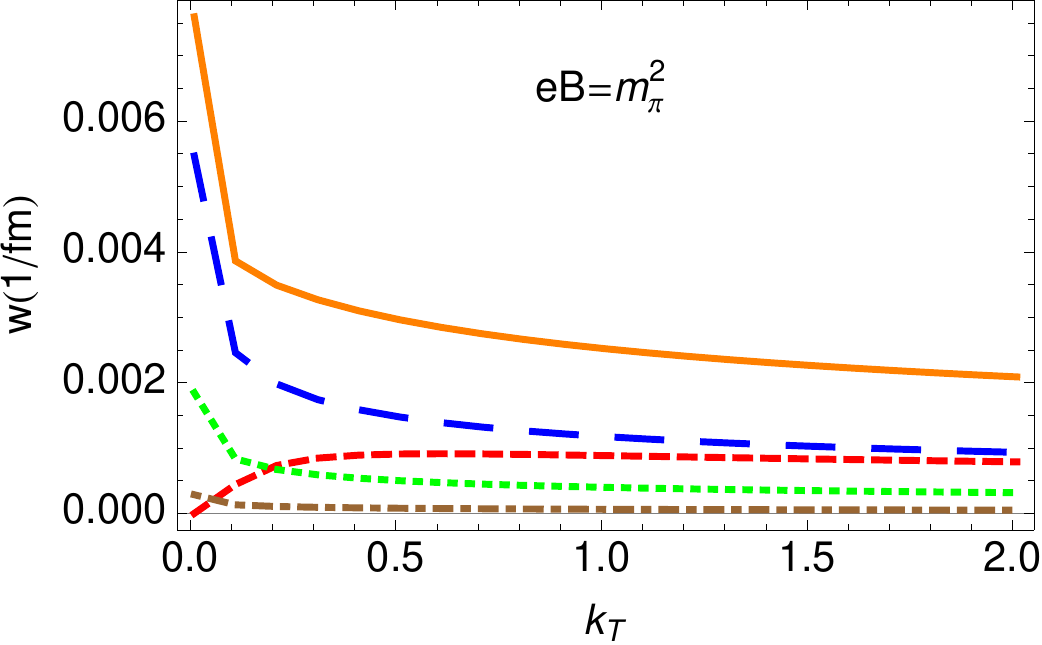} &
      \includegraphics[height=5cm]{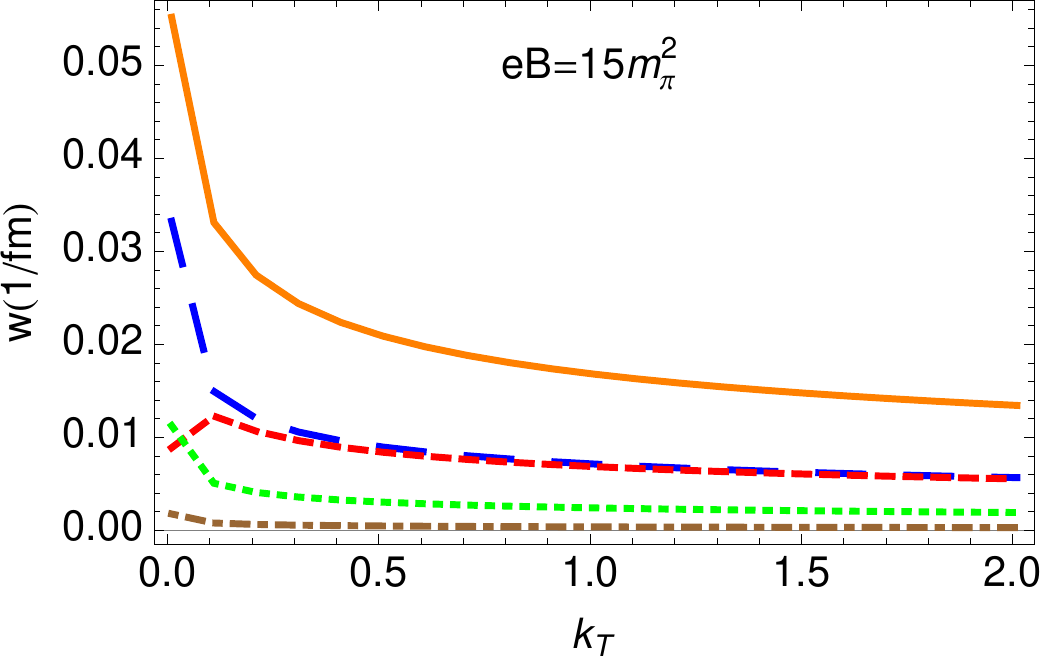}\\
      (a) & (b)
   \end{tabular}   
  \caption{Decay rate of photons moving in reaction plane in magnetic field as a function of transverse momentum $k_T$: (a) at RHIC, (b) at LHC. Broken lines from bottom to top give contributions of $\gamma\to d\bar d$, $\gamma\to u\bar u$, $\gamma\to \mu^+\mu^-$ and  $\gamma\to e^+e^-$ channels. Upper solid line is the total rate.   }
\label{rate-kt}
\end{figure}

Plotted in \fig{rate-kt}  is the  photon decay rate  \eq{main} for RHIC and LHC.
The survival probability of photons in magnetic field is $P= 1-w\Delta t $, where $\Delta t$ is the time spent by a photon in plasma.  Estimating $\Delta t = 10$~fm we determine that photon survives with probability $P_\text{RHIC}\approx 97$\% at RHIC, while only  $P_\text{LHC}\approx 80$\% at LHC. Such strong depletion can certainly be observed in heavy-ion collisions at LHC. 

\begin{figure}[ht]
      \includegraphics[height=8cm]{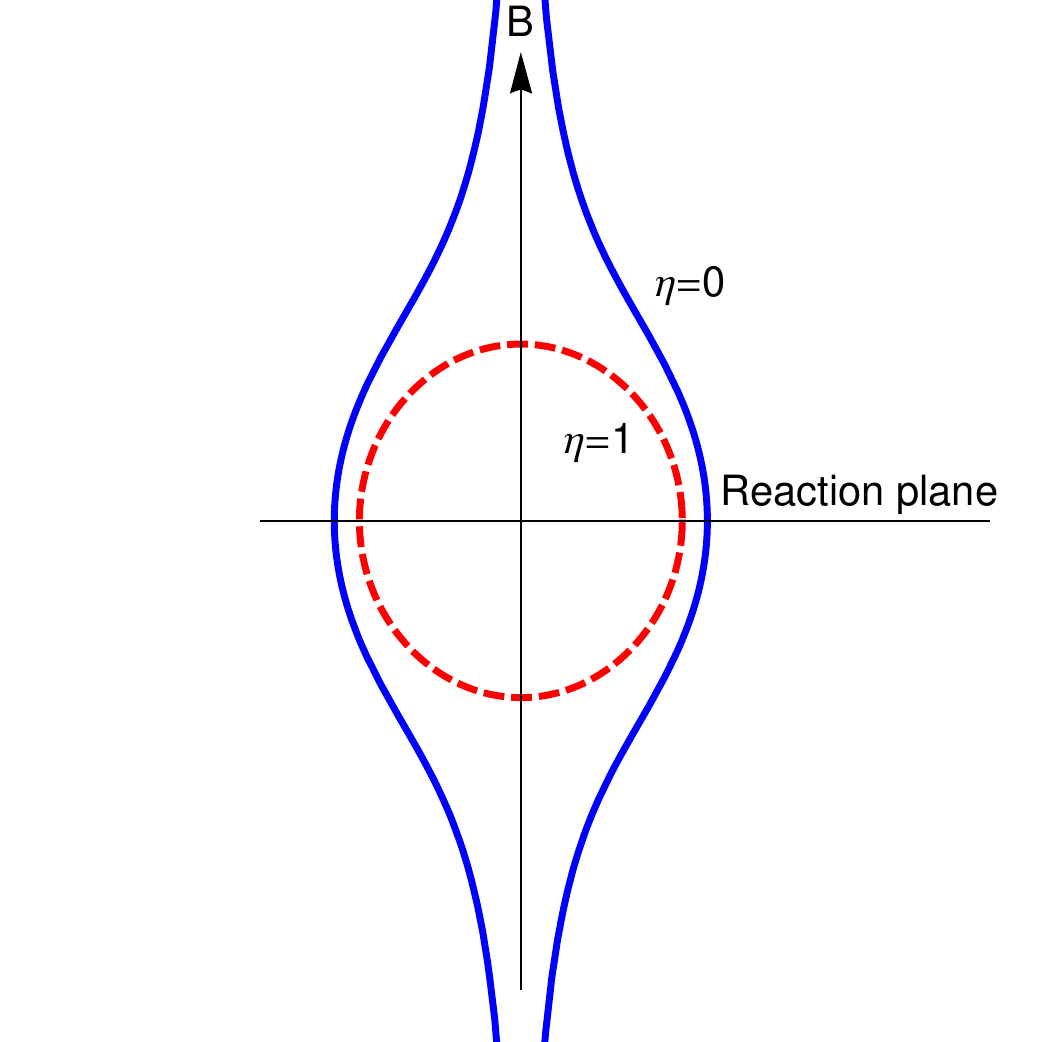} 
  \caption{Azimuthal distribution of the decay rate of photons at different rapidities at LHC. Only contribution of the $\gamma\to e^+e^-$ channel is shown.   }
\label{rate-azimuth}
\end{figure}

Azimuthal distribution of the decay rate of photons at LHC is azimuthally asymmetric as can be seen in \fig{rate-azimuth} \cite{Tuchin:2010gx}. The strongest suppression is in the $B$ field direction, i.e.\ in the direction orthogonal to the reaction plane. At $\eta\gtrsim 1$ the $\varphi$ dependence of $\varkappa_a$ is very weak which is reflected in nearly symmetric azimuthal shape of the  dashed line in \fig{rate-azimuth}.

To quantify the azimuthal asymmetry it is customary to expand
the decay rate in Fourier series with respect to the azimuthal angle. Noting that $w$ is an even function of $\varphi$ we have
\beql{four}
w(\varphi)= \frac{1}{2}w_0+\sum_{n=1}^\infty w_n\,\cos(n\varphi)\,,\quad w_n= \frac{1}{\pi}\int_{-\pi}^{\pi}w(\varphi)\,\cos(n\varphi)\,d\varphi\,.
\eeq
In strong fields  $\varkappa_a\gg 1$. For example, for $\gamma\to \mu^+\mu^-$ at RHIC at $\varphi=\eta=0$ and $k_T=1$~GeV we get $\varkappa_\mu = 19$. Therefore, we can expand the rate  \eq{main}  at large $\varkappa_a$ as \cite{Nikishov:1964zza}
\beql{appr}
w\approx \frac{3^{1/6}\, 5\, \Gamma^2\left(\frac{2}{3}\right)}{2^{4/3}\,7\, \pi^{1/2}\, \Gamma\left(\frac{7}{6}\right)}\sum_a\frac{\alpha_\mathrm{em}eB z_a^3}{m_a\varkappa_a^{1/3}}\equiv \frac{A}{(\sinh^2\eta +\cos^2\varphi)^{1/6}}\,,\quad \varkappa_a \gg 1\,.
\eeq
At $\eta =0$ the Fourier coefficients $w_n$ can be calculated analytically using formula 3.631.9 of \cite{GR}
\beql{fc}
w_{2k}=\frac{3\,2^{1/3}\,A}{B\left( \frac{5}{6}+k,\frac{5}{6}-k\right)}\,,\quad   w_{2k+1}=0\,,\quad k=0,1,2,\ldots\,,
\eeq
where $B$ is the Euler's Beta-function and $A$ is defined in \eq{appr}. 
Substituting these expressions into \eq{four} we find
\beql{f22}
w=\frac{1}{2}w_0\left[ 1-\sum_{k=1}^\infty \frac{\sqrt{\pi}\Gamma\left(-\frac{1}{6}\right)}{2^{2/3}B\left( \frac{5}{6}+k,\frac{5}{6}-k\right)}\cos(2k\varphi)\right]
\eeq
The first few terms in this expansion read
\beql{f1}
w= \frac{1}{2}w_0\left( 1-\frac{2}{5}\cos(2\phi)+\frac{14}{55}\cos(4\phi)-\ldots\right)\,,
\eeq

What is measured experimentally is not the decay rate, but rather the photon spectrum. 
This spectrum is modified by the survival probability $P$ which is obviously azimuthally asymmetric. To quantify this asymmetry we write using \eq{four}
\beql{pphi}
P= \bar P \left( 1+ \sum_{k=1}^\infty v_{2k}\cos(2\varphi k)\right)\,,\quad v_{2k}= -\frac{1-\bar P}{\bar P}\,\frac{2\,w_{2k}}{w_0}\,,
\eeq
where $\bar P = \aver{1-w \Delta t}_\varphi=1-w_0\Delta t$ is the survival probability averaged over the azimuthal angle. Since $w_0\Delta t \ll 1$, as can be seen in \fig{rate-kt}, we can estimate using \eq{appr} and \eq{fc}
\beql{vs}
v_{2k}\approx  - \frac{2w_{2k}}{w_0}\,w_0 \Delta t =- \frac{2w_{2k}}{w_0}\Delta t\,\frac{5\, 6^{2/3}\Gamma\left(\frac{2}{3}\right)}{7\pi}\,\sum_a\frac{\alpha_\mathrm{em} (eB)^{2/3}z_a^{8/3}}{(k_T)^{1/3}}\,.
\eeq
In particular, the ``elliptic flow" coefficient is \cite{Tuchin:2010gx}
\beql{v2P}
v_2= \Delta t\,\frac{2\, 6^{2/3}\Gamma\left(\frac{2}{3}\right)}{7\pi}\,\sum_a\frac{\alpha_\mathrm{em} (eB)^{2/3}z_a^{8/3}}{(k_T)^{1/3}}\,.
\eeq
For example, at $k_T=1$~GeV and $\Delta t \sim 10$~fm/c   one expects $v_2\simeq   2$\% at RHIC and  $v_2  \simeq 14$\% at LHC only due to the presence of  magnetic field.  
We see that decay of photons in external magnetic field significantly contributes to the photon asymmetry in heavy-ion collisions  along with other possible effects. 

\bigskip

{\bf In summary},  I calculated photon pair-production rate in external magnetic field created in off-central heavy-ion collisions. Photon decay leads to depletion of the photon yield by a few percent at RHIC and by as much as 20\% at the LHC. The decay rate depends on the rapidity and azimuthal angle. At mid-rapidity the azimuthal asymmetry of the decay rate translates into asymmetric photon yield and contributes to the ``elliptic flow". 
Let me also quote a known result  that photons polarized parallel to the field are 3/2 times more likely to decay than those polarized transversely \cite{Nikishov:1964zza}. Therefore, polarization of the final photon spectrum perpendicular to the field is a signature of existence of strong magnetic field. Finally, photon decay necessarily leads to enhancement of dilepton yield.


\setcounter{equation}{0}
\newpage
\section{Quarkonium dissociation in magnetic field }\label{sec:h}

\subsection{Effects of magnetic field on quarkonium}\label{sec:ha}

Strong magnetic field created in heavy-ion collisions generates a number of remarkable effects on quarkonium production some of which I will describe in this section.  
Magnetic field can be treated as static if the distance $\lambda$ over which it significantly varies is much larger than the quarkonium radius.   If $\Delta t$ is magnetic field life-time, then $\lambda\sim c\Delta t$. For a  quarkonium with binding energy $\e_b$ and radius $\as/\e_b$,  the quasi-static approximation applies when $\e_b \lambda  /\as\gg 1$. Estimating conservatively $\lambda  \sim 2$~fm we get for $\jpsi$: $\e_b\lambda/\as\approx 23$, which is comfortably large  to justify the quasi-static approximation, where I assumed that $\e_b$ is given by its vacuum value.  As temperature $T$ increases $\e_b$ drops. Temperature dependence of $\e_b$ is model dependent, however it is certain that eventually it vanishes at some finite temperature $T_0$. Therefore, only in  the close vicinity of $T_0$, i.e.\ at very small binding energies, the quasi-static approximation is not applicable.  I thus rely on  the quasi-static approximation to calculate $\jpsi$ dissociation \cite{Marasinghe:2011bt,Tuchin:2011cg}. 

Magnetic field has a three-fold effect on quarkonium:
\begin{enumerate}
\item \emph{Lorentz ionization}. Consider quarkonium traveling with constant velocity in  magnetic field in the laboratory frame. Boosting to the quarkonium  comoving  frame, we find mutually orthogonal electric and magnetic fields given by Eqs.~\eq{boost},\eq{lt}. In the presence of  an electric field quark and antiquark have a finite probability to tunnel through the potential barrier thereby causing  quarkonium dissociation. 
In atomic physics such  a process is referred to as Lorentz ionization. In the non-relativistic approximation, the tunneling probability is of order unity when the electric field $E$ in the comoving frame satisfies $eE\gtrsim m^{1/2}\e_b^{3/2}$ (for weakly bound states), where $m$ is quark mass, see \eq{fnr-gsmall}. This effect causes a significant increase in quarkonium dissociation rate; numerical calculation for $\jpsi$ is shown in \fig{fig:pt}.

\item \emph{Zeeman effect.} Energy of a quarkonium state depends on  spin $S$, orbital angular momentum $L$, and total angular momentum $J$. In a magnetic field these states split; the splitting energy in a weak field is  $\Delta M = \frac{eB_0}{2m}g J_z$, where $J_z=-J,-J+1,\ldots, J$ is  projection of the total angular momentum on the direction of magnetic field, $m$ is quark mass and $g$ is Land\'e factor depending on $J$, $L$ and $S$ in a well-known way, see e.g.\ \cite{LL3-113}.  For example, $\jpsi$ with $S=1$, $L=0$ and $J=1$ ($g\approx 2$) splits into three states with $J_z= \pm 1, 0$ and with mass difference $\Delta M=0.15$~GeV, 
where we used $eB_0=15m_\pi^2$. Thus,  the Zeeman  effect leads to the emergence of new quarkonium states in plasma. 

\item \emph{Distortion of the quarkonium potential} in magnetic field. This effect arises in higher order  perturbation theory and becomes important at field strengths of order $B\sim 3\pi m^2/e^3$  \cite{Machet:2010yg}. This  is $3\pi/\alpha$ times stronger than the critical Schwinger's field. Therefore, this effect can be neglected at the present collider energies. 

\end{enumerate}

Some of the notational definitions used in this section:   $\b V$ and $\b P$ are velocity and momentum of quarkonium in the lab frame;  $M$ is its mass; $\b p$ is the momentum of quark or anti-quark  in the comoving frame;  $m$ is its mass;  $\b B_0$ is the magnetic field in the lab frame, $\b E$ and $\b B$ are  electric and magnetic fields in the comoving frame; $\gamma_L$ is the quarkonium Lorentz factor; and  $\gamma$ is a parameter defined in \eq{gamma}. I use Gauss units throughout the section; note that expressions $eB$, $eE$ and $eB_0$ are the same in Gauss and Lorentz-Heaviside units.  

\subsection{Lorentz ionization: physical picture}

In this section I  focus on Lorentz ionization, which is an important mechanism of $\jpsi$ suppression in heavy-ion collisions \cite{Marasinghe:2011bt,Tuchin:2011cg}. Before we proceed to analytical calculations it is worthwhile to discuss the physics picture in more detail in two reference frames: the quarkonium proper frame and the lab frame.  
In the quarkonium proper frame the potential energy of, say, antiquark (with $e<0$) is a sum of its potential energy in the binding potential and its energy in the electric field $-eEx$, where $x$ is the electric field direction, see \fig{fig:V}.  Since $|e|Ex$ becomes large and negative at large and negative $x$ (far away from the bound state)  and because the quarkonium potential has finite radius, this region opens up for the motion of the antiquark. Thus  there is a quantum mechanical probability to tunnel through the potential barrier formed on one side by the vanishing quarkonium potential and on the other by increasing absolute value of the antiquark energy in electric field. Of course the total energy of the antiquark (not counting its mass) is negative after tunneling. However, its kinetic energy grows proportionally to $eEx$ as it goes away. By picking up a light quark out of vacuum it can hadronize into a $D$-meson. 
 
If we now go to the reference frame where $E=0$ and there is only magnetic field $B$ (we can always do so since $E<B$), then the entire process looks quite different. An energetic quarkonium travels in external magnetic field and decays into quark-antiquark pair that can late hadronize into $D$-mesons. This happens in spite of the fact that $\jpsi$ mass is smaller than masses of two $D$-mesons due to additional momentum $e\b A$ supplied by the magnetic field. Similarly a photon can decay into electron-positron pair in external magnetic field.

\subsection{Quarkonium ionization rate}\label{sec:hb}

\subsubsection{Comoving frame}\label{sec:com}

Consider a quarkonium traveling with velocity $\b V$ in constant magnetic field $\b B_0$.  Let $\b B $ and $\b E$ be magnetic and electric fields in the comoving frame, and let  subscripts $\parallel$ and $\bot$ denote  field components parallel and perpendicular  to $\b V$ correspondingly. Then,
\begin{subequations}\label{boost}
\begin{align}
&E_\parallel = 0\,,    & \b E_ \bot=\gamma_L \b V\times \b B_0\,,  \\
& B_\parallel = \frac{\b B_0\cdot \b V}{V}\,, & \b B_ \bot =\gamma_L \frac{(\b V\times \b B_0)\times \b V}{V^2}\,,
\end{align}
\end{subequations}
where $\gamma_L=(1-V^2)^{-1/2}$. Clearly, in the comoving frame $\b B\cdot \b E=0$. If quarkonium travels at angle $\phi$ with respect to the magnetic field in the laboratory frame, then
\begin{align}\label{lt}
& B= B_0\sqrt{\cos^2\phi(1-\gamma_L^2) +\gamma_L^2}\,, &E=B_0\gamma_L V\sin\phi\,.
\end{align}
We choose $y$ and $x$ axes of the comoving frame such that $\b B= B\unit y$ and $\b E= E\unit x$. A convenient gauge choice is $\b A= -B x\, \unit z$ and $\varphi=-Ex$. For a future reference we also define a useful dimensionless parameter $\rho$ \cite{Popov:1997-A}
\begin{align}\label{rho}
\rho= \frac{E}{B}= \frac{\gamma_L V\sin\phi}{\sqrt{\cos^2\phi(1-\gamma_L^2) +\gamma_L^2}}\,.
\end{align}
Note, that (i) $0\le \rho\le 1$ because $B^2- E^2 =  B_0^2 \ge 0$ and (ii) when quarkonium moves perpendicularly to the magnetic field $\b B_0$, $\rho=V$. 

\subsubsection{WKB method}\label{sec:wkb}

I assume that the force binding $q$ and $\bar q$ into quarkonium as a short-range one i.e.\  $(M \e_b)^{1/2} R \ll1$, where $\e_b$ and  $M$ are  binding energy and mass of  quarkonium, respectively,    and $R$ is the radius of the nuclear force  given by $R\approx (\as/\sigma)^{1/2}$, where $\sigma= 1$~GeV/fm is the string tension. For example, the binding energy of $c$ and $\bar c$ in $\jpsi$ in vacuum is  $\e_b=0.64~\text{GeV}\ll M/R^2= M\sigma/\as \approx 3$~GeV. This approximation is even better at finite temperature on account of $\e_b$ decrease.  Regarding $\jpsi$ as being bound by a short-range force enables us to 
 calculate the dissociation probability $w$ with exponential accuracy $w\approx e^{-f}$, independently of the precise form of the quarkonium wave function.  This is especially important since solutions of the relativistic two-body problem for quarkonium are not readily available.

It is natural to study quarkonium ionization in the comoving frame \cite{Marasinghe:2011bt}. As explained in the Introduction, ionization is quantum tunneling through the potential barrier caused by the electric field $\b E$. In this subsection I employ the quasi-classical WKB approximation to calculate the quarkonium decay probability $w$.  For the gauge choice specified in \sec{sec:com} quark energy $\e_0$ ($\e_0<m$) in electromagnetic field can be written as
\beql{hamilt}
 \e_0= \sqrt{m^2+(\bm p-e \bm A)^2}+e\varphi=\sqrt{m^2+(p_z+eBx)^2+p_x^2+p_y^2}-eEx\,.
\eeq
In terms of $\e_0$, quarkonium binding energy is  $\e_b=m-\e_0$. To simplify notations, we will set $p_x=0$, because the quark moves constant momentum along the direction of magnetic field.

\begin{figure}[ht]
      \includegraphics[height=5.5cm]{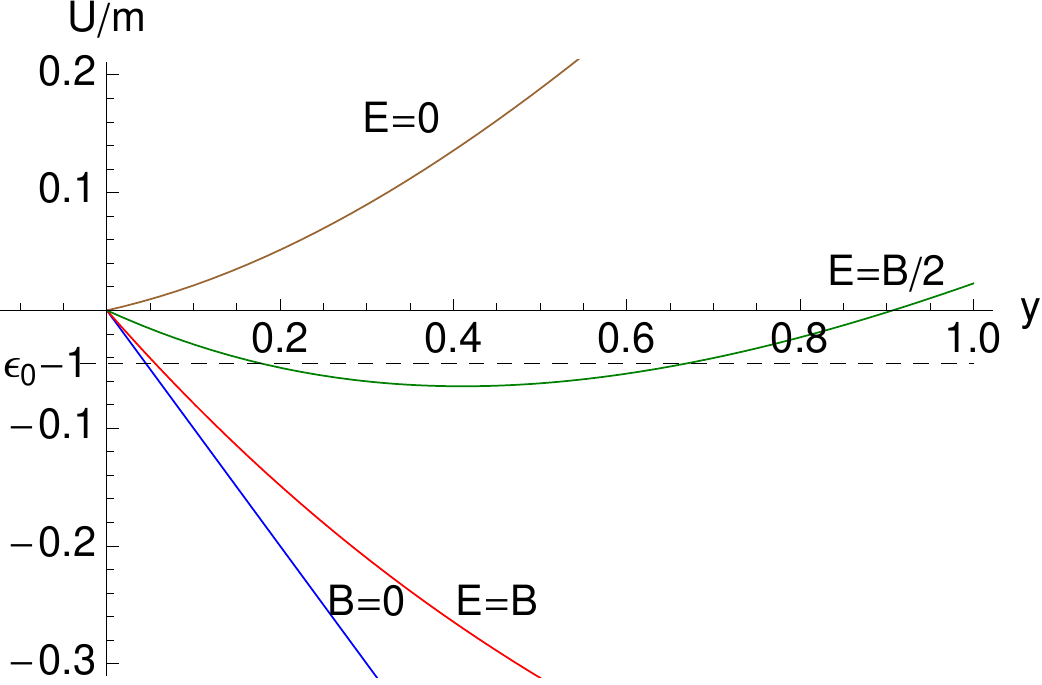} 
        \caption{Effective potential $U(x)=\sqrt{m^2+(p_z+eBx)^2+p_x^2}-eEx-\sqrt{m^2+p_x^2+p_x^2}$ for $p_x=0$, $p_z=m/6$, $B=m$ (except the blue line where $B=0$). The width of the potential barrier decreases with $E$ and increases with $B$. $1-\epsilon_0$ corresponds to the binding energy in units of $m$. }
\label{fig:V}
\end{figure}

The effective potential $U(x)= \e_0(x)-\sqrt{m^2+\b p^2}$ corresponding to  \eq{hamilt} is plotted in \fig{fig:V}.
We can see that the tunneling probability is finite only if $E>0$. It is largest when $B=0$. It has been already noted before in \cite{Popov:1998aw,Popov:1997-A,Popov:1998-A} that the effect of the magnetic field is to stabilize the bound state. In spite of the linearly rising potential (at $B>E$) tunneling probability is finite as the result of rearrangement of the QED vacuum in electric field. 

Ionization probability of quarkonium equals its tunneling probability through  the potential barrier. The later is given by the transmission coefficient 
\beql{ww}
w= e^{-2\int_0^{y_1}\sqrt{-p_y^2}dy}\equiv e^{-f}\,.
\eeq
In the non-relativistic approximation one can also calculate the pre-exponential factor, which appears due to the deviation of the quark wave function from the quasi-classical approximation. This is discussed later in \sec{secII}. 
We now proceed with the calculation of function $f$. Since $B>E$ Eq.~\eq{hamilt} can be written as \cite{Marasinghe:2011bt} 
\beql{py1}
p_x^2=-e^2(B^2-E^2)(x-x_1)(x-x_2)\,,
\eeq
where 
\beql{y12}
x_{1,2}=\frac{\e_0 E-p_z B\mp \sqrt{(\e_0 E-p_zB)^2-(B^2-E^2)(-\e_0^2+m^2+p_z^2)}}{e(B^2-E^2)}\,.
\eeq
Define dimensionless variables $\epsilon_0 = \e_0/m$ and $q=p_z/m$. Integration in \eq{w} gives:
\begin{align}\label{f}
\frac{f}{m^2}&=\frac{\sqrt{-\epsilon_0^2+1+q^2}(\epsilon_0 E-q B)}{e(B^2-E^2)}\nonumber\\
&-\frac{(\epsilon_0 E-q B)^2-(B^2-E^2)(-\epsilon_0^2+1+q^2)}{e(B^2-E^2)^{3/2}}
\ln\left\{
\frac{\epsilon_0 E-q B+\sqrt{(B^2-E^2)(-\epsilon_0^2+1+q^2)}}{\sqrt{(\epsilon_0 E-qB)^2-(B^2-E^2)(\epsilon_0^2+1+q^2)}} 
\right\}\,.
\end{align}
For different $q$'s $w=e^f$ gives the corresponding ionization probabilities. The largest probability corresponds to smallest $f$, which occurs at  momentum $q_m$ determined by equation \cite{Popov:1998aw}
\beql{q-min}
\frac{\partial f(q_m)}{\partial q_m}=0\,.
\eeq
Using \eq{f} and  parameter $\rho$ defined in \eq{rho} we find \cite{Marasinghe:2011bt} 
\beql{f'}
\frac{\rho(\epsilon_0 -\rho q_m)}{1-\rho^2}\ln\left\{\frac{\epsilon_0\rho-q_m+\sqrt{1-\rho^2}\sqrt{-\epsilon_0^2 +1 +q_m^2}}{\sqrt{(\epsilon_0-\rho q_m)^2-1+\rho^2}}\right\}=\frac{\sqrt{-\epsilon_0^2 +1+q_m^2}}{\sqrt{1-\rho^2}}\,.
\eeq
This is an implicit equation for  the extremal momentum $q_m=q_m(\epsilon_0,\rho)$. Substituting $q_m$ into \eq{f} one obtains $f=f(\epsilon_0,\rho)$, which by means of \eq{ww} yields the ionization probability.  The quasi-classical approximation that we employed in this section is valid inasmuch as  $f(q_m)\gg 1$. 

In order to compare with the results obtained in  \cite{Popov:1998aw} using the imaginary time method, we can re-write Eq.~\eq{f'}  in terms of an auxiliary parameter  $\tau_0$ as
\begin{subequations}
\begin{align}
&
\label{tau}
\tau_0=\frac{\sqrt{1-\rho^2}\sqrt{-\epsilon_0^2 +1 +q_m^2}}{\rho(\epsilon_0 -\rho q_m)}\,,\\
&\label{f'2}
\frac{\tanh\tau_0}{\tau_0}=\rho\,\frac{\epsilon_0-\rho q_m}{\epsilon_0 \rho -q_m}\,.
\end{align}
\end{subequations}
Taking advantage of these equations, Eq.~\eq{f} can be cast into a more compact form 
\beql{f2}
f_m=\frac{m^2\tau_0\rho}{eE\sqrt{1-\rho^2}}[1-\epsilon_0(\epsilon_0-q_m\rho)]\,,
\eeq
where we denoted $f_m=f(q_m)$.
 This agrees with results of Ref.~\cite{Popov:1998aw}.

\subsubsection{Special case:  Crossed fields}
An important limiting case is crossed fields $E=B$. Since also  $\b E\perp \b B$, see \sec{sec:com},
both field invariants vanish. Nevertheless, quarkonium ionization probability
is finite \cite{Popov:1998aw}. This limit is obtained by  taking $\rho\to 1$ in the equations from the previous section. Employing \eq{tau} and \eq{f'2} we get the following condition for extremum
\beql{r1-ext}
\epsilon_0^2-1+2q_m^2-3\epsilon_0 q_m=0\,,
\eeq
with the solution
\beql{r1-e-s}
q_m= \frac{1}{4}\left( 3\epsilon_0-\sqrt{\epsilon_0^2+8}\right)\,.
\eeq
Substituting into \eq{f2} produces
\beql{fr-1}
f_m= \frac{2}{3}\frac{m^2}{eE}\frac{(-\epsilon_0^2+1+q_m^2)^{3/2}}{\epsilon_0-q_m}\,.
\eeq

\subsection{Non-relativistic approximation}\label{sec:nr}

A very useful approximation of the relativistic formulas derived in the previous section is  the non-relativistic limit because (i) it provides a very good numerical estimate, see \fig{fig:phi}, (ii) it allows 
us to eliminate the parametric dependence in \eq{f},\eq{f'} and write $f(q_m)$ explicitly in terms of $\rho$ and $\epsilon_0$, and (iii) spin effects can be accounted for \cite{Marasinghe:2011bt,Tuchin:2011cg}.

\subsubsection{Arbitrary binding}\label{sec:arb}

Motion of a particle can be treated non-relativistically if its momentum is much less than its mass. In such a case  $\e_0\approx m$ or \ $\e_b=m-\e_0\ll m$.  Additionally, motion of a charged particle in electromagnetic field is non-relativistic if  $E\ll B$. Indeed, the average velocity of a non-relativistic particle is of order $v\sim E/B=\rho$. Thus,  the non-relativistic limit is obtained by taking the limits $\epsilon_b=\e_b/m\ll 1$ and $\rho\ll 1$. In these limits the extremum conditions \eq{tau},\eq{f'2} reduce to 
\begin{subequations}
\begin{align}
&\tau_0= \frac{\sqrt{2\epsilon_b+q_m^2}}{\rho}\,,\label{nr-tau-1}\\
& \frac{\tanh\tau_0}{\tau_0}=\frac{\rho}{\rho-q_m}\label{nr-tau-2}\,.
\end{align}
\end{subequations}
Out of two solution to \eq{nr-tau-1} we pick the following one
\beql{qm}
q_m= -\sqrt{\tau_0^2\rho^2-2\epsilon_b}\,.
\eeq
The sign of $q_m$ is fixed using \eq{nr-tau-2} by noticing that $\tanh \tau_0/\tau_0<1$.  Eliminating $q_m$ gives:
\beql{nr2}
\tau_0^2-(\tau_0\coth\tau_0-1)^2=\gamma^2\,,
\eeq
where 
\beql{gamma}
\gamma = \frac{\sqrt{2\epsilon_b}}{\rho}\,.
\eeq
$\gamma$ is analogous to the  adiabaticity parameter of Keldysh \cite{Keldysh-ioniz}. Taking the non-relativistic limit of \eq{f2} and using \eq{qm} yields
\beql{fnr}
f_m=\frac{2m^2(2\epsilon_b)^{3/2}}{3eE}g(\gamma)\,,
\eeq
where $g(\gamma)$ is  the  Keldysh function \cite{Keldysh-ioniz} 
\beql{keldysh}
g(\gamma)=\frac{3\tau_0}{2\gamma}\left[ 1-\frac{1}{\gamma}\left( \frac{\tau_0^2}{\gamma^2}-1\right)^{1/2}\right]\,.
\eeq

\begin{figure}[ht]
      \includegraphics[height=6cm]{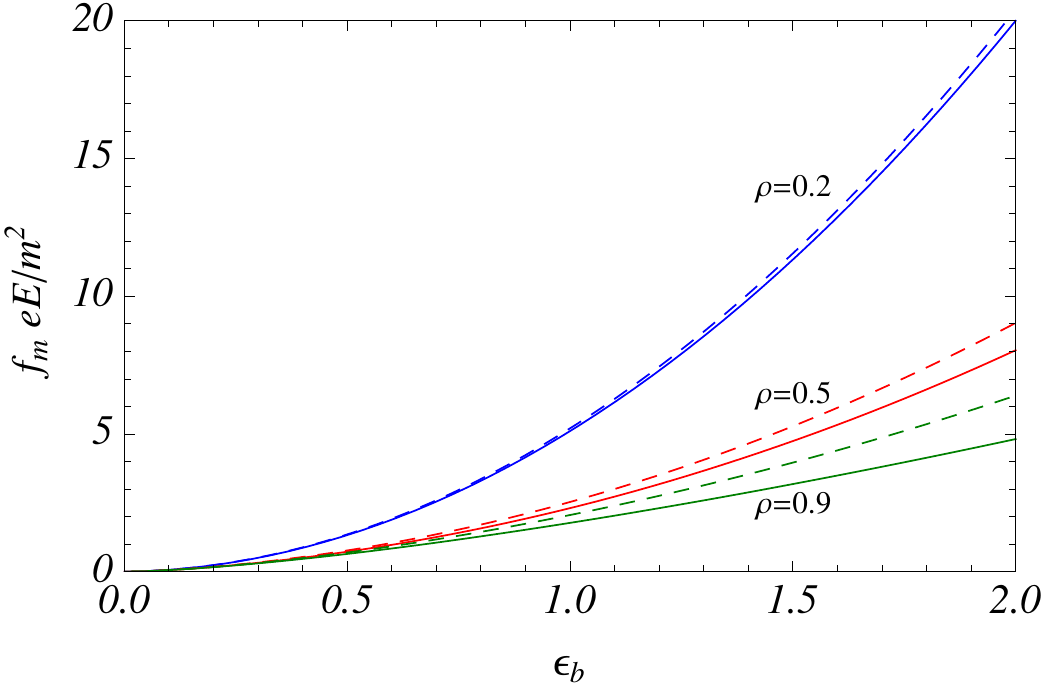} 
        \caption{Dimensionless function $f_m eE/m^2$ versus $\epsilon_b$ for different values of $\rho$. The solid line is the full relativistic calculation, the dashed line is the non-relativistic approximation. $\jpsi$ binding energy in vacuum corresponds to $\epsilon_b=0.68$. }
\label{fig:phi}
\end{figure}

In \fig{fig:phi} we show the dimensionless ratio $f_meE/m^2$ as a function of the binding energy $\epsilon_b$ (in units of $m$) for several values of  $\rho$. The vacuum binding energy of $\jpsi$ corresponds to $\epsilon_b = 0.68$. We observe an excellent agreement between the full relativistic calculation and  the non-relativistic approximation.  At $\rho=0.9$ and $\epsilon_b = 0.68$ the difference between the two lines is 10\% and can be further improved by considering higher order corrections to $f_m$ \cite{Popov:1998-A}.

\subsubsection{Weak binding}\label{sec:weak}

Of special interest is the limit of  weak binding $\gamma\ll 1$, i.e.\ $\sqrt{2\epsilon_b}\ll \rho$. Expanding \eq{nr2} at small $\gamma$  and $\tau_0$ yields
\beql{nr-gsmall}
\tau_0=\gamma\left(1+\frac{1}{18}\gamma^2\right)\,
\eeq
and substituting into \eq{keldysh} and subsequently into  \eq{fnr} yields
\beql{fnr-gsmall-0}
f_m=\frac{2}{3}\frac{m^2}{eE}(2\epsilon_b)^{3/2}\,.
\eeq
Hence, the quarkonium dissociation probability reads \cite{LL3-77}
\beql{fnr-gsmall}
w=   \exp\left\{-\frac{2}{3}\frac{(2\e_b m)^{3/2}}{meE}\right\}\,.
\eeq
Since the quasi-classical approximation employed in this paper is valid if $f(q_m)\gg 1$, it follows that 
the binding energy must satisfy 
\beql{cond-1}
 \frac{(eE)^{2/3}}{m^{1/3}}\ll \e_b\ll \rho^2m\,.
\eeq
Note also that we work in the approximation of the short-range binding potential meaning that  $\sqrt{2\epsilon_b} \ll 1/(mR)$, see \sec{sec:ha}.

\subsubsection{Strong binding}\label{sec:strong}

In the limit $\gamma\gg 1$,  \eq{nr2} and \eq{keldysh} imply that 
\beql{str-g}
\tau_0=\frac{\gamma^2}{2}\,,\quad g(\gamma)= \frac{3\gamma}{8}\,.
\eeq
Substituting \eq{str-g} into  \eq{fnr} we derive 
\beql{str-f}
f_m= \frac{\e_b^2}{eE}\frac{B}{E}\,.
\eeq
Thus, quarkonium dissociation  probability in the case of strong binding is 
\beql{w-sb}
w=\exp\left\{-\frac{\e_b^2}{eE}\frac{B}{E} \right\}\,.
\eeq
This formula is valid when
\beql{cond-2}
\rho^2m\,,\,\sqrt{eE\rho}\ll \e_b\ll 1/R
\eeq

\subsubsection{Contribution of quark spin}\label{sec:spin}

So far I have neglected the contribution of quark spin.  In order to take into account the effect of spin interaction with the external field, we can use  squared Dirac equation for a bi-spinor $\psi$: 
\beql{dirac}
\left[ (\e-e\varphi)^2-(\b p-e\b A)^2-m^2+e\b\Sigma\cdot \b B-i e \b \alpha\cdot \b E\right]\psi=0\,,
\eeq
where
\begin{align}
&
\b \Sigma = \left( 
  \begin{array}{cc}
    \b \sigma & 0 \\ 
    0 & \b\sigma 
  \end{array}
  \right)
  &
  \b \alpha = \left( 
  \begin{array}{cc}
    \b \sigma & 0 \\ 
    0 & -\b\sigma 
  \end{array}
  \right)
  \end{align}
Operators $\Sigma_y$ and $\alpha_x$ do not commute. Therefore, in order to apply the WKB method for calculation of the ionization probability one actually needs to square \eq{dirac}, which leads to a differential equation of the fourth order in derivatives. The problem becomes more tractable in the non-relativistic case and for crossed fields. Spin effects in crossed fields were discussed in \cite{popov-review}. 

With  quark spin taken into account, the non-relativistic version of \eq{hamilt} becomes:
\beql{nr-hamilt}
\frac{1}{2m}\left[ (p_z+eBx)^2+p_x^2\right] -eEx-\frac{\mu}{s}\b s\cdot \b B=-\e_b\,,
\eeq
and hence
\beql{py1-nr}
p_x^2= 2m\left(-\e_b+\frac{\mu}{s}\b s\cdot \b B+e E x\right)-(p_z+eBx)^2\,,
\eeq  
where $\mu$ is the quark magnetic moment and $s$ is the projection of spin in the direction of the magnetic field. For a point quark, $\mu=\mu_B= \frac{e\hbar}{2mc}$. 
The effect of quark spin on quarkonium dissociation probability can be taken into account by replacing $\e_b\to \e_b'=\e_b-\frac{\mu}{s}\b s\cdot \b B$ in formulas for $f_m$. With this replacement, all results of this section apply  to a particle with spin. Note that  effective binding energy $\e_b'$ decreases if spin is parallel to the magnetic field and increases if it is antiparallel. In particular, in the case of weak binding
\beql{weak-spin}
w= \sum_{s=\pm 1/2}  \exp\left\{-\frac{2}{3}\frac{(2\e_b m+2seB)^{3/2}}{meE}\right\}\,.
\eeq
Since  the non-relativistic limit provide a good approximation of the full relativistic formulas, we will implement the quark spin dependence using  the non-relativistic prescription \cite{Marasinghe:2011bt,Tuchin:2011cg}. 

\subsection{Effect of electric field produced in the lab frame}\label{sec:j}

\subsubsection{Origin of electric field in the lab frame}\label{sec:ja}

So far I have entirely neglected possible existence of electric field in the lab frame. This field, which we will denote by $\b E_0$, can have two origins: (i) Asymmetry of nucleon distributions in the colliding heavy ions, see \fig{fig:aa5}(b) and (ii) Chiral Magnetic Effect (CME) \cite{Kharzeev:2004ey,Kharzeev:2007jp,Kharzeev:2007tn,Fukushima:2008xe,Kharzeev:2009fn,Basar:2010zd,Asakawa:2010bu}, which has recently attracted a lot of attention. In a nutshell, if a metastable $P$ and $CP$-odd bubble is induced by axial anomaly in the hot nuclear matter, then in the presence of external magnetic field $\bm B_0$ the bubble generates an electric field which is \emph{parallel} to the magnetic one. According to \cite{Kharzeev:2007tn} the value of the electric field $\bm E_0$ in the bubble is 
\begin{align}\label{cme}
\bm E_0 = - N_c\sum_f\frac{e_f^2}{4\pi^2}\, \frac{\Theta}{N_f}\, \bm B_0 = -\frac{2}{3}\frac{\alpha\, \Theta}{\pi}\bm B_0
\end{align}
where the sum runs over quark flavors $f$ and it is  assumed that only three lightest flavors contribute. The value of the $\Theta$-angle fluctuates from event to event.  CME refers to the  macroscopic manifestation of this effect --  separation of electric charges with respect to the reaction plane. This effect is a possible explanation of  experimentally observed charge asymmetry fluctuations \cite{:2009uh,:2009txa,Ajitanand:2010rc}. 

No matter what is the origin of electric field in the lab frame, it  averages to zero  over an ensemble of events.  We are interested to know the effect of this field on quarkonium dissociation -- this is the problem we are turning to now \cite{Tuchin:2011cg}. 

\subsubsection{Quarkonium dissociation rate}\label{secII}

Ionization probability of quarkonium equals its tunneling probability through  the potential barrier. In the WKB approximation the later is given by the transmission coefficient and was calculated in \sec{sec:hb}. 
In this method contribution of the quark spin can be easily taken into account. 
Another method of calculating the ionization probability, the imaginary time method  \cite{ITM1,ITM2}, was employed in \cite{Popov:1997-A,Popov:1998-A,Popov:1998aw}. It also yielded  in the non-relativistic approximation the pre-exponential factor that appears due to the deviation of the quark wave function from the quasi-classical approximation. Such a calculation requires matching quark wave function inside and outside the potential barrier \cite{LL3-77}. Extension of this approach to the relativistic case is challenging due to analytical difficulties of the relativistic two-body problem. Fortunately, it was argued in \sec{sec:hb}, that the non-relativistic approximation provides a very good accuracy in the $\e_b\ll m$ region, which is relevant in the quarkiononium dissociation problem \cite{Popov:1998aw,Marasinghe:2011bt}. 

Given the  electromagnetic field in the laboratory frame $\bm B_0$, $\bm E_0$, the electromagnetic field $\bm B$, $\bm E$ in the comoving frame moving with velocity $\bm V$   is given by 
\begin{subequations}\label{em-cf-0}
\begin{align}
\bm E=& E_0\left\{ \gamma_L(\bm b_0+\rho_0^{-1}\bm V\times \bm b_0)- (\gamma_L-1)\bm V\frac{\bm V\cdot \bm b_0}{V^2}\right\}\\
\bm B=& B_0\left\{ \gamma_L(\bm b_0-\rho_0\bm V\times \bm b_0)- (\gamma_L-1)\bm V\frac{\bm V\cdot \bm b_0}{V^2}\right\}
\end{align}
\end{subequations}
where $\bm b_0=\bm B_0/B_0$ is a unit vector in the magnetic field direction, $\rho_0 = E_0/B_0= 2\alpha |\Theta|/3\pi$ (see \eq{cme}) and $\gamma_L= 1/\sqrt{1-V^2}$. It follows from \eq{em-cf-0} that
\begin{subequations}\label{em-cf}
\begin{align}
E&=E_0\sqrt{1+\gamma_L^2(\bm b_0\times \bm V)^2(1+\rho_0^{-2})}\\
B&= B_0\sqrt{1+\gamma_L^2(\bm b_0\times \bm V)^2(1+\rho_0^2)}
\end{align}
\end{subequations}
Using \eq{em-cf} we find that the angle $\theta$  between the electric and magnetic field in the comoving frame is 
\begin{align}\label{theta}
\cos\theta  = \frac{\bm E\cdot \bm B}{EB} = \frac{1}{\sqrt{[1+\gamma_L^2(\bm b_0\times \bm V)^2(1+\rho_0^{-2}) ][1+\gamma_L^2(\bm b_0\times \bm V)^2(1+\rho_0^2)}]}&
\end{align}
where  we used the relativistic invariance of $\bm E\cdot \bm B$.

It is useful to introduce dimensionless parameters  $\gamma$, $\epsilon$  and $\rho$ as \cite{Popov:1998aw}
\begin{align}\label{param}
\gamma= \frac{1}{\rho}\sqrt{\frac{2\e_b}{m}}\,,\quad \rho= \frac{E}{B}\,,\quad \epsilon = \frac{eE}{m^2}\left( \frac{m}{2\e_b}\right)^{3/2}
\end{align}
where $m$ is quark mass and  $\e_b$ is quarkonium binding energy.
I will treat the quarkonium binding potential in the non-relativistic approximation, which provides a very good accuracy to the dissociation rate  \cite{Popov:1998aw,Marasinghe:2011bt}.
The quarkonium dissociation rate in the comoving frame in the non-relativistic approximation 
 is given by \cite{Popov:1997-A}
\begin{align}\label{www}
w= \frac{8\e_b}{\epsilon}\, P(\gamma,\theta)\,C^2(\gamma,\theta)\, e^{-\frac{2}{3\epsilon}g(\gamma,\theta)} 
\end{align}
where  function $g$ reads
\begin{align}
g&=\frac{3\tau_0}{2\gamma}\left[ 1-\frac{1}{\gamma}\left( \frac{\tau_0^2}{\gamma^2}-1\right)^{1/2}\sin\theta -\frac{\tau_0^2}{3\gamma^2}\cos^2\theta\right] 
\end{align}
and functions $P$ and $C$ are given be the following formulas:
\begin{align}
P& = \frac{\gamma^2}{\tau_0}\left[ (\tau_0\coth\tau_0+\frac{\sinh\tau_0\cosh\tau_0}{\tau_0}-2)\sin^2\theta+\sinh^2\tau_0\cos^2\theta\right]^{-1/2}\\
C&= \exp\left[ \ln \frac{\tau_0}{2\gamma}+\int_0^{\tau_0}d\tau \left( \frac{\gamma}{\xi(\tau)}-\frac{1}{\tau_0-\tau}\right)\right]\\
\xi(\tau)&= \left\{ \frac{1}{4}(\tau_0^2-\tau^2)^2\cos^2\theta+\tau_0^2\left[ \left( \frac{\cosh\tau_0-\cos\tau}{\sinh\tau_0}\right)^2-\left( \frac{\sinh\tau}{\sinh\tau_0}-\frac{\tau}{\tau_0}\right)^2\right]\sin^2\theta\right\}^{1/2}\,.
\end{align}
  The contribution of quark spin is taken into account by replacing $\e_b\to \e_b'=\e_b-\frac{e}{m}\bm s\cdot \bm B$ \cite{Marasinghe:2011bt}.  Function $g$ represents the leading quasi-classical exponent, $P$ is the pre-factor for the $S$-wave state of quarkonium and $C$ accounts for the   Coulomb interaction between the  valence quarks.  Parameter $\tau_0$ satisfies the following equation 
\begin{align}
\tau_0^2-\sin^2\theta(\tau_0\coth\tau_0-1)^2=\gamma^2
\end{align}
which establishes its dependence on $\theta$ and $\gamma$. Note, that  in the limit $E\to 0$ the dissociation rate \eq{www} exponentially vanishes. This is because  pure magnetic field cannot force a charge to tunnel through a potential barrier.  

In the case that  mechanism (i) is responsible for generation of electric field, $\b E_0$ is the field permitting the entire plasma in a single event. Event average is then obtained by averaging \eq{www} over an ensemble of events. In the case that mechanism (ii) is operative, averaging is more complicated. Eq.~\eq{www} gives the quarkonium  dissociation rate in a bubble with a given value of $\Theta$. Its derivation assumes that  the dissociation process happens entirely inside a bubble and that $\Theta$ is constant inside the bubble. Since in a relativistic heavy ion collision many bubbles can be produced with a certain distribution of $\Theta$'s (with average $\aver{\Theta}=0$) more than one bubble can affect the dissociation process. This will result in a distractive interference leading to reduction of the $CP$-odd effect on quarkonium dissociation. However, if a typical bubble  size $R_0$ is much larger than the size of quarkonium $R_J$, then  the dissociation is affected by one bubble at a time independently of others, and hence the interference effect can be neglected. In this case \eq{w} provides, upon a proper average, a reasonable estimate of quarkonium dissociation in a heavy ion collision. We can estimate the bubble size as the size of the sphaleron, which is of the order of the chromo-magnetic screening length $\sim 1/g^2T$, whereas the quarkonium size is of the order $\as/\e_b$. Consequently, at small coupling and below the zero-field dissociation temperature  (i.e.\ when $\e_b$ is not too small) $R_0$ is parametrically much larger than $R_J$. A more quantitative estimate of the sphaleron size is   $R_0\simeq 1.2/\as N_c T\simeq 0.4$~fm  \cite{Moore:2010jd}; whereas  for $\jpsi$ $R_J\simeq \as/\e_b\simeq 0.1-0.2$~fm. Thus, based on this estimate bubble interference can be neglected in the first approximation.  However, since the ratio  $R_J/R_0$ is actually not so small this effect nevertheless warrants further investigation. 

To obtain the experimentally observed $\jpsi$ dissociation rate we need to average \eq{w} over the bubbles 
produced in a given event and then over all events. To this end it is important to note that because the dissociation rate depends only on $\rho_0^2$ it is insensitive to the sign of the $\b E_0$ field or, in other words, it depends only on absolute value of $\Theta$ but not on its sign. Therefore, it  stands to reason that although the precise distribution of $\Theta$'s is not known, \eq{w} gives an approximate event average  with parameter $\Theta$ representing a characteristic absolute value of the theta-angle.

\subsubsection{Limiting cases}

Before I proceed with the numerical calculations, let us consider for illustration several limiting cases.
 If  quarkonium moves with non-relativistic velocity, then in the comoving frame electric and magnetic fields are approximately parallel $\theta\approx 0$, whereas in the ultra-relativistic case they are orthogonal $\theta\approx \pi/2$, see \eq{theta}. In the later case the electromagnetic field in the comoving frame does not depend on $E_0$ as seen in \eq{em-cf} and therefore the dissociation rate becomes insensitive to the CME. In our estimates I will assume that $\rho_0<1$ which is the relevant phenomenological situation. Indeed, it was proposed in \cite{Kharzeev:2007tn} that  $\rho_0\sim \alpha\ll 1$ produces charge fluctuations with respect to  the reaction plane of the magnitude consistent with experimental data. 

1) $\theta\gtrsim 0$, i.e.\ electric and magnetic fields are approximately parallel. This situation is realized in the following two cases.  (i) Non-relativistic quarkonium velocities: $V\ll \rho_0$ or (ii)  motion of quarkonium at small angle $\phi$ to the direction of the magnetic field $\bm b_0$: $\phi\ll \rho_0/\gamma_LV$. In both cases $E\approx E_0$ and $B\approx B_0$. This is precisely the case where the dissociation rate exhibits its strongest sensitivity to the strength of the electric field $\bm E_0$ generated by the local parity violating QCD effects. Depending on the value of the $\gamma$ parameter  defined in \eq{param} we can distinguish the case of strong electric field $\gamma\gg 1$ and weak electric field $\gamma\ll 1$ \cite{Popov:1998-A}. In the former case, $g= (3/8)\gamma$, $P= (8/e)^{1/2}\gamma e^{-\gamma^2/2}$, $C= e^{\pi \gamma/2}/\gamma$. Substituting into \eq{w} the dissociation rate reads 
\begin{align}\label{gbig}
w= \frac{8\e_b}{\epsilon\gamma}\sqrt{\frac{8}{e}}  e^{-\gamma^2/2}e^{-\frac{\gamma}{4\epsilon}}
= \frac{16\e_b^2 m}{eB_0}\sqrt{\frac{8}{e}}  e^{-\frac{\e_b}{\rho_0^2m}}e^{-\frac{\e_b^2}{\rho_0eE_0}}\,,\quad \gamma\gg 1
\end{align}
In the later case, $g= P=C=1$ and
\begin{align}\label{zz}
w= \frac{8\e_b}{\epsilon}e^{-\frac{2}{3\epsilon}}=\frac{8\e_b m^2}{eE_0}\left(\frac{2\e_b}{m} \right)^{3/2}  
e^{-\frac{2m^2}{3eE_0}\left(\frac{2\e_b}{m} \right)^{3/2}}\,,\quad \gamma\ll 1
\end{align}
where the electromagnetic field in the comoving frame equals one in the laboratory frame as was mentioned before.

2) $\theta\sim \pi/2$, i.e.\ electric and magnetic fields are approximately orthogonal.\footnote{Note, that the limit $\gamma\gg 1$ is different in $\theta = \frac{\pi}{2}$ and $\theta<\frac{\pi}{2}$ cases \cite{Popov:1997-A}. } This occurs for an ultra-relativistic motion of quarkonium $V\to 1$. In this  case 
\begin{align}\label{gg}
B= E= B_0\gamma_L|\bm b_0\times \bm V|\sqrt{1+\rho_0^2}
\end{align}
This case was discussed in detail in our previous paper \cite{Marasinghe:2011bt}. In particular for $\gamma\ll 1$ we get
\begin{align}\label{zz1}
w = \frac{8\e_b m^2}{eE}\left(\frac{2\e_b}{m} \right)^{3/2}  
e^{-\frac{2m^2}{3eE}\left(\frac{2\e_b}{m} \right)^{3/2}}
\end{align}
Due to \eq{gbig} and \eq{zz1} dependence of $w$ on $E_0$ is  weak unless $\rho_0\gg 1$. 
 
\subsection{Dissociation rate of $\jpsi$}\label{sec:jd}
One of the most interesting applications of this formalism  is calculation of  the dissociation rate of $\jpsi$ which is considered a litmus test of the quark-gluon plasma \cite{Matsui:1986dk}. 
Let $z$ be the heavy ions collision axis; heavy-ion collision geometry implies  that $\bm b_0\cdot \unit z=0$. The plane containing $z$-axis and perpendicular to the magnetic field direction is the reaction plane.  We have
\begin{align}
(\bm b_0\times \bm V)^2&= V_z^2+V_\bot^2\sin^2\phi
\end{align}
where   $\phi$ is the angle between the directions of $\bm B_0$ and $\bm V_\bot$ and I denoted vector components in the $xy$-plane by the subscript $\bot$.
We can express the components of the quarkonium velocity $\bm V$ in terms of the rapidity $\eta$ as $V_z= \tanh \eta$, $V_\bot= P_\bot/(M_\bot\cosh\eta)$, where $\bm P$ and $M$ are the quarkonium momentum and mass and $M_\bot^2= M^2+P_\bot^2$.

\begin{figure}[ht]
\begin{tabular}{cc}
      \includegraphics[height=5cm]{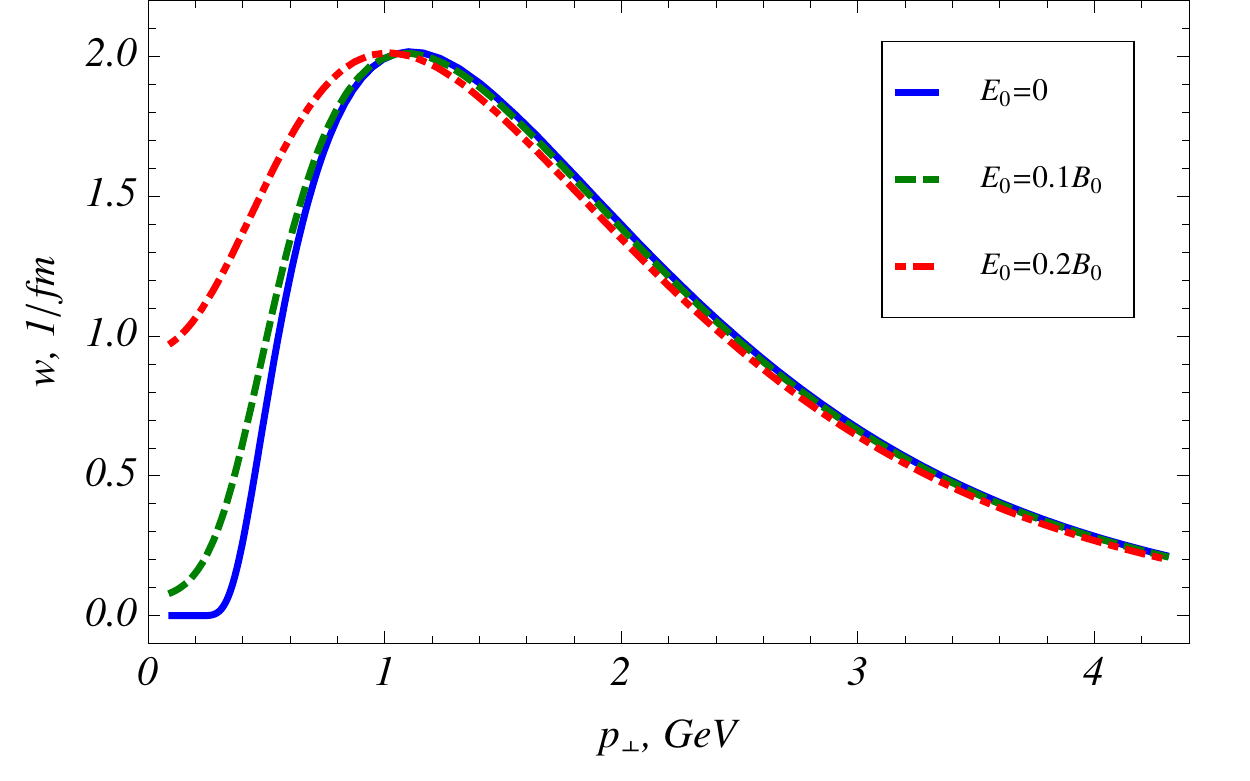} &
      \includegraphics[height=5cm]{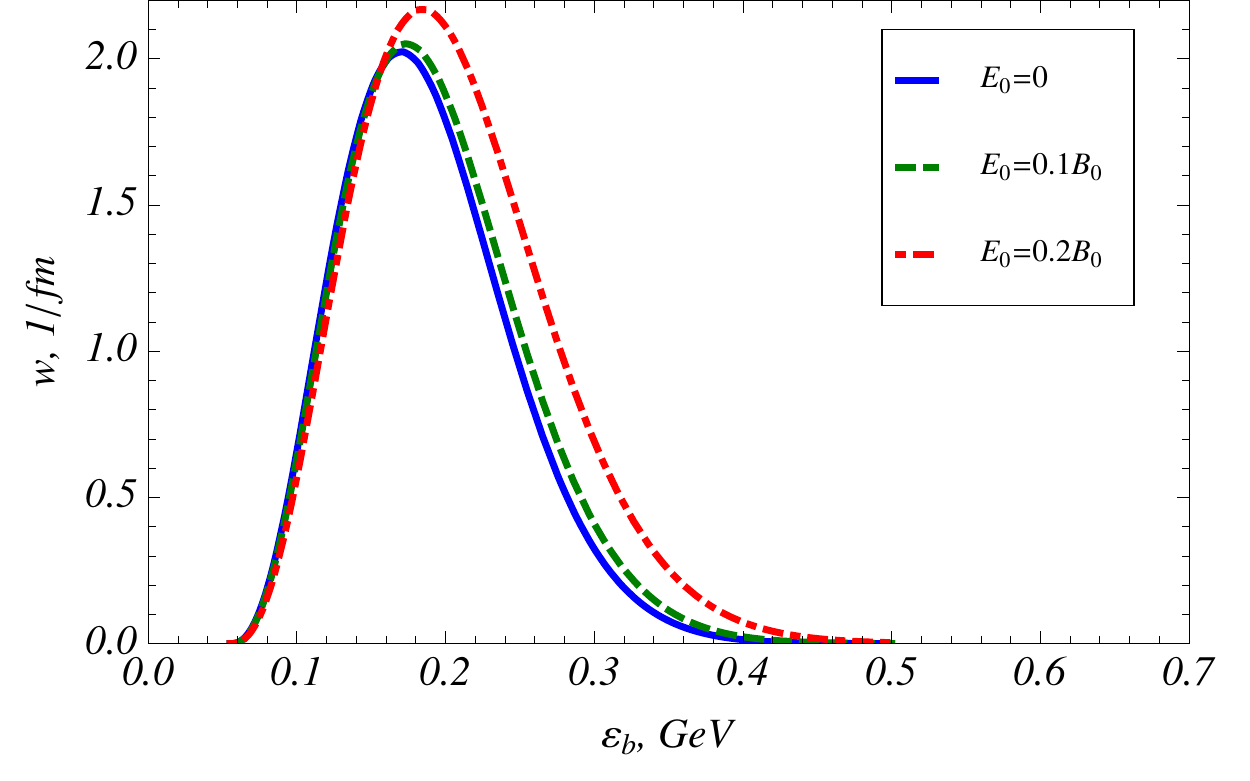}\\
      (a) & (b) 
      \end{tabular}
  \caption{ Dissociation rate of $\jpsi$ at $eB_0=15m_\pi^2$, $\phi=\pi/2$ (in the reaction plane), $\eta=0$ (midrapidity) as a function of (a) $P_\bot$  at $\e_b= 0.16$~GeV and (b) $\e_b$ at $P_\bot=1$~GeV. }
\label{fig:pt}
\end{figure}
\begin{figure}[ht]
      \includegraphics[height=7cm]{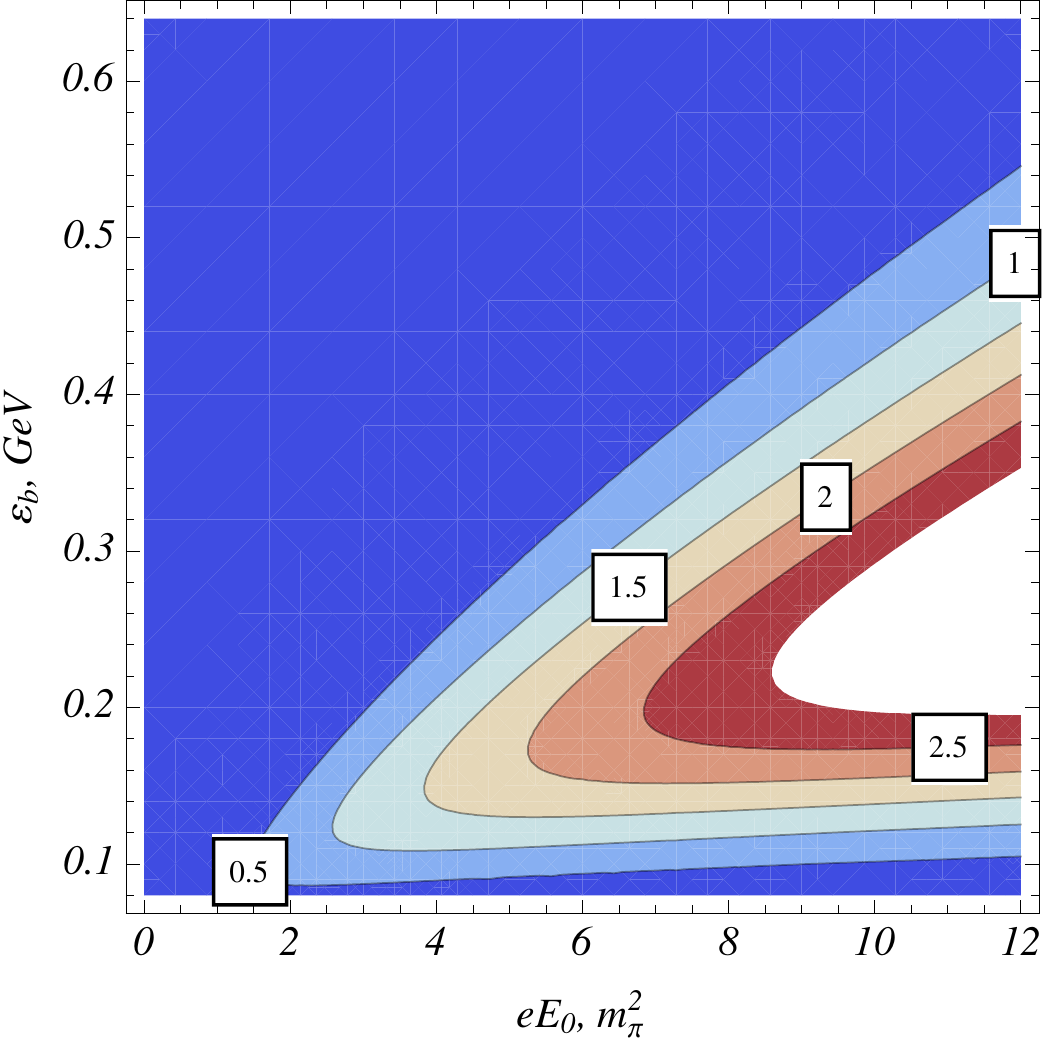} 
  \caption{Contour plot of the dissociation rate of $\jpsi$ as a function of $\e_b$ and $eE_0$ at   $eB_0=15m_\pi^2$, $\phi=\pi/2$ (in the reaction plane), $\eta=0$ (midrapidity) and $P_\bot=0.1$~GeV. Numbers inside boxes indicate the values of $w$ in 1/fm.}
\label{fig:contour}
\end{figure}
\begin{figure}[ht]
\begin{tabular}{cc}
      \includegraphics[height=4.5cm]{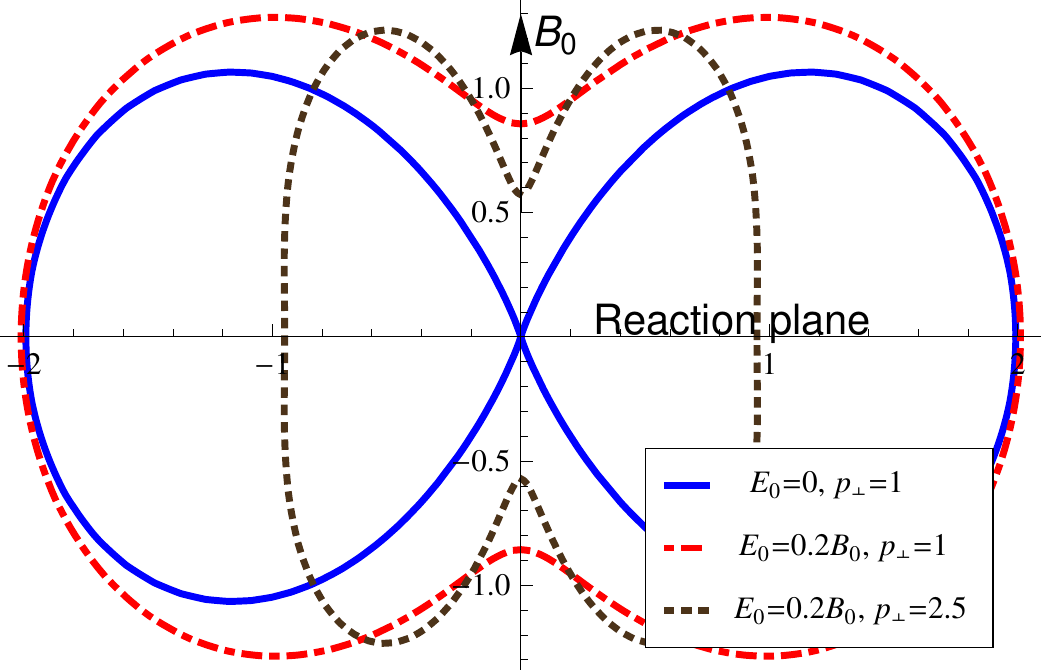} &
      \includegraphics[height=5.cm]{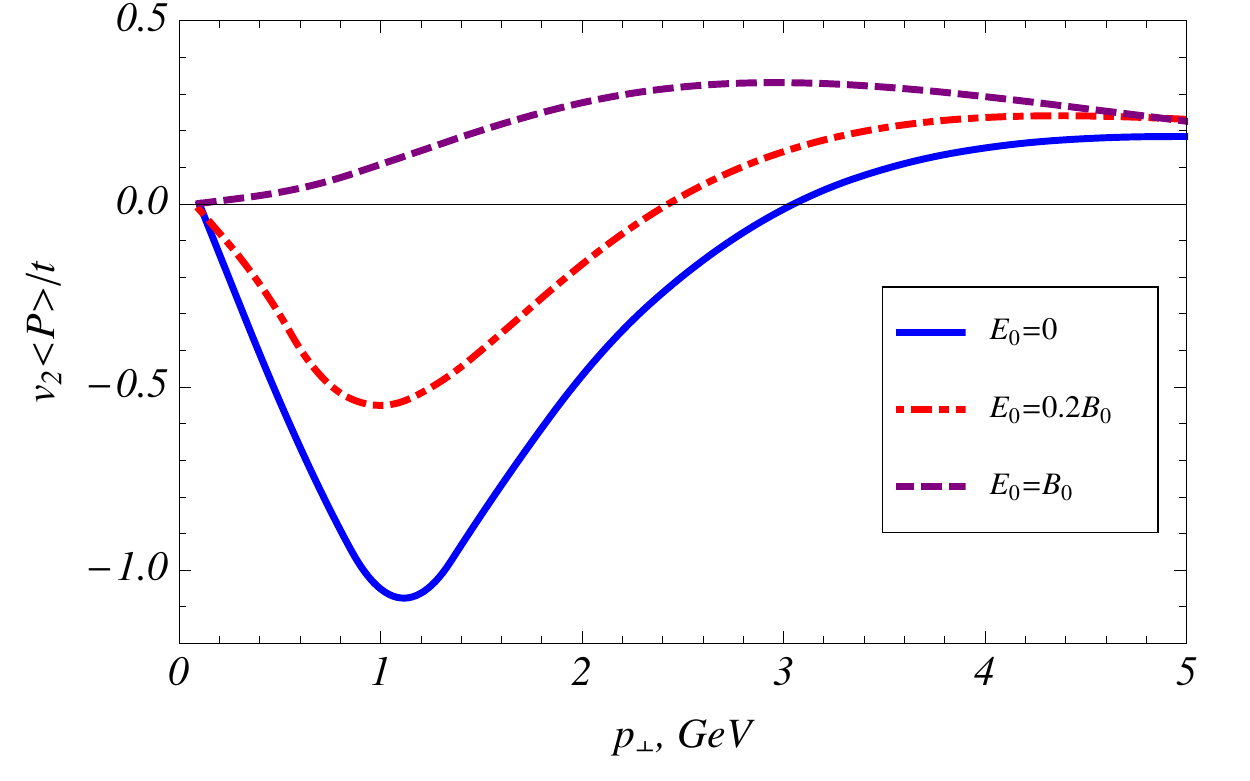}
      \end{tabular}
  \caption{ (a) Angular distribution of $\jpsi$ dissociation rate  at $eB_0=15m_\pi^2$,  $\eta=0$ at different $E_0$ and $P_\bot$ (in GeV's). Magnetic field $\bm B_0$ points in the positive vertical direction. Reaction plane coincides with the horizontal plane. (b) Rescaled second Fourier-harmonic $v_2$  of the azimuthal distribution  as a function of $P_\bot$. $\aver{P}$ is the azimuthal average of the survival probability and $t$ is the time spent by $\jpsi$ in the $P$-odd bubble. }
\label{fig:azimuth}
\end{figure}

Results of  numerical calculations are exhibited in Figs.~\ref{fig:pt}--\ref{fig:azimuth} \cite{Tuchin:2011cg}.
In \fig{fig:pt} I show the  dissociation rate of $\jpsi$ for several values of the electric field $\bm E_0$ induced by the Chiral Magnetic Effect.  Note, that the typical size of the medium traversed by a quarkonium in magnetic field can be estimated very conservatively as a few fm. Therefore, $w\sim 0.3-0.5$ fm$^{-1}$ corresponds to complete destruction of $\jpsi$'s. 
This means that in the magnetic field of strength $eB_0\sim 15m_\pi^2$ all $\jpsi$'s with $P_\bot\gtrsim0.5$~GeV  are destroyed   independently of the strength of $E_0$. Since  magnetic field strength decreases towards the QGP periphery, most of $\jpsi$ surviving at later times originate from that region. 
Effect of  electric field $\bm E_0$   is strongest at low $P_\bot$, which is consistent with our discussion in the previous section. The dissociation rate at low $P_\bot$ exponentially decreases with decrease of $E_0$. Probability of quarkonium ionization by the fields below $E_0\lesssim 0.1 B_0$ (i.e.\ $\rho_0\lesssim 0.1$) is exponentially small. This  is an order of magnitude higher than the estimate $\rho_0\sim \alpha$ of electric field due to CME effect as proposed in \cite{Kharzeev:2007tn}.

As the plasma temperature varies, so is the binding energy of quarkonium although the precise form of the function $\e_b(T)$ is model-dependent. 
The dissociation rate picks at some $\e_b^0<\e_b^\text{vac}$ (see \fig{fig:pt}(b)), where $\e_b^\text{vac}$ is the binding energy in vacuum,  indicating that $\jpsi$ breaks down even before $\e_b$ drops to zero, which is the case at $\bm B_0=0$. This $\e_b^0$ is a strong function of $E_0$ as can be seen in \fig{fig:contour}. 
It satisfies the equation $\partial w/\partial \e_b = 0$. In the case $\gamma\ll 1$  \eq{zz} and \eq{zz1} imply 
that 
\beql{e0-1}
\e_b^0= \frac{m}{2}\left( \frac{5eE}{2m^2}\right)^{2/3}\,,\quad \gamma\ll 1
\eeq
At $\gamma\gg 1$ and $\theta= \pi/2$ we employ \eq{gbig} to derive the condition $(\e_b^0)^2+eB\e_b^0/2m-eE^2/B=0$. In view of \eq{gg}  $E\approx B$ and we obtain
\beql{e0-2}
\e_b^0=\frac{eB}{4m}\left( \sqrt{\frac{16m^2}{eB}+1}-1\right)\approx \sqrt{eB}\,,\quad \gamma\gg 1
\eeq
where in the last step I used that $eB\ll m^2$. For a given function $\e_b(T)$ one can convert $\e_b^0$ into the dissociation temperature, which is an important phenomenological parameter.  

In the absence of electric fied $\b E_0$, the dissociation probability peaks in the direction perpendicular to the direction of magnetic field $\bm b_0$, i.e.\ in the reaction plane. Dissociation rate vanishes in the $\bm b_0$ direction. Indeed,  for $\bm V\cdot \bm b_0=0$ \eq{em-cf} implies that $E=0$.  This feature is seen in the left panel of \fig{fig:azimuth}. At finite $\bm E_0$ the  dissociation probability is finite in the $\bm b_0$ direction making the azimuthal distribution more symmetric. The shape of the azimuthal distribution strongly depends on quarkonium velocity: while at low $V$ the strongest dissociation is in the direction of the reaction plane, at higher $V$ the maximum shifts towards small angles around the $\bm b_0$ direction. 
Extrema of the azimuthal distribution are roots of the equation $\partial w/\partial \phi=0$. At  $\gamma\ll 1$ it yields minimum at  $\phi_0=0$, maximum at $\phi_0=\pi/2$ and another maximum that satisfies 
the condition (neglecting the spin-dependence of $\e_b$) 
\beql{max=phi}
eE_0\sqrt{1+\gamma_L^2(V_z^2+V_\bot^2\sin^2\phi_0)(1+\rho_0^{-2})}= \frac{2m^2}{3}\left( \frac{2\e_b}{m}\right)^{3/2}
\eeq
In order to satisfy  \eq{max=phi} $\phi_0$ must decrease when $V$ increases and visa versa. This features are seen in the left panel of \fig{fig:azimuth}.

Spectrum of quarkonia  surviving in the electromagnetic field is proportional to the survival probability $P=1-w t$, where $t$ is the time spent by the quarkonium in the field. Consider $P$ as a function of the angle $\chi$ between the quarkonium velocity and the reaction plane $\chi= \pi/2-\phi$. Fourier expansion of $P$ in $\chi$ reads
\beql{fourP}
P(\chi)= \frac{1}{2}P_0+\sum_{n=1}^\infty P_n\,\cos(n\chi)\,,\quad P_n= \frac{1}{\pi}\int_{-\pi}^{\pi}P(\chi)\,\cos(n\chi)\,d\chi\,.
\eeq
Ellipticity of the distribution is characterized  by the ``elliptic flow" coefficient $v_2$ defined as 
\beq\label{v2}
v_2=\frac{P_2}{\frac{1}{2}P_0}= \frac{\int_{-\pi}^{\pi} (1-wt)\,\cos2\chi\, d\chi}{\pi\aver{P}} = - \frac{t}{\pi\aver{P}}\int_{-\pi}^{\pi}  w\,\cos2\chi \, d\chi
\eeq
where $\aver{P}$ denotes average of $P$ over the azimuthal angle. These formulas are applicable only as long as $wt<1$ because otherwise there are no surviving quarkonia. 
In  the right panel of \fig{fig:azimuth} \cite{Tuchin:2011cg} I show $v_2\aver{P}/t$, which is independent of $t$, as a function of $P_\bot$. As expected, in the absence of the CME, $v_2$ is negative at low $P_\bot$ and positive at high $P_\bot$. $v_2$ changes sign at $P_\bot$ that depends on the strength of the electric field. It decreases as  $E_0$ increases until at $E_0\simeq B_0$ it becomes positive at all $P_\bot$.  
\fig{fig:azimuth}(b) provides the low bound for $v_2$  because $\aver{P}<1$ and $t\gtrsim 1$~fm  . We thus expect that  magnetic field strongly modifies the azimuthal distribution of the produced $\jpsi$'s. Role of the magnetic field in generation of azimuthal anisotropies in heavy-ion collisions has been pointed out before in \cite{Tuchin:2010gx,Mohapatra:2011ku}.


{\bf In summary}, we observed that $\jpsi$ dissociation energy increases  with magnetic field strength and quarkonium momentum. As a consequence, \emph{quarkonia dissociate at lower temperature} than one would have expected based on calculations neglecting magnetic field \cite{Marasinghe:2011bt,Tuchin:2011cg}. \fig{fig:pt} indicates that in heavy-ion collisions at the LHC, all $\jpsi$'s moving with $P_\bot > 0.5$~GeV  in the reaction plane  would dissociate with probability of order unity even if the QGP effect were completely negligible. If electric field fluctuations shown in \fig{fig:aa5} are taken into account,  then even low $P_\bot$ $\jpsi$'s are destroyed. However, Chiral Magnetic Effect has negligible effect on $\jpsi$ dissociation. 
 
 Although magnetic fields in $pp$ and $pA$ collisions are much weaker than in $AA$ collisions, they are still strong enough to cause $\jpsi$ dissociation at sufficiently high momenta $P_\bot$. A truly spectacular feature of such process would be $\jpsi$ decay into two heavier $D$-mesons. 

The effect of $\jpsi$  dissociation in a magnetic field  vanishes in the direction parallel to the magnetic field, i.e.\ perpendicular to the reaction plane. Therefore,  $\jpsi$ dissociation gives negative contribution to the total azimuthal asymmetry coefficient $v_2$. It remarkable that  presence of electric field reverses this effect making $v_2$ positive.

\setcounter{equation}{0}
\newpage
\section{Electromagnetic radiation by quark-gluon plasma in magnetic field}\label{sec:k}

\subsection{Necessity to quantize fermion motion }\label{sec:ka}

In \sec{sec:d} we discussed synchrotron radiation of \emph{gluons} by \emph{fast} quarks.
Our main interest was the energy loss problem. In this section we turn to the problem of  electromagnetic radiation by QGP, viz.\ radiation of \emph{photons} by thermal fermions \cite{Tuchin:2012mf}. In this case quasi-classical approximation that we employed in \sec{sec:d} and \sec{sec:f} is no longer applicable and one has to take into account quantization of fermion motion in magnetic field.

Electromagnetic radiation by quarks and antiquarks of QGP moving in external magnetic field originates from two sources: (i) synchrotron radiation and (ii) quark and antiquark annihilation. QGP is transparent to the emitted electromagnetic radiation because its absorption coefficient  is suppressed by $\alpha^2$. Thus,  QGP is shinning in magnetic field. The main goal of this paper is to calculate the spectrum and angular distribution of this radiation. In strong magnetic field it is essential to account for quantization of fermion spectra. Indeed, spacing between the Landau levels is of the order $e B/\e$ ($\e$ being quark energy), while their thermal width is of the order $T$.  Spectrum quantization is negligible only if $eB/\e\ll T$ which is barely the case at RHIC and certainly not the case at LHC (at least during the first few fm's of the evolution). Fermion spectrum quantization is important not only for hard and electromagnetic probes but also for the bulk properties of QGP.

\subsection{Synchrotron radiation}\label{sec:synch}

Motion of charged fermions in external magnetic field, which I will approximately treat as spatially homogeneous,  is quasi-classical in the field  direction and quantized in the \emph{reaction plane}, which is  perpendicular to the magnetic field and span by the impact parameter and the heavy ion collision axis.   In high energy physics  one usually distinguishes the \emph{transverse plane}, which is  perpendicular to the collision axis and span by the magnetic field and the impact parameter.  In this section I use  notation in which three-vectors are discriminated by the bold face and their  component along the field direction by the plain face. Momentum projections onto the transverse plane are denoted by subscript $\bot$.

In the configuration space, charged fermions move along spiral trajectories with the symmetry axis aligned with the field direction. Synchrotron radiation is a process of photon $\gamma$ radiation by a fermion $f$ with electric charge $e_f=z_f e$  in external magnetic field $B$:
\beql{srad}
f(e_f,j,p)\to f(e_f,k,q)+\gamma(\b k)\,,
\eeq
where $\b k$ is the photon momentum, $p,q$ are the momentum components along the magnetic field direction and 
indicies $j,k=0,1,2,\ldots$  label the discrete Landau levels in the reaction plane.  
The Landau levels  are given  by
\begin{align}\label{mass-shell}
\e_j= \sqrt{m^2+p^2+2j e_f  B}\,,\quad \e_k= \sqrt{m^2+q^2+2ke_f  B}\,,
\end{align}
In the constant magnetic field only momentum component along the field direction is conserved. Thus, the  conservation laws  for synchrotron radiation read
\begin{align}\label{conserv}
\e_j= \omega +\e_k\,,\quad p=q+\omega\cos\theta\,,
\end{align}
where $\omega$ is the photon energy and $\theta$ is the photon emission angle with respect to the magnetic field. Intensity of the synchrotron radiation was derived in \cite{Sokolov:1968a}. In \cite{Herold:1982a,Harding:1987a,Latal:1986a,Baring:1988a}  it was thoroughly investigated as a possible mechanism for $\gamma$-ray bursts. In particular, synchrotron radiation in electromagnetic plasmas was calculated. 
Spectral intensity of angular distribution of synchrotron radiation by a fermion in the $j$'th Landau state  is given by 
\begin{align}\label{intj}
\frac{dI^j}{d\omega d\Omega}=\sum_f\frac{z_f^2 \alpha}{\pi}\omega^2\sum_{k=0}^j\Gamma_{jk}\left\{
|\mathcal{M}_\bot|^2+|\mathcal{M}_\parallel|^2\right\}\,\delta(\omega-\e_j+\e_k)
\end{align}
where $\Gamma_{jk} = (1+\delta_{j0})(1+\delta_{k0})$ accounts for the double degeneration of all Landau levels except the ground one.  The squares of matrix elements $\mathcal{M}$, which  appear in \eq{intj}, corresponding to photon polarization perpendicular and parallel to the magnetic field are given by, respectively,
\begin{align}
4\e_j\e_k|\mathcal{M}_\bot|^2= &(\e_j\e_k-pq-m^2)[I_{j,k-1}^2+I_{j-1,k}^2]+2\sqrt{2j e_f   B}\sqrt{2k e_f   B}[I_{j,k-1}I_{j-1,k}]\label{Mmat1}\,.\\
 4\e_j\e_k|\mathcal{M}_\parallel|^2=& \cos^2\theta\big\{ ( \e_j\e_k-pq-m^2)[I_{j,k-1}^2+I_{j-1,k}^2]-2\sqrt{2je_f   B}\sqrt{2k e_f   B}[I_{j,k-1}I_{j-1,k}]\big\}\nonumber\\
 &-2\cos\theta\sin\theta\big\{ p\sqrt{2 k e_f   B}[I_{j-1,k}I_{j-1,k-1}+I_{j,k-1}I_{j,k}]\nonumber\\
  &  +
 q\sqrt{2je_f   B}[I_{j,k}I_{j-1,k}+I_{j-1,k-1}I_{j,k-1}]\big\}\nonumber\\
 &+ \sin^2\theta\big\{ (\e_j\e_k+pq-m^2)[I_{j-1,k-1}^2+I_{j,k}^2]+2\sqrt{2je_f   B}\sqrt{2ke_f   B}(I_{j-1,k-1}I_{j,k})\big\}\,, \label{Mmat2}
\end{align}
where for $j\ge k$, 
\beql{ijk} 
I_{j,k}\equiv I_{j,k}(x)= (-1)^{j-k}\sqrt{\frac{k!}{j!}}e^{-\frac{x}{2}}x^{\frac{j-k}{2}}L_k^{j-k}(x).
\eeq
and $I_{j,k}(x)= I_{k,j}(x)$ when $k>j$. ($I_{j,-1}$ are identically zero).
The functions $L_k^{j-k}(x)$ are the generalized Laguerre polynomials. Their argument is 
\beql{xa}
x=\frac{\omega^2}{2e_f   B}\sin^2\theta\,.
\eeq

Angular distribution of radiation is obtained by  integrating over the photon energies and remembering that $\e_k$ also depends on $\omega$ by virtue of \eq{mass-shell} and \eq{conserv}:
\begin{align}\label{dIdO}
\frac{dI^j}{d\Omega}=\sum_f\frac{z_f  ^2 \alpha}{\pi}\sum_{k=0}^j \frac{\omega^*(\e_j-\omega^*)}{\e_j-p\cos\theta -\omega^*\sin^2\theta}\Gamma_{jk} \left\{
|\mathcal{M}_\bot|^2+|\mathcal{M}_\parallel|^2\right\}\,,
\end{align}
where photon energy $\omega$ is fixed to be 
\beql{omega*}
\omega^* = \frac{1}{\sin^{2}\theta}\left\{(\e_j-p\cos\theta)-\big[(\e_j-p\cos\theta)^2-2e_f   B(j-k)\sin^2\theta\big]^{1/2}\right\}\,.
\eeq

In the context  of heavy-ion collisions the relevant  observable is the differential photon spectrum. For ideal plasma in equilibrium each quark flavor gives the following contribution to the photon spectrum:
\begin{align}\label{def-sp}
\frac{dN^\text{synch}}{dt d\Omega d\omega} =\sum_f \int_{-\infty}^\infty dp \frac{e_f  B (2N_c) V}{2\pi^2} \sum_{j=0}^\infty\sum_{k=0}^j \frac{dI^j}{\omega d\omega d\Omega}(2-\delta_{j,0}) f(\e_j)[1-f(\e_k)]\,,
\end{align}
where $2N_c$ accounts for quarks and antiquarks each of  $N_c$ possible colors, and $(2-\delta_{j,0})$ sums over the initial quark spin. Index $f$ indicates different quark flavors. $V$ stands for the plasma volume. 
The statistical factor $f(\e)$ is
\beql{stat}
f(\e)= \frac{1}{e^{\e/T}+1}\,.
\eeq
The $\delta$-function appearing in \eq{intj} can be re-written 
using \eq{mass-shell} and \eq{conserv}  as 
\beql{re1}
\delta(\omega-\e_j+\e_k) = \sum_{\pm}\frac{\delta(p-p^*_\pm)}{\big| \frac{p}{\e_j}-\frac{q}{\e_k}\big|}\,,
\eeq
where 
\begin{align}\label{p*}
p^*_\pm= &\bigg\{\cos\theta (m_j^2-m_k^2+\omega^2\sin^2\theta)\nonumber\\
&\pm \sqrt{[(m_j+m_k)^2-\omega^2\sin^2\theta][(m_j-m_k)^2-\omega^2\sin^2\theta]}\,\bigg\}/(2\omega\sin^2\theta)\,.
\end{align}
The following convenient notation was introduced:
\beql{mm}
m_j^2= m^2+2je_f  B\,,\quad m_k^2= m^2+2ke_f  B\,.
\eeq
The physical meaning of \eq{p*} is that synchrotron radiation of a  photon with energy $\omega$ at angle $\theta$ by a fermion undergoing transition from $j$'th to $k$'th Landau level is possible only if the initial quark momentum along the field direction equals $p^*_\pm$.  

Another consequence of the conservation laws  \eq{conserv} is that 
for a given $j$ and $k$ the photon energy cannot exceed a certain maximal value that will be denoted by $\omega_{s,jk}$.  Indeed, inspection of \eq{p*} reveals that this equation has a real solution only in two cases 
\beql{cases}
\text{(i)}\,\, m_j-m_k\ge\omega \sin\theta\,,\quad \text{or}\quad \text{(ii)}\,\, m_j+m_k\le \omega \sin\theta\,.
\eeq
The first case is relevant for the synchrotron radiation while the second one for the one-photon pair annihilation as discussed in the next section. Accordingly, allowed photon energies in the $j\to k$ transition satisfy  
\beql{omegac}
\omega\le \omega_{s,jk} \equiv \frac{m_j-m_k}{\sin\theta}= \frac{\sqrt{m^2+2je_f   B}-\sqrt{m^2+2ke_f  B}}{\sin\theta}\,.
\eeq 
No synchrotron radiation is possible for  $\omega>\omega_{s,jk}$. In particular, when $j=k$, $\omega_{s,jk}=0$, i.e.\ no photon is emitted, which is also evident in   \eq{omega*}.  The reason is clearly seen in the frame where $p=0$: since $\e_j\ge \e_k$, constraints \eq{mass-shell} and \eq{conserv} hold only if $\omega=0$. 

Substitution of \eq{intj} into \eq{def-sp} yields the spectral distribution of the synchrotron radiation rate per unit volume
\begin{align}\label{spec1}
\frac{dN^\text{synch}}{Vdt d\Omega d\omega}= \sum_f\frac{2N_cz_f^2\alpha}{\pi^3}e_f  B 
\sum_{j=0}^\infty \sum_{k=0}^j \omega (1+\delta_{k0}) \,\vartheta(\omega_{s,ij}-\omega)
\int dp \sum_\pm\frac{\delta(p-p^*_\pm)}
{\big| \frac{p}{\e_j}-\frac{q}{\e_k}\big|}
&\nonumber\\
\times\left\{
|\mathcal{M}_\bot|^2+|\mathcal{M}_\parallel|^2\right\} f(\e_j)[1-f(\e_k)] \,,&
\end{align} 
where $\vartheta$ is the step-function. 

The natural variables to study the synchrotron radiation are the photon energy $\omega$ and its emission angle $\theta$ with respect to the magnetic field. However, in high energy physics particle spectra are traditionally presented in terms of  rapidity $y$ (which for photons is equivalent to pseudo-rapidity)  and transverse momentum $k_\bot$. $k_\bot$ is a projection of three-momentum $\b k$ onto the transverse plane. These variables are not convenient to study electromagnetic processes in external magnetic field. In particular, they conceal the azimuthal symmetry with respect to the magnetic field direction. To change variables, let $z$ be the collision axis and $\unit y$ be the direction of the magnetic field. In spherical coordinates photon momentum is given by $\b k =\omega( \sin\alpha\cos\phi\unit x+\sin\alpha\sin\phi\unit y+\cos\alpha\unit z)$, where $\alpha$ and $\phi$ are the polar and azimuthal angles with respect to $z$-axis. The plane $xz$ is  the reaction plane.  By definition, $\unit k\cdot \unit y = \cos\theta$ implying that $\cos\theta= \sin\alpha\sin\phi$. Thus, 
\beql{coor1}
k_\bot=\sqrt{ k_x^2+k_y^2} = \frac{\omega\cos\theta}{\sin\phi}\,,\quad y= -\ln\tan\frac{\alpha}{2}\,.
\eeq  
The second of these equations is the definition of (pseudo)-rapidity. Inverting \eq{coor1} yields
\beql{coor2}
\omega= k_\bot \cosh y\,,\quad \cos\theta= \frac{\sin\phi}{\cosh y}\,.
\eeq
Because $dy=dk_z/\omega$ the photon multiplicity in a unit volume per unit time reads
\beql{mult2}
\frac{dN^\text{synch}}{dVdt\,d^2k_\bot dy}=\omega\frac{dN^\text{synch}}{dVdt\,d^3k}= \frac{dN^\text{synch}}{dVdt\,\omega d\omega d\Omega}
\eeq

\begin{figure}[ht]
\begin{tabular}{cc}
      \includegraphics[height=5cm]{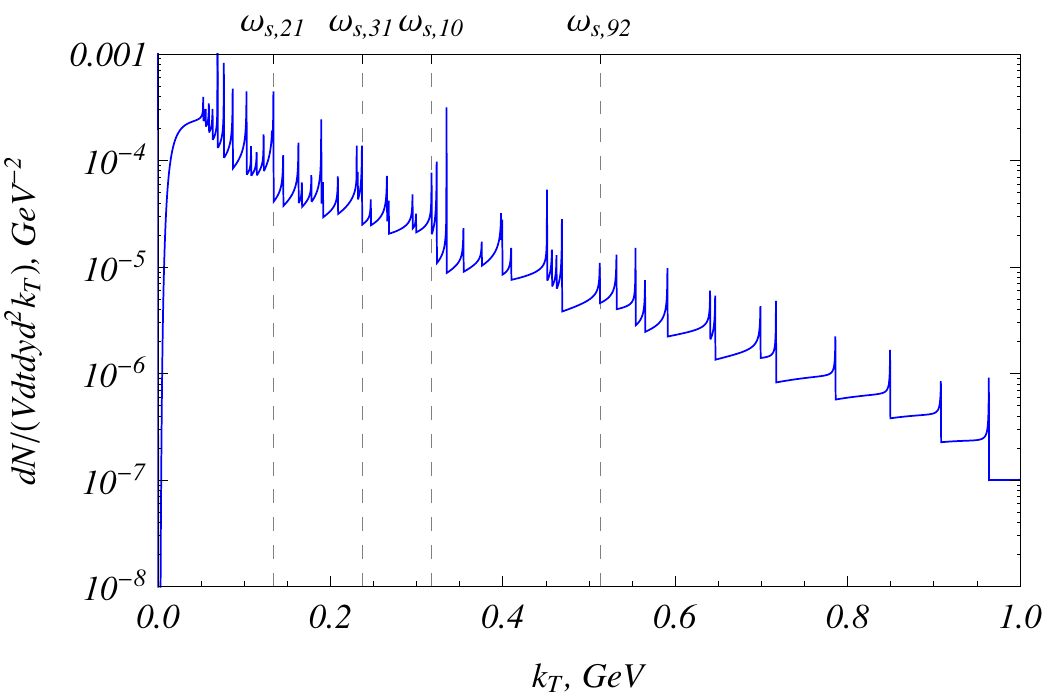} &
      \includegraphics[height=5cm]{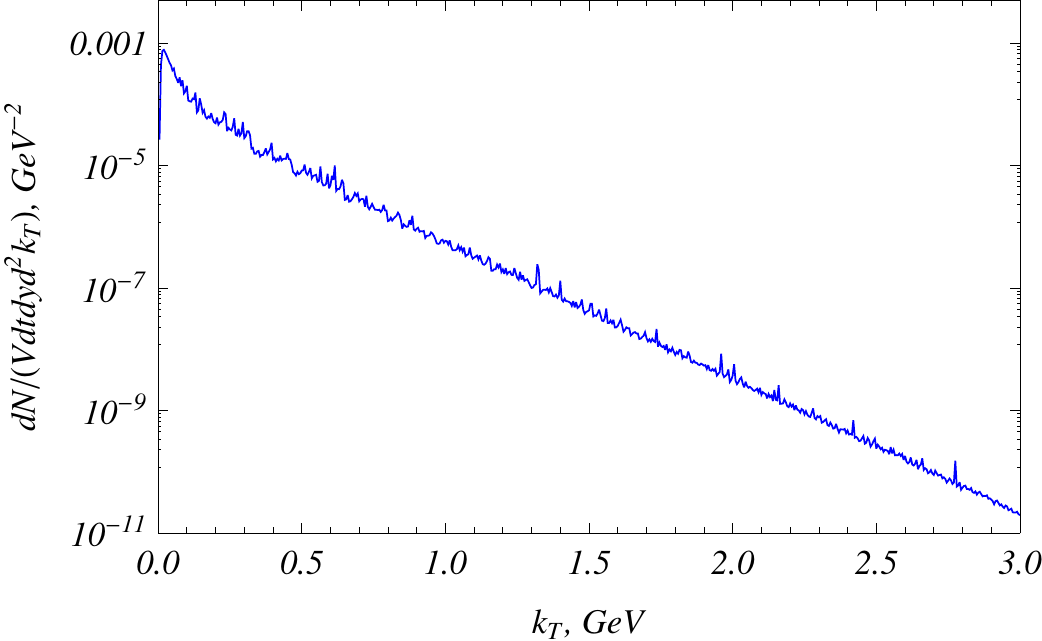}\\
      $(a)$ & $(b)$ 
      \end{tabular}
  \caption{Spectrum of synchrotron radiation by $u$ quarks at $eB=m_\pi^2$, $y=0$, $\phi=\pi/3$: (a) contribution of 10 lowest Landau levels $j\le 10$; several cutoff frequencies are indicated; (b) summed over all Landau levels. $m_u=3$~MeV, $T=200$~MeV. Adopted from \cite{Tuchin:2012mf}.}
\label{synch}
\end{figure}
\begin{figure}[ht]
      \includegraphics[height=5cm]{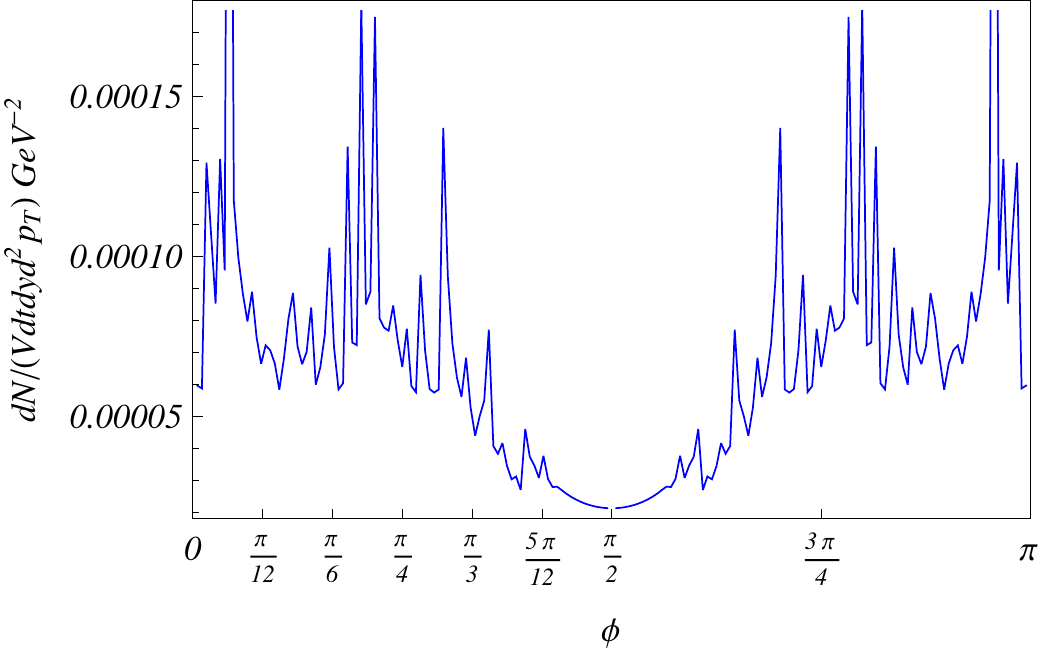} 
  \caption{Azimuthal distribution of synchrotron radiation by $u$-quarks at  $k_\bot=0.2$~GeV, $eB=m_\pi^2$, $y=0$. $m_u=3$~MeV. Adopted from \cite{Tuchin:2012mf}}
\label{synch2}
\end{figure}

\fig{synch} displays the spectrum of synchrotron radiation by $u$ quarks as a function of $k_\bot$ at fixed $\phi$ \cite{Tuchin:2012mf}. At midrapidity $y=0$ \eq{coor2} implies that $k_\bot=\omega$.  Contribution of $d$ and $s$ quarks is qualitatively similar. At $eB\gg m^2$, quark masses do not  affect the spectrum much.  The main difference stems from the difference in electric charge. In panel (a) only the contributions of the first ten Landau levels are displayed. The cutoff frequencies $\omega_{s,jk}$ can be clearly seen and some of them are indicated on the plot for convenience. The azimuthal distribution is shown in \fig{synch2}. Note, that  at midrapidity $\phi= \pi/2-\theta$. Therefore, the figure indicates that photon production in the direction of magnetic field (at $\phi= \pi/2$) is suppressed. More photons are produced in the direction of the reaction plane $\phi=0$. This results in the ellipticity of the photon spectrum that translates into the positive ``elliptic flow" coefficient $v_2$. It should be noted, that the classical synchrotron radiation has a similar angular  distribution. 

\begin{figure}[ht]
\begin{tabular}{cc}
      \includegraphics[height=5cm]{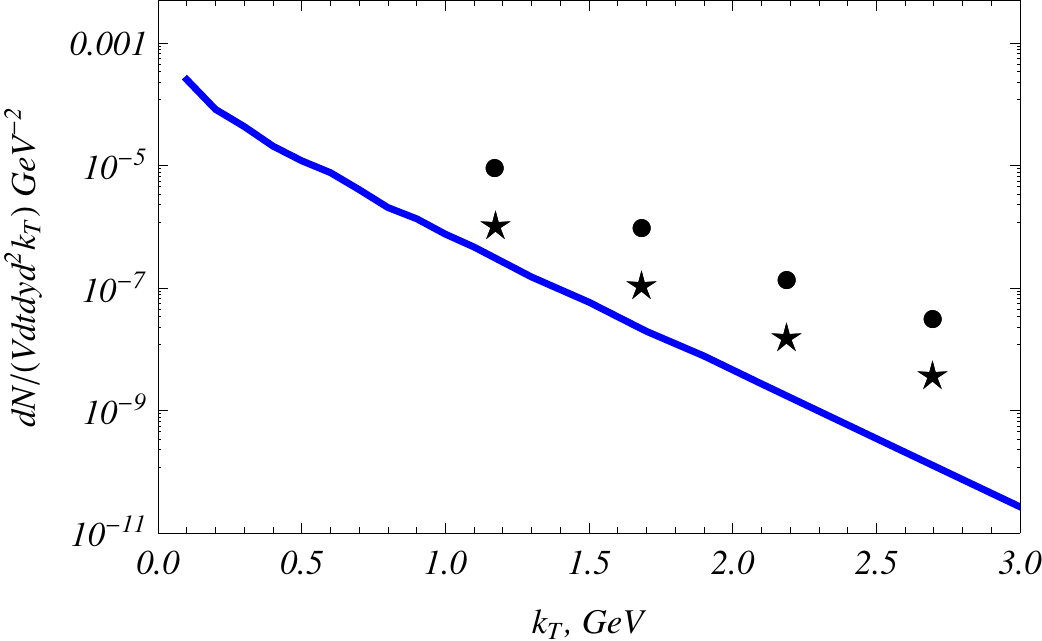} &
      \includegraphics[height=5.cm]{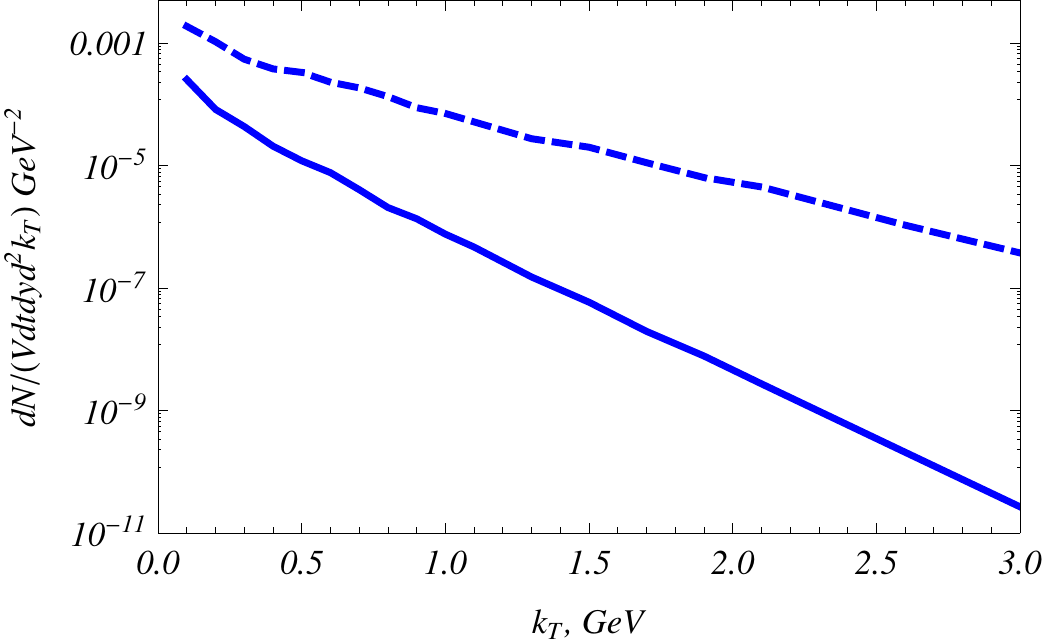}\\
      $(a)$ & $(b)$ 
      \end{tabular}
  \caption{ Azimuthal average of the synchrotron radiation spectrum of $u$,$d$,$s$ quarks and their corresponding antiquarks.  (a) $eB=m_\pi^2$, $y=0$ compared to the experimental data from \cite{Adare:2008ab} divided by $Vt=25\pi$~fm$^4$ (dots) and 
 $Vt=9\times 25\pi$~fm$^4$ (stars), (b)  $eB=m_\pi^2$, $T=200$~MeV, $y=0$ (solid line) compared to $eB=15m_\pi^2$, $T=400$~MeV, $y=0$ (dashed line). $m_u=3$~MeV, $m_d=5$~MeV, $m_s=92$~MeV. Adopted from \cite{Tuchin:2012mf}.}
\label{tot-synch}
\end{figure}

In order to compare the photon spectrum produced by synchrotron radiation to the photon spectrum measured in heavy-ion collisions, the $u$, $d$ and $s$ quarks contributions were summed up. Furthermore, the experimental data from \cite{Adare:2008ab} was divided by $Vt$, where $t$ is the magnetic field relaxation time.  The volume of the plasma can be estimated as $V=\pi R^2t$ with $R\approx 5$~fm being the nuclear radius. Therefore, 
\beql{mult4}
\frac{dN^\gamma_\text{exp}}{dVdt\, d^2k_\bot dy}=\frac{dN^\gamma_\text{exp}}{d^2k_\bot dy}\,\frac{1}{\pi R^2t^2}=\frac{dN^\gamma_\text{exp}}{d^2k_\bot dy}\,\left(\frac{\text{GeV}}{14.9}\right)^4\,\left( \frac{1\,\text{fm}}{t}\right)^2\,.
\eeq
 The results are plotted in \fig{tot-synch}. In panel (a) it is seen that synchrotron radiation gives a significant contribution to the photon production in heavy-ion collisions at RHIC energy. This contribution is larger at small transverse momenta. This may explain enhancement of  photon production observed in \cite{Adare:2008ab}. Panel (b) indicates the increase of the photon spectrum produced by the synchrotron radiation mechanism at the LHC energy. This increase is due to enhancement of the magnetic field strength, but mostly because of  increase of plasma temperature.  This qualitative features can be better understood by considering the limiting cases of low and high photon energies.

One possible way to ascertain the contribution of electromagnetic radiation in external magnetic field is to isolate the azimuthally symmetric component with respect to the direction of the magnetic field. It seems that synchrotron radiation dominates the photon spectrum at low $k_\bot$. Thus, azimuthal symmetry can be easily checked by simply plotting the multiplicity vs $\omega$, $\theta$ and $\varphi$, where $\omega$ is photon energy, $\theta$ is emission angle with respect to the magnetic field and $\varphi$ is azimuthal angle around the magnetic field direction (which is perpendicular both to the collision axis and to the impact parameter). In \fig{synch}(a) it is also seen that in these variables it may be possible to discern the cutoff frequencies $\omega_{s,jk}$ that appear as resonances (in  \fig{synch} $y=0$ so $k_\bot = \omega$). Note that averaging over the azimuthal angle $\alpha$ around the collision axis direction destroys these features, see \fig{tot-synch}. 

\subsubsection{Low photon energy}

The low energy  part of the photon spectrum satisfies the condition  $\omega\ll \sqrt{e_fB}$. The corresponding initial quark momentum component along the field $p$ and energy $\e_j$ follow from \eq{p*} and \eq{mass-shell} and are given by
\beql{papprox}
 p_\pm^*\approx \frac{(j-k)e_fB(\cos\theta\pm 1)}{\omega\sin^2\theta}+\mathcal{O}(\omega)\,,\qquad \e_j\approx |p_\pm^*|+\mathcal{O}(\omega)\,.
\eeq
Evidently, $\e_j\gg eB$. In practice, magnetic field strength satisfies $\sqrt{eB}\gtrsim T$, so that $\e_j\gg T$. Therefore, synchrotron radiation is dominated by fermion transitions from low Landau levels due to the statistical factors appearing in \eq{def-sp}. 

For a qualitative discussion it is sufficient to consider the $1\to 0$ transition. In this case the matrix elements \eq{Mmat1} and \eq{Mmat2} read
\beql{m10}
|\mathcal{M}^{1,0}|^2=\frac{1}{2\e_1\e_0}\left\{ I_{1,0}^2(\e_1\e_0 -pq\cos^2\theta-m^2)+\cos\theta\sin\theta q\sqrt{2e_f   B}I_{1,0}I_{0,0}\right\}\,.
\eeq
Assuming that the field strength is supercritical, i.e.\  $e_f   B\gg m^2$, but keeping all powers of  $\omega$ (for future reference) \eq{p*} reduces to 
\beql{p*sc}
p_\pm^*\approx \frac{1}{2\omega \sin^2\theta}\left\{ 2e_f  B(\cos\theta\pm 1)+\omega^2 \sin^2\theta (\cos\theta\mp 1)\right\}\,.
\eeq
Furthermore, using the conservation laws \eq{conserv} we obtain in this approximation
\begin{align}
\e_{1\pm}= &\frac{1}{2\omega \sin^2\theta}\left| 2e_f   B(\cos\theta\pm 1)-\omega^2 \sin^2\theta (\cos\theta \mp 1)\right|\label{en-mom1}\,,\\
q_\pm=& \frac{1}{2\omega \sin^2\theta}(2e_f  B-\omega^2\sin^2\theta)(\cos\theta\pm 1)\,,
\label{en-mom2}\\
\e_{0\pm}=&|q|\,.\label{en-mom3}
\end{align}
The values of the non-vanishing matrix elements $I_{j,k}$ defined by \eq{ijk} are
\begin{align}\label{is}
I_{1,0}(x)&= -x^{1/2}e^{-x/2}\,,\qquad I_{0,0}(x)= e^{-x/2}\,.
\end{align}
For $j=1$, $k=0$ we write using \eq{omegac} $\omega_{s,10}= \sqrt{2e_f  B}/\sin\theta$. Then \eq{xa} implies $x=\omega^2/\omega_{s,10}^2$. 
Substituting \eq{p*sc}--\eq{is} into \eq{m10} gives
\begin{align}
|\mathcal{M}^{1,0}_\pm|^2= \frac{1}{2}xe^{-x}\left[ 1-
\frac{\cos\theta(1+x)\pm (1-x)}{\cos\theta(1-x)\pm (1+x)}\cos^2\theta
-\frac{2(1-x)\cos\theta\sin^2\theta}{\cos\theta(1-x)\pm (1+x)} 
\right]\,.
\end{align}
According to \eq{spec1} the contribution of the $1\to 0$ transition to the synchrotron radiation reads \cite{Tuchin:2012mf}
\begin{align}\label{ap1}
\frac{dN^{\text{synch},10}}{Vdt d\Omega d\omega }= \sum_f\frac{2N_cz_f^2\alpha}{\pi}\omega \Gamma\frac{e_f  B}{2\pi^2}
\sum_\pm  f(\e_1)[1-f(\e_0)]|\mathcal{M}^{1,0}_\pm|^2\,&\nonumber\\
\times\frac{(1-x)\cos\theta\pm (1+x)}{-2x(\cos\theta\mp 1)}\,\vartheta(\omega_{s,10}-\omega)\,.&
\end{align}
Consider radiation spectrum at $\theta=\pi/2$, i.e.\ perpendicular to the magnetic field. The spectrum increases with $x$ and reaches maximum at $x=1$. Since  $x=\omega^2/(2e_f   B)$, spectrum decreases with increase of $B$ at fixed $\omega$. This feature holds at low $x$ part of the spectrum for other emission angles and even for transitions form higher excited states. However, at high energies, it is no longer possible to approximate the spectrum by the contribution of a few low Landau levels. In that case the typical values of quantum numbers are $j,k\gg 1$. For example, to achieve the numerical accuracy of 5\%, sum over $j$ must run up to a certain $j_\text{max}$. Some values of $j_\text{max}$ are listed in Table~\ref{table} \cite{Tuchin:2012mf}.  
\begin{table}[ht]
\begin{center}
\begin{tabular}{|c|cccccccccc|}
\hline 
$f$ & $u$ & $u$ & $u$ & $u$ & $u$ & $u$ & $s$ & $u$ & $u$ & $s$ \\ 
$eB/m_\pi^2$ & 1 & 1 & 1 & 1 & 1 & 1 & 1 & 15 & 15 & 15\\
$T$, GeV & 0.2 & 0.2 & 0.2 & 0.2 & 0.2 & 0.2 & 0.2 & 0.4 & 0.4 & 0.4\\
$\phi$ & $\frac{\pi}{3}$ & $\frac{\pi}{3}$ &  $\frac{\pi}{3}$ &  $\frac{\pi}{3}$ & $\frac{\pi}{6}$ & $\frac{\pi}{12}$ &$\frac{\pi}{3}$ & $\frac{\pi}{3}$ & $\frac{\pi}{3}$ & $\frac{\pi}{3}$ \\ 
$k_\bot$,~GeV & 0.1 & 1 & 2& 3 & 1 & 1& 1& 1 & 2 & 1\\
$x$ &0.096 & 9.6 & 38& 86& 29& 35& 19& 0.64& 2.6& 1.3\\ \hline
$j_\text{max}$ & 30 & 40 & 90 & 150 & 120 & 200& 90& 8 &  12& 16 \\
\hline
\end{tabular}
\end{center}
\caption{The upper summation limit in  \eq{spec1} that yields the 5\% accuracy. $j_\text{max}$ is the highest Landau level  of the initial quark that is taken into account at this accuracy. Throughout the table $y=0$.}\label{table}
\end{table}
%

\subsubsection{High  photon energy}

\begin{figure}[ht]
      \includegraphics[height=5cm]{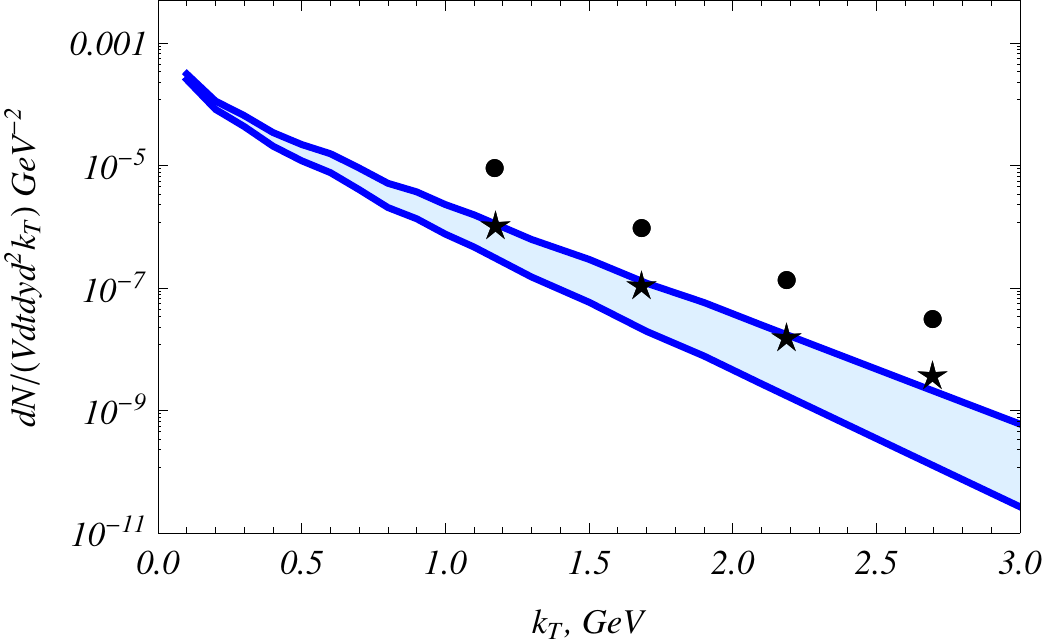} 
  \caption{Variation of the synchrotron spectrum with plasma temperature. Lower line: $T=200$~MeV, upper line: $T=250$~MeV. Other parameters are the same as in \fig{tot-synch}(a). Adopted from \cite{Tuchin:2012mf}.}
\label{synch-err}
\end{figure}

The high energy tail of the photon spectrum is quasi-classical and approximately continuous. In this case the Laguerre polynomials can be approximated by the Airy functions or the corresponding  modified Bessel functions. The angular distribution of the spectrum can be found in \cite{Baring:1988a}:
\beql{hel}
\frac{dN^\text{synch}}{Vdt d\Omega d\omega }=\sum_f\frac{z_f  ^2\alpha}{\pi}\frac{n_f \omega m^2}{4T^3}\sqrt{\frac{e_fBT\sin\theta}{m^3}} e^{-\omega/T}\,,
\eeq
provided that $\omega\gg m\sqrt{mT/e_fB\sin\theta}$. Here $n_f$ is number density of flavor $f$, which is independent of $B$:
\beql{n.den}
n_f= \frac{2\cdot 2N_c\,e_fB}{4\pi^2}\sum_{j=0}^\infty\int_{-\infty}^\infty dp\, e^{-\e_j/T}\approx \frac{4N_c}{\pi^2}\,T^3\,.
\eeq
Here summation over $j$ was replaced by integration. It follows that this part of the spectrum increases with magnetic field strength  as $\sqrt{B}$ and and with temperature as $\sqrt{T}e^{-\omega/T}$. Therefore, variation of the spectrum with $T$ is much stronger than with $B$. The $T$ dependence is shown in \fig{synch-err}.

Unlike time-dependence of magnetic field, time-dependence of  temperature is non-negligible even during the  first few fm/c. Final synchrotron spectrum, which is an average over all temperatures, 
is dominated by high temperatures/early times. However, the precise form of time-dependence of temperature is model-dependent. Therefore,  the spectrum is presented at fixed temperatures, so that a reader can appreciate  its qualitative features in a model-independent way.

\subsection{Pair annihilation}\label{sec:annih}

The theory of  one-photon pair annihilation was developed in \cite{Harding:1986a,Wunner:1986a}.  It was shown in \cite{Wunner:1979a} that in the super-critical regime $eB\gg m^2$ one-photon annihilations is much larger than the two-photon annihilation.  In this section the one-photon annihilation of $q$ and $\bar q$ pairs in the QGP is calculated. 

For $q\bar q$ pair annihilation the conservation of energy and momentum is given by
\beql{conserv2}
\e_j+\e_k=\omega \,,\quad p+q= \omega\cos\theta\,.
\eeq
The spectral density of the annihilation rate per unit volume reads
\begin{align}\label{anih1}
\frac{dN^\text{annih}}{Vdtd\omega d\Omega}= \sum_f\frac{\alpha z_f^2 \omega N_c}{4\pi e_f  B}
\sum_{j=0}^\infty \sum_{k=0}^\infty \int dp\, \frac{2e_f   B}{2\pi^2}f(\e_j)
\int dq\, \frac{2e_f   B}{2\pi^2}f(\e_k)&\nonumber\\
\times \delta(p+q-\omega \cos\theta)\delta(\e_j+\e_k-\omega)\{ |\mathcal{T}_\bot|^2 + |\mathcal{T}_\parallel|^2\}\,, &
\end{align}
where the matrix elements $\mathcal{T}$ can be obtained from \eq{Mmat1},\eq{Mmat2} by making substitutions $\e_k\to -\e_k$, $q\to -q$ and  are given by
\begin{align}
4\e_j\e_k|\mathcal{T}_\bot|^2=& (\e_j\e_k-pq+m^2)[I_{j,k-1}^2+I_{j-1,k}^2]-2\sqrt{2j e_f   B}\sqrt{2k e_f   B}[I_{j,k-1}I_{j-1,k}]\label{Tmat1}\,.\\
 4\e_j\e_k|\mathcal{T}_\parallel|^2=& \cos^2\theta\big\{ ( \e_j\e_k-pq+m^2)[I_{j,k-1}^2+I_{j-1,k}^2]+2\sqrt{2je_f   B}\sqrt{2k e_f   B}[I_{j,k-1}I_{j-1,k}]\big\}\nonumber\\
 &-2\cos\theta\sin\theta\big\{ -p\sqrt{2 k e_f   B}[I_{j-1,k}I_{j-1,k-1}+I_{j,k-1}I_{j,k}]
 \nonumber\\
 &+q\sqrt{2je_f   B}[I_{j,k}I_{j-1,k}+I_{j-1,k-1}I_{j,k-1}]\big\}\nonumber\\
 &+ \sin^2\theta\big\{ (\e_j\e_k+pq+m^2)[I_{j-1,k-1}^2+I_{j,k}^2]-2\sqrt{2je_f   B}\sqrt{2ke_f   B}(I_{j-1,k-1}I_{j,k})\big\}\,, \label{Tmat2}
\end{align}
with the same functions $I_{i,j}$ as in \eq{ijk}. Integration over $q$ removes the  delta function responsible for the conservation of momentum along the field direction. The remaining delta function is responsible for  energy conservation and can be written in exactly the same form as in \eq{re1} with particle energies and momenta now obeying the conservation laws \eq{conserv2}.  It is straightforward to see that  momentum $p^*_\pm$ is still given by \eq{p*},\eq{mm}. The photon spectrum produced  by annihilation of quark in state $j$ with antiquark in state $k$ has a threshold 
$\omega_{a,ij}$ that is given by the  case (ii) in \eq{cases}:
\beql{cuta}
\omega\ge \omega_{a,ij}=\frac{m_j+m_k}{\sin\theta}= \frac{\sqrt{m^2+2j e_f   B}+\sqrt{m^2+2ke_f   B}}{\sin\theta}\,.
\eeq
Thus, the spectral density of the annihilation rate per unit volume is 
\begin{align}\label{spec2}
\frac{dN^\text{annih}}{Vdt d\omega d\Omega}= \sum_f\frac{\alpha z_f^2 \omega N_c}{4\pi^5}e_f  B 
\sum_{j=0}^\infty \sum_{k=0}^\infty 
\vartheta(\omega-\omega_{a,ij})
\int dp \sum_\pm\frac{\delta(p-p^*_\pm)}
{\big| \frac{p}{\e_j}-\frac{q}{\e_k}\big|}
&\nonumber\\
\times\left\{
|\mathcal{T}_\bot|^2+|\mathcal{T}_\parallel|^2\right\}f(\e_j)f(\e_k)\,. &
\end{align}
Passing to $y$ and $p_\bot$ variables in place of $\omega$ and $\theta$ is similar to  \eq{mult2}.

\begin{figure}[ht]
\begin{tabular}{cc}
      \includegraphics[height=5cm]{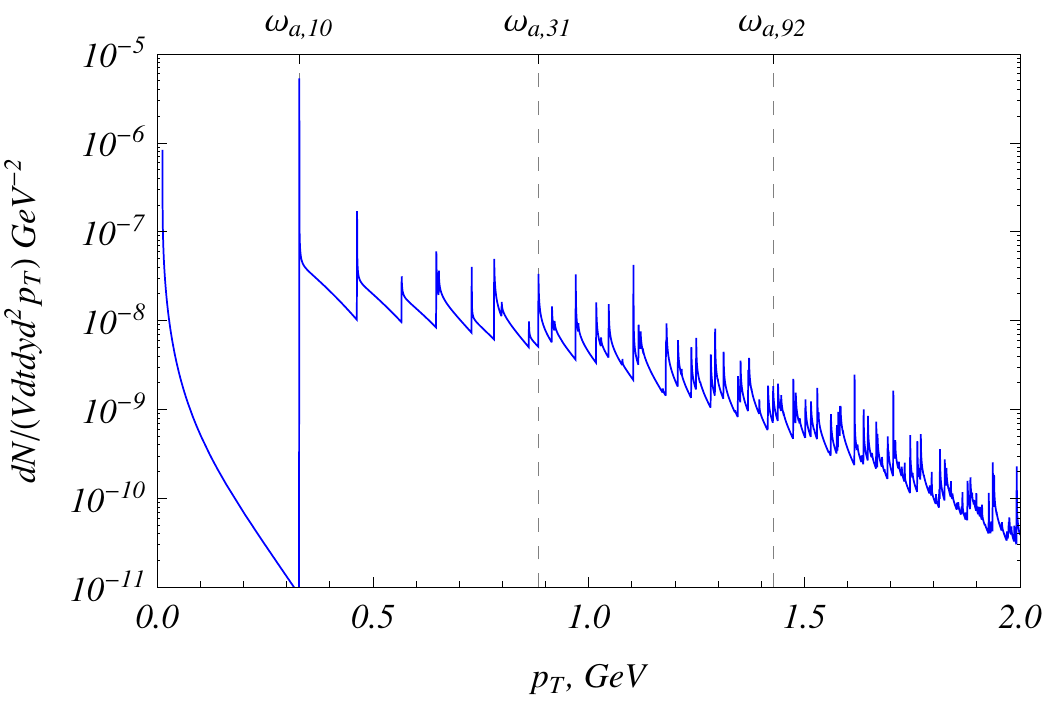} &
      \includegraphics[height=5.cm]{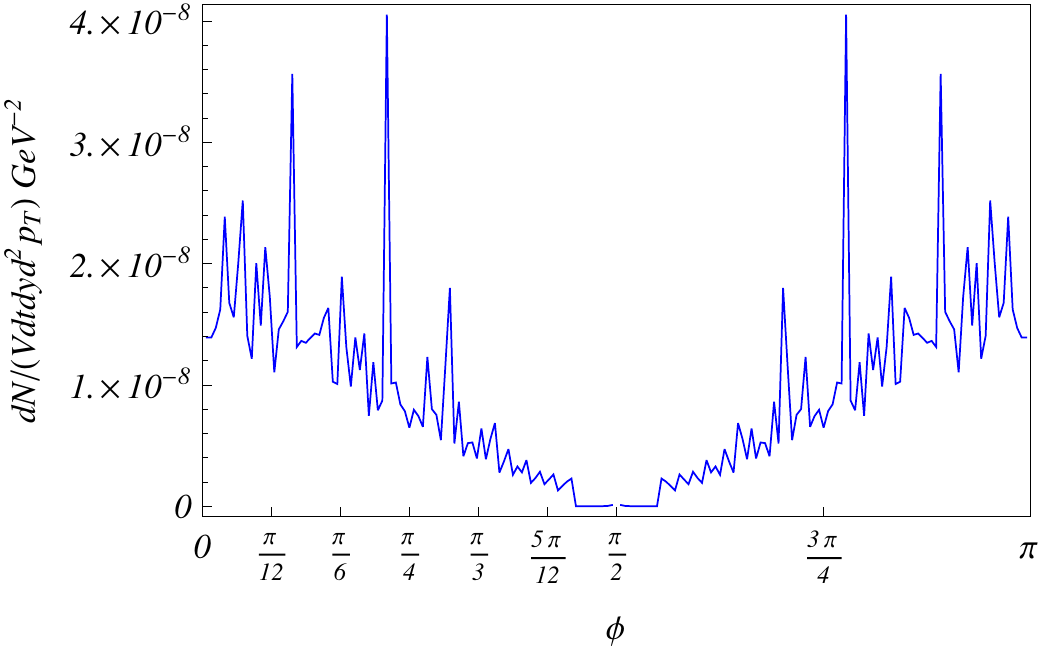}\\
      $(a)$ & $(b)$ 
      \end{tabular}
  \caption{ Photon spectrum in one-photon annihilation of $u$ and $\bar u$ quarks. $eB= m_\pi^2$, $y=0$. (a) $k_\bot$-spectrum  at  $\phi=\pi/3$, (b) azimuthal angule distribution at $k_\bot=1$~GeV.  Adopted from \cite{Tuchin:2012mf}.}
\label{annih}
\end{figure}

The results of the numerical calculations are represented in \fig{annih}. Panel (a) shows the spectrum of photons radiated in annihilation of $u$ and $\bar u$. We conclude that contribution of the annihilation channel is negligible as compared to the synchrotron radiation. 

\bigskip

{\bf In summary}, results of the calculations presented in this section indicate that photon production by QGP due to its interaction with external magnetic field give a considerable contribution to the total photon multiplicity in heavy-ion collisions. This is seen in \fig{tot-synch} were the model calculation is compared with the experimental data \cite{Adare:2008ab}.  The largest contribution to the photon multiplicity arises from photon momenta of the order of $\sqrt{eB}$. This may provide an explanation of the photon excess  observed by the PHENIX experiment \cite{Adare:2008ab}. Similar mechanism may also be responsible for enhancement 
of low mass di-lepton production that proceeds via emission of virtual photon which subsequently decays into di-lepton pair.

\setcounter{equation}{0}
\newpage
\section{Summary}\label{sec:y}

Analytical and numerical calculations indicate existence of extremely powerful electromagnetic fields in relativistic heavy ion collisions. They are the strongest  electromagnetic fields that exist in nature. They evolve slowly on characteristic QGP time scale and therefore have a profound effect on dynamics of QGP. In this review I described the recent progress in understanding of particle production in presence of these fields. Treating the fields as quasi-static and spatially homogeneous allowed us to use analytical results derived over the past half-century. This is however the main source of uncertainty that can be clarified only in comprehensive numerical approach based on  relativistic magnetohydrodynamics.

I discussed  many spectacular effects caused by magnetic field. All of them have direct phenomenological relevance. Breaking of spherical symmetry by magnetic field  in the direction perpendicular to the collision axis results in azimuthal asymmetry of particle production in the reaction plane. Fast quarks moving in magnetic field radiate a significant fraction of their energy. All electromagnetic probes are also naturally affected by magnetic field. Therefore, all experimental processes that are being used to study the properties of QGP have strong magnetic field dependence. In addition, the QCD phase diagram is modified by magnetic field as  
 has been extensively studied using model calculations \cite{Gusynin:1995nb,Chernodub:2010qx,Hidaka:2012mz,Chernodub:2012zx,Mizher:2010zb,Fraga:2008um,Fraga:2012fs,Gatto:2010pt,Gatto:2010qs,Osipov:2007je,Kashiwa:2011js,Johnson:2008vna,Kanemura:1997vi,Alexandre:2000yf,Agasian:2008tb,Preis:2010cq,Miransky:2002rp,Boomsma:2009yk,Shushpanov:1997sf,Cohen:2007bt,Agasian:2001hv,Galilo:2011nh} and lattice simulations \cite{Cea:2005td,Cea:2007yv,Cea:2002wx,Bali:2011qj,Bali:2011uf,Bali:2012zg,Ilgenfritz:2012fw,Aoki:2006we,Nam:2011vn,Bruckmann:2011zx,D'Elia:2010nq,D'Elia:2011zu,Buividovich:2010tn}. 
 Entanglement of effects produced by magnetic field with conventional QGP ones makes it difficult to quantify the role of magnetic field in QGP dynamics. A unique observable is polarization of leptons escaping from QGP, which can be induced only by magnetic field, see \sec{sec:dc}. 

Profound influence of  magnetic field on properties of QGP is truly remarkable. Hopefully, progress in theory will soon be matched by experimental investigations that will eventually discover  properties of QCD at high temperatures and  strong electromagnetic fields.

\acknowledgments
This work  was supported in part by the U.S. Department of Energy under Grant No.\ DE-FG02-87ER40371.



\begin{thebibliography}{80}



\bibitem{Kouveliotou:2003tb} 
  C.~Kouveliotou, R.~C.~Duncan and C.~Thompson,
  Sci.\ Am.\  {\bf 288N2}, 24 (2003).
  
\bibitem  {StrongestMagField}
B.~A.~Boyko {\it at. al.},  Pulsed Power Conference. Digest of Technical Papers. 12th IEEE International {\bf 2}, 746 (1999).



\bibitem{Schwinger:1951nm}
J.~S.~Schwinger,
Phys.\ Rev.\  {\bf 82}, 664 (1951).

\bibitem{Kharzeev:2007jp}
  D.~E.~Kharzeev, L.~D.~McLerran and H.~J.~Warringa,
  Nucl.\ Phys.\  A {\bf 803}, 227 (2008).

\bibitem{Ambjorn:1990jg} 
  J.~Ambjorn and P.~Olesen,
  Phys.\ Lett.\ B {\bf 257}, 201 (1991).

\bibitem{Olesen:2012zb} 
  P.~Olesen,
  arXiv:1207.7045 [hep-ph].
 
\bibitem{Bzdak:2011yy} 
  A.~Bzdak and V.~Skokov,
  Phys.\ Lett.\ B {\bf 710}, 171 (2012).

\bibitem{Skokov:2009qp}
  V.~Skokov, A.~Y.~Illarionov and V.~Toneev,
  Int.\ J.\ Mod.\ Phys.\  A {\bf 24}, 5925 (2009).

\bibitem{Voronyuk:2011jd}
  V.~Voronyuk, V.~D.~Toneev, W.~Cassing, E.~L.~Bratkovskaya, V.~P.~Konchakovski, S.~A.~Voloshin,
  Phys.\ Rev.\  {\bf C83}, 054911 (2011).
  
\bibitem{Deng:2012pc} 
  W.~-T.~Deng and X.~-G.~Huang,
  Phys.\ Rev.\ C {\bf 85}, 044907 (2012).
  
\bibitem{Lappi:2006fp} 
  T.~Lappi and L.~McLerran,
  Nucl.\ Phys.\ A {\bf 772}, 200 (2006)
  [hep-ph/0602189].
  
\bibitem{Blaizot:2012qd} 
  J.~-P.~Blaizot, F.~Gelis, J.~Liao, L.~McLerran and R.~Venugopalan,
  arXiv:1210.6838 [hep-ph].
  
  \bibitem{Jackson's_text}
  J.~D.~Jackson, ``Classical Electrodynamics", 3rd edition.
 
\bibitem{Ding:2010ga}
  H.~T.~Ding, A.~Francis, O.~Kaczmarek, F.~Karsch, E.~Laermann and W.~Soeldner,
  arXiv:1012.4963 [hep-lat].
   

 
\bibitem{Aarts:2007wj}
  G.~Aarts, C.~Allton, J.~Foley, S.~Hands and S.~Kim,
  Phys.\ Rev.\ Lett.\  {\bf 99}, 022002 (2007)
  [arXiv:hep-lat/0703008].
  
\bibitem{Gupta:2003zh}
  S.~Gupta,
  Phys.\ Lett.\  B {\bf 597}, 57 (2004)
  [arXiv:hep-lat/0301006].
  
\bibitem{Bjorken:1982qr} 
  J.~D.~Bjorken,
  Phys.\ Rev.\ D {\bf 27}, 140 (1983).
  
\bibitem{Tuchin:2010vs}
  K.~Tuchin,
  Phys.\ Rev.\  C {\bf 82}, 034904 (2010)
  [Erratum-ibid.\  C {\bf 83}, 039903 (2011)].
  
\bibitem{Dunne:2005sx} 
  G.~V.~Dunne and C.~Schubert,
  Phys.\ Rev.\ D {\bf 72}, 105004 (2005)
  [hep-th/0507174].
  
\bibitem{Wang:1988ct} 
  R.~-C.~Wang and C.~Y.~Wong,
  Phys.\ Rev.\ D {\bf 38}, 348 (1988).
  
\bibitem{Martin:1988gr} 
  C.~Martin and D.~Vautherin,
  Phys.\ Rev.\ D {\bf 40}, 1667 (1989).
  
\bibitem{Kim:2007pm}
  S.~P.~Kim and D.~N.~Page,
  Phys.\ Rev.\ D {\bf 75} (2007) 045013
  [hep-th/0701047].
  
\bibitem{Kluger:1992gb} 
  Y.~Kluger, J.~M.~Eisenberg, B.~Svetitsky, F.~Cooper and E.~Mottola,
  Phys.\ Rev.\ D {\bf 45}, 4659 (1992).

\bibitem{Cooper:1992hw} 
  F.~Cooper, J.~M.~Eisenberg, Y.~Kluger, E.~Mottola and B.~Svetitsky,
  Phys.\ Rev.\ D {\bf 48}, 190 (1993)
  [hep-ph/9212206].

\bibitem{Kluger:1991ib} 
  Y.~Kluger, J.~M.~Eisenberg, B.~Svetitsky, F.~Cooper and E.~Mottola,
  Phys.\ Rev.\ Lett.\  {\bf 67}, 2427 (1991).
  
\bibitem{Tanji:2008ku} 
  N.~Tanji,
  Annals Phys.\  {\bf 324}, 1691 (2009)
  [arXiv:0810.4429 [hep-ph]].
  
\bibitem{Mohapatra:2011ku} 
  R.~K.~Mohapatra, P.~S.~Saumia and A.~M.~Srivastava,
  Mod.\ Phys.\ Lett.\ A {\bf 26}, 2477 (2011).
 
 
\bibitem{LLX}
E.~M.~Lifshitz, L.~P.~Pitaevskiy, ``{\it Physical kinetics}", Pergamon Press, 1981, \S59. 

  
\bibitem{Erkelens:1977}
H. Van Erkelens and W. A. Van Leeuwen, 
Physica A {\bf 89}, 113-126 (1977); ibid.\  {\bf 89}, 225-244 (1977), ibid.\  {\bf 90}, 97-108 (1978).


  
\bibitem{Tuchin:2011jw} 
  K.~Tuchin,
  J.\ Phys.\ G  {\bf 39}, 025010 (2012).

\bibitem{Gribov:1984tu}
  L.~V.~Gribov, E.~M.~Levin and M.~G.~Ryskin,
  Phys.\ Rept.\  {\bf 100}, 1 (1983).
  
  
\bibitem{Blaizot:1987nc}
  J.~P.~Blaizot and A.~H.~Mueller,
  Nucl.\ Phys.\  B {\bf 289}, 847 (1987).

\bibitem{Song:2007fn}
  H.~Song and U.~W.~Heinz,
  Phys.\ Lett.\  B {\bf 658}, 279 (2008)
  [arXiv:0709.0742 [nucl-th]].
  
\bibitem{Romatschke:2007mq}
  P.~Romatschke and U.~Romatschke,
  Phys.\ Rev.\ Lett.\  {\bf 99}, 172301 (2007)
  [arXiv:0706.1522 [nucl-th]].

\bibitem{Dusling:2007gi}
  K.~Dusling and D.~Teaney,
  Phys.\ Rev.\  C {\bf 77}, 034905 (2008)
  [arXiv:0710.5932 [nucl-th]].
  
\bibitem{Romatschke:2009im}
  P.~Romatschke,
  Int.\ J.\ Mod.\ Phys.\  {\bf E19}, 1-53 (2010).
  [arXiv:0902.3663 [hep-ph]].

\bibitem{Weibel}
E.~S.~Weibel, Phys.\ Rev.\ Lett.\ {\bf 2}, 83 (1959).

\bibitem{Mrowczynski:1994xv}
  S.~Mrowczynski,
  Phys.\ Rev.\  {\bf C49}, 2191-2197 (1994).

\bibitem{Arnold:2003zc}
  P.~B.~Arnold, G.~D.~Moore and L.~G.~Yaffe,
  JHEP {\bf 0305}, 051 (2003)
  [arXiv:hep-ph/0302165].
 

\bibitem{Baym:1990uj}
  G.~Baym, H.~Monien, C.~J.~Pethick and D.~G.~Ravenhall,
  Phys.\ Rev.\ Lett.\  {\bf 64}, 1867 (1990).
  
\bibitem{Nikishov:1964zza}
  A.~I.~Nikishov and V.~I.~Ritus,
  Sov.\ Phys.\ JETP {\bf 19}, 529 (1964)
  [Zh.\ Eksp.\ Teor.\ Fiz.\  {\bf 46}, 776 (1964)].
  
\bibitem{Ritus-dissertation}
V.~I.~Ritus, 
J.\ Sov.\ Laser Res.\ {\bf 6}, 497 (1985).

\bibitem{Sokolov:1963zn}
  A.~A.~Sokolov and I.~M.~Ternov,
  Sov.\ Phys.\ Dokl.\  {\bf 8}, 1203 (1964)
  [Dokl.\ Akad.\ Nauk Ser.\ Fiz.\  {\bf 153}, 1052 (1964\ PHDOE,8,1203-1205.1964)].

\bibitem{Berestetsky:1982aq}
  V.~B.~Berestetsky, E.~M.~Lifshitz and L.~P.~Pitaevsky,
 ``Quantum Electrodynamics,'' \S90, 
{\it  Oxford, Uk: Pergamon (1982) 652 P. ( Course Of Theoretical Physics, 4)}


\bibitem{Jackson:1975qi}
  J.~D.~Jackson,
  Rev.\ Mod.\ Phys.\  {\bf 48}, 417 (1976).


\bibitem{Gyulassy:1993hr}
  M.~Gyulassy and X.~n.~Wang,
  Nucl.\ Phys.\  B {\bf 420}, 583 (1994)
  [arXiv:nucl-th/9306003].

\bibitem{Baier:1994bd}
  R.~Baier, Y.~L.~Dokshitzer, S.~Peigne and D.~Schiff,
  Phys.\ Lett.\  B {\bf 345}, 277 (1995)
  [arXiv:hep-ph/9411409].

\bibitem{Shuryak:2002ai}
  E.~V.~Shuryak and I.~Zahed,
  Phys.\ Rev.\  D {\bf 67}, 054025 (2003)
  [arXiv:hep-ph/0207163].

\bibitem{Kharzeev:2008qr}
  D.~E.~Kharzeev,
  arXiv:0806.0358 [hep-ph].

\bibitem{Zakharov:2008uk}
  B.~G.~Zakharov,
  JETP Lett.\  {\bf 88}, 475 (2008)
  [arXiv:0809.0599 [hep-ph]].

 \bibitem{GR}
  I.S. Gradshteyn, I.M. Ryzhik; Alan Jeffrey, Daniel Zwillinger, editors. Table of Integrals, Series, and Products, 7th edition. Academic Press, 2007. 
  
  
\bibitem{Tuchin:2010gx}
  K.~Tuchin,
  Phys.\ Rev.\   C {\bf 83}, 017901 (2011).


\bibitem{Baier:1964}
  V.~N.~Baier and V.~M.~Katkov, 
  Phys.\ Lett.\ A {\bf 25}, 492 (1967).
  
  
\bibitem{Matsui:1986dk}
T.~Matsui and H.~Satz,
Phys.\ Lett.\ B {\bf 178}, 416 (1986).




  
\bibitem{Marasinghe:2011bt} 
  K.~Marasinghe and K.~Tuchin,
  Phys.\ Rev.\ C {\bf 84}, 044908 (2011).
  
\bibitem{Tuchin:2011cg} 
  K.~Tuchin,
  Phys.\ Lett.\ B {\bf 705}, 482 (2011).
  

\bibitem {LL3-113}
L. D. Landau and L. E. Lifshitz, ``Quantum Mechanics Non-Relativistic Theory", Butterworth-Heinemann, 3rd.\ ed., \S113.


\bibitem{Machet:2010yg}
  B.~Machet and M.~I.~Vysotsky,
  Phys.\ Rev.\  D {\bf 83}, 025022 (2011)
  [arXiv:1011.1762 [hep-ph]].
  
  
\bibitem{Popov:1997-A}
  V.~S.~Popov, B.~M.~Karnakov and V.~D.~Mur,
  Phys.\ Lett.\  A {\bf 229}, 306 (1997).
  
\bibitem{Popov:1998aw}
  V.~S.~Popov, B.~M.~Karnakov and V.~D.~Mur,
  Phys.\ Lett.\  A {\bf 250}, 20 (1998).

  
\bibitem{Popov:1998-A}
  V.~S.~Popov, B.~M.~Karnakov and V.~D.~Mur,
  JETP {\bf 86}, 860 (1998).
  
\bibitem {LL3-77}
L. D. Landau and L. E. Lifshitz, ``Quantum Mechanics Non-Relativistic Theory", Butterworth-Heinemann, 3rd.\ ed., \S77.

\bibitem{scott}
T.C. Scott, J. Shertzer, R.A. Moore, 
Phys.\ Rev.\ A {\bf 45}, 4393 (1992).


\bibitem{Keldysh-ioniz}
L. V. Keldysh, 
Sov.\ Phys.\ JETP {\bf 20}  1307 (1965).

\bibitem{popov-review}
V.~S. Popov. Physics Uspekhi, {\bf 47}, 855 (2004).

\bibitem{Kharzeev:2004ey}
  D.~Kharzeev,
  Phys.\ Lett.\  {\bf B633}, 260-264 (2006).
  
\bibitem{Kharzeev:2007tn}
  D.~Kharzeev and A.~Zhitnitsky,
  Nucl.\ Phys.\  A {\bf 797}, 67 (2007)
  [arXiv:0706.1026 [hep-ph]].
  
   
\bibitem{Fukushima:2008xe}
  K.~Fukushima, D.~E.~Kharzeev and H.~J.~Warringa,
  Phys.\ Rev.\  D {\bf 78}, 074033 (2008)
  [arXiv:0808.3382 [hep-ph]].
    
 
\bibitem{Kharzeev:2009fn}
  D.~E.~Kharzeev,
  Annals Phys.\  {\bf 325}, 205-218 (2010).
  [arXiv:0911.3715 [hep-ph]].
  
\bibitem{Basar:2010zd}
  G.~Basar, G.~V.~Dunne and D.~E.~Kharzeev,
  arXiv:1003.3464 [hep-ph].

\bibitem{Asakawa:2010bu}
  M.~Asakawa, A.~Majumder and B.~Muller,
  arXiv:1003.2436 [hep-ph].
  
\bibitem{:2009uh}
  B.~I.~Abelev {\it et al.}  [STAR Collaboration],
  Phys.\ Rev.\ Lett.\  {\bf 103}, 251601 (2009)
  [arXiv:0909.1739 [nucl-ex]].

\bibitem{:2009txa}
  B.~I.~Abelev {\it et al.}  [STAR Collaboration],
  Phys.\ Rev.\  C {\bf 81}, 054908 (2010)
  [arXiv:0909.1717 [nucl-ex]].
  
  
\bibitem{Ajitanand:2010rc}
  N.~N.~Ajitanand, R.~A.~Lacey, A.~Taranenko and J.~M.~Alexander,
  Phys.\ Rev.\ C {\bf 83} (2011) 011901
  [arXiv:1009.5624 [nucl-ex]].
  
  
\bibitem{ITM1}
A.M. Perelomov, V.S. Popov and M.V. TerentÕev, Zh.\ Eksp.\ Tear.\ Fiz.\ {\bf 50} (1966) 1393; {\bf 51} (1966) 309.

\bibitem{ITM2}
V.S. Popov, V.P. Kuznetzov and A.M. Perelomov, Zh.\ Eksp.\ Teor.\ Fiz.\ {\bf 53} (1967) 331.



  
  
\bibitem{Moore:2010jd} 
  G.~D.~Moore and M.~Tassler,
  JHEP {\bf 1102}, 105 (2011)
  [arXiv:1011.1167 [hep-ph]].
  
  
\bibitem{Tuchin:2012mf} 
  K.~Tuchin,
  arXiv:1206.0485 [hep-ph].
    
  \bibitem{Sokolov:1968a}
A. A. Sokolov and I. M. Ternov, ``\emph{Synchrotron radiation}", Pergamon Press, Oxford, (1968).

\bibitem{Baring:1988a}
M.~G.~Baring, 
Mon.\ Not.\ R.\ ast.\ Soc.\ {\bf 235}, 57, (1988). 

\bibitem{Herold:1982a}
H. Herold, H. Ruder and G. Wunner, 
Astron.\ Astrophys.\ {\bf 115}, 90, (1982).

\bibitem{Harding:1987a}
A. K. Harding and R. Preece,
Ap.\ J.\  {\bf 319}, 939, (1987).

\bibitem{Latal:1986a}
H. G. Latal,
Ap.\ J.\ {\bf 309}, 372, (1986).

\bibitem{Adare:2008ab} 
  A.~Adare {\it et al.}  [PHENIX Collaboration],
  Phys.\ Rev.\ Lett.\  {\bf 104}, 132301 (2010).
     
\bibitem{Harding:1986a}
A. K. Harding, 
Ap.\ J.\ {\bf 300}, 167, (1986).

\bibitem{Wunner:1986a}
G. Wunner, J. Paez, H. Herold and H. Ruder,
Astron.\ Astrophys.\ {\bf 170}, 179, (1986)

\bibitem{Wunner:1979a}
G. Wunner,
Phys.\ Rev.\ Lett. {\bf 42}, 79, (1979).
     
     

  

  




  







  
  
\bibitem{Mizher:2010zb}
  A.~J.~Mizher, M.~N.~Chernodub and E.~S.~Fraga,
  Phys.\ Rev.\  D {\bf 82}, 105016 (2010)

\bibitem{Fraga:2008um}
  E.~S.~Fraga and A.~J.~Mizher,
  Nucl.\ Phys.\  A {\bf 820}, 103C (2009)
  
\bibitem{Fraga:2012fs} 
  E.~S.~Fraga and L.~F.~Palhares,
  Phys.\ Rev.\ D {\bf 86}, 016008 (2012).
  
\bibitem{Gatto:2010pt} 
  R.~Gatto and M.~Ruggieri,
  Phys.\ Rev.\ D {\bf 83}, 034016 (2011).
  
\bibitem{Gatto:2010qs} 
  R.~Gatto and M.~Ruggieri,
  Phys.\ Rev.\ D {\bf 82}, 054027 (2010).
  
\bibitem{Osipov:2007je} 
  A.~A.~Osipov, B.~Hiller, A.~H.~Blin and J.~da Providencia,
  Phys.\ Lett.\ B {\bf 650}, 262 (2007).
  
\bibitem{Kashiwa:2011js} 
  K.~Kashiwa,
  Phys.\ Rev.\ D {\bf 83}, 117901 (2011).
  
\bibitem{Johnson:2008vna} 
  C.~V.~Johnson and A.~Kundu,
  JHEP {\bf 0812}, 053 (2008).
  
\bibitem{Kanemura:1997vi} 
  S.~Kanemura, H.~-T.~Sato and H.~Tochimura,
  Nucl.\ Phys.\ B {\bf 517}, 567 (1998).
  
\bibitem{Alexandre:2000yf} 
  J.~Alexandre, K.~Farakos and G.~Koutsoumbas,
  Phys.\ Rev.\ D {\bf 63}, 065015 (2001).
  
\bibitem{Agasian:2008tb} 
  N.~O.~Agasian and S.~M.~Fedorov,
  Phys.\ Lett.\ B {\bf 663}, 445 (2008).
  
\bibitem{Preis:2010cq} 
  F.~Preis, A.~Rebhan and A.~Schmitt,
  JHEP {\bf 1103}, 033 (2011).
  
\bibitem{Gusynin:1995nb} 
  V.~P.~Gusynin, V.~A.~Miransky and I.~A.~Shovkovy,
  Nucl.\ Phys.\ B {\bf 462}, 249 (1996).
  
\bibitem{Miransky:2002rp} 
  V.~A.~Miransky and I.~A.~Shovkovy,
  Phys.\ Rev.\ D {\bf 66}, 045006 (2002).
  
\bibitem{Chernodub:2010qx} 
  M.~N.~Chernodub,
  Phys.\ Rev.\ D {\bf 82}, 085011 (2010)
  [arXiv:1008.1055 [hep-ph]].

\bibitem{Hidaka:2012mz} 
  Y.~Hidaka and A.~Yamamoto,
  arXiv:1209.0007 [hep-ph].

\bibitem{Chernodub:2012zx} 
  M.~N.~Chernodub,
  Phys.\ Rev.\ D {\bf 86}, 107703 (2012)
  [arXiv:1209.3587 [hep-ph]].
  
\bibitem{Boomsma:2009yk} 
  J.~K.~Boomsma and D.~Boer,
  Phys.\ Rev.\ D {\bf 81}, 074005 (2010).
  
\bibitem{Shushpanov:1997sf} 
  I.~A.~Shushpanov and A.~V.~Smilga,
  Phys.\ Lett.\ B {\bf 402}, 351 (1997).
  
\bibitem{Cohen:2007bt} 
  T.~D.~Cohen, D.~A.~McGady and E.~S.~Werbos,
  Phys.\ Rev.\ C {\bf 76}, 055201 (2007).
  
\bibitem{Agasian:2001hv} 
  N.~O.~Agasian,
  Phys.\ Atom.\ Nucl.\  {\bf 64}, 554 (2001)
  [Yad.\ Fiz.\  {\bf 64}, 608 (2001)].
  
\bibitem{Galilo:2011nh} 
  B.~V.~Galilo and S.~N.~Nedelko,
  Phys.\ Rev.\ D {\bf 84}, 094017 (2011)
  [arXiv:1107.4737 [hep-ph]].
  
  
  
  

  
\bibitem{D'Elia:2010nq} 
  M.~D'Elia, S.~Mukherjee and F.~Sanfilippo,
  Phys.\ Rev.\ D {\bf 82}, 051501 (2010).

\bibitem{Cea:2005td} 
  P.~Cea and L.~Cosmai,
  JHEP {\bf 0508}, 079 (2005).
  
\bibitem{Cea:2002wx}
  P.~Cea and L.~Cosmai,
  JHEP {\bf 0302}, 031 (2003).
  
\bibitem{Cea:2007yv} 
  P.~Cea, L.~Cosmai and M.~D'Elia,
  JHEP {\bf 0712}, 097 (2007).

  
\bibitem{Bali:2011qj} 
  G.~S.~Bali \emph{et.\ al}.,
  JHEP {\bf 1202}, 044 (2012).
  
\bibitem{Bali:2011uf} 
  G.~S.~Bali, F.~Bruckmann, G.~Endrodi, Z.~Fodor, S.~D.~Katz, S.~Krieg, A.~Schafer and K.~K.~Szabo,
  PoS LATTICE {\bf 2011}, 192 (2011)
  [arXiv:1111.5155 [hep-lat]].
  
\bibitem{Bali:2012zg} 
  G.~S.~Bali, F.~Bruckmann, G.~Endrodi, Z.~Fodor, S.~D.~Katz and A.~Schafer,
  Phys.\ Rev.\ D {\bf 86}, 071502 (2012)
  [arXiv:1206.4205 [hep-lat]].
  
\bibitem{Ilgenfritz:2012fw} 
  E.~-M.~Ilgenfritz, M.~Kalinowski, M.~Muller-Preussker, B.~Petersson and A.~Schreiber,
  Phys.\ Rev.\ D {\bf 85}, 114504 (2012)
  [arXiv:1203.3360 [hep-lat]].

\bibitem{Aoki:2006we} 
  Y.~Aoki, G.~Endrodi, Z.~Fodor, S.~D.~Katz and K.~K.~Szabo,
  Nature {\bf 443}, 675 (2006)
  [hep-lat/0611014].
  
\bibitem{Nam:2011vn} 
  S.~-i.~Nam and C.~-W.~Kao,
  Phys.\ Rev.\ D {\bf 83}, 096009 (2011)
  [arXiv:1103.6057 [hep-ph]].
  

  
\bibitem{D'Elia:2011zu} 
  M.~D'Elia and F.~Negro,
  Phys.\ Rev.\ D {\bf 83}, 114028 (2011)
  [arXiv:1103.2080 [hep-lat]].
  
\bibitem{Buividovich:2010tn}
  P.~V.~Buividovich, M.~N.~Chernodub, D.~E.~Kharzeev, T.~Kalaydzhyan, E.~V.~Luschevskaya and M.~I.~Polikarpov,
  arXiv:1003.2180 [hep-lat].
  
\bibitem{Bruckmann:2011zx} 
  F.~Bruckmann and G.~Endrodi,
  Phys.\ Rev.\ D {\bf 84}, 074506 (2011)
  [arXiv:1104.5664 [hep-lat]].
 


\end{thebibliography}
\end{document}